\documentclass[aps,prd,twocolumn,reprint,10pt,superscriptaddress,longbibliography,amsfonts,amssymb,amsmath,preprintnumbers,nobibnotes,nofootinbib,noshowpacs,noshowkeyws,tightenlines,floats,notitlepage,final,letterpaper,oneside,noeqsecnum,balancelastpage,flushbottom,noraggedfooter]{revtex4-1}
\pdfoutput=1
\usepackage[utf8]{inputenc}
\usepackage[T1]{fontenc}
\usepackage[british]{babel}
\usepackage[british,cleanlook,printdayon]{isodate}
\usepackage[Symbolsmallscale]{upgreek}
\usepackage{mathtools}
\usepackage{isomath}
\usepackage[protrusion,tracking,kerning,spacing,final,babel]{microtype}
\usepackage{physics}
\usepackage[dvipsnames]{xcolor}
\usepackage[final]{hyperref}
\hypersetup{colorlinks=true,citecolor=MidnightBlue,linkcolor=MidnightBlue,urlcolor=MidnightBlue}
\usepackage{cleveref}

\newcommand{\rme}{\mathrm{e}}
\newcommand{\rmi}{\mathrm{i}}
\newcommand{\rmc}{\mathrm{c}}
\newcommand{\rmk}{\mathrm{k}}
\newcommand{\rmb}{\mathrm{b}}
\newcommand{\Pl}{\mathrm{Pl}}

\begin{document}
\title{Nonlinear gravitational-wave memory from cusps and kinks on cosmic strings}

\author{Alexander~C.~Jenkins}
\email{alexander.jenkins@kcl.ac.uk}
\affiliation{Theoretical Particle Physics and Cosmology Group, Physics Department, King's College London, University of London, Strand, London WC2R 2LS, UK}

\author{Mairi~Sakellariadou}
\email{mairi.sakellariadou@kcl.ac.uk}
\affiliation{Theoretical Particle Physics and Cosmology Group, Physics Department, King's College London, University of London, Strand, London WC2R 2LS, UK}
\affiliation{Theoretical Physics Department, CERN, Geneva, Switzerland}

\begin{abstract}
    The nonlinear memory effect is a fascinating prediction of general relativity (GR), in which oscillatory gravitational-wave (GW) signals are generically accompanied by a monotonically-increasing strain which persists in the detector long after the signal has passed.
    This effect is directly accessible to GW observatories, and presents a unique opportunity to test GR in the dynamical and nonlinear regime.
    In this article we calculate, for the first time, the nonlinear memory signal associated with GW bursts from cusps and kinks on cosmic string loops, which are an important target for current and future GW observatories.
    We obtain analytical waveforms for the GW memory from cusps and kinks, and use these to calculate the `memory of the memory' and other higher-order memory effects.
    These are among the first memory observables computed for a cosmological source of GWs, with previous literature having focused almost entirely on astrophysical sources.
    Surprisingly, we find that the cusp GW signal \emph{diverges} for sufficiently large loops, and argue that the most plausible explanation for this divergence is a breakdown in the weak-field treatment of GW emission from the cusp.
    This shows that previously-neglected strong gravity effects must play an important r\^ole near cusps, although the exact mechanism by which they cure the divergence is not currently understood.
    We show that one possible resolution is for these cusps to collapse to form primordial black holes (PBHs); the kink memory signal does not diverge, in agreement with the fact that kinks are not predicted to form PBHs.
    Finally, we investigate the prospects for detecting memory from cusps and kinks with current and future GW observatories, considering both individual memory bursts and the contribution of many such bursts to the stochastic GW background.
    We find that in the scenario where the cusp memory divergence is cured by PBH formation, the memory signal is strongly suppressed and is not likely to be detected.
    However, alternative resolutions of the cusp divergence may in principle lead to much more favourable observational prospects.
\end{abstract}

\date{\today}
\preprint{KCL-PH-TH/2021-04}
\preprint{CERN-TH-2021-016}

\maketitle
\tableofcontents

\section{Introduction}

The advent of gravitational-wave (GW) astronomy has given us unprecedented observational access to gravity in the dynamical, nonlinear regime, allowing us to test the predictions of Einstein's General Relativity (GR) in this regime as never before~\cite{TheLIGOScientific:2016src,Yunes:2016jcc,LIGOScientific:2020tif}.
One such prediction is the \emph{GW memory effect}~\cite{Braginsky:1986ia,Braginsky:1987gw,Favata:2010zu}, in which the passage of a GW signal causes a permanent offset in the distance between two nearby freely-falling test masses; i.e., the system `remembers' the passage of the GW signal, rather than just oscillating and returning to its initial configuration.

Effects like this are a generic prediction for any GW signal whose source has unbound components that escape to infinity---for example, hyperbolic binary encounters~\cite{Zeldovich:1974gvh,Smarr:1977fy,Turner:1977hyp,Turner:1978zz,Kovacs:1978eu,Bontz:1979zfl}, core-collapse supernovae~\cite{Epstein:1978dv,Turner:1978jj,Burrows:1995bb,Kotake:2005zn,Ott:2008wt}, gamma-ray bursts~\cite{Segalis:2001ns,Sago:2004pn,Birnholtz:2013bea,Akiba:2013qwa}, and particle decay~\cite{Tolish:2014bka,Tolish:2014oda,Allen:2019hnd}.
The resulting displacement is then called the \emph{linear} memory, as it is already present in the GW signal obtained by solving the linearised Einstein equation.
Gravitationally-bound systems, such as the compact binary coalescences (CBCs) observed by LIGO/Virgo~\cite{Harry:2010zz,TheLIGOScientific:2014jea,TheVirgo:2014hva,LIGOScientific:2020ibl}, do not source significant amounts of linear GW memory.\footnote{%
CBCs do generate some linear memory if the final black hole has a non-zero recoil velocity, but this linear memory signal is likely far too weak to be detected~\cite{Favata:2008ti}, even for systems with maximal recoil velocities (the so-called `super-kick' configuration)~\cite{Campanelli:2007cga}.}

In a now-classic paper, Christodoulou~\cite{Christodoulou:1991cr} used rigorous asymptotic methods to show that there is also a \emph{nonlinear} memory effect, in which essentially \emph{any} GW signal is accompanied by a permanent strain offset (the same effect was also discovered soon after by Blanchet and Damour~\cite{Blanchet:1992br} in the very different context of post-Minkowskian theory).
Intuitively, we can understand this by appreciating that linearised GWs carry energy and momentum, and their effective stress-energy can therefore source further perturbations to the spacetime metric beyond linear order.
By heuristically treating the gravitons sourced at linear order as GW sources in their own right, Thorne~\cite{Thorne:1992sdb} showed soon after Christodoulou's paper that the nonlinear memory can be interpreted as the cumulative \emph{linear} memory associated with these gravitons escaping to null infinity.
By searching for nonlinear memory with GW observatories, we can therefore directly probe the `ability of gravity to gravitate' (in the words of reference~\cite{Favata:2009ii}).

While originally derived in the context of classical gravitational physics, it has since been recognised that effects analogous to the nonlinear GW memory are a generic feature of field theories with massless degrees of freedom~\cite{Bieri:2013hqa,Tolish:2014bka,Tolish:2014oda,Susskind:2015hpa,Pate:2017vwa}, and that these memory effects are intimately related to both the asymptotic symmetry group of spacetime and so-called `soft theorems' which govern the production of low-energy massless particles in the quantum theory~\cite{He:2014laa,Strominger:2014pwa,Pasterski:2015tva,Pasterski:2015zua,Kapec:2015vwa,Flanagan:2015pxa,Strominger:2017zoo}.
These deep theoretical links add to the appeal of nonlinear GW memory as an observational probe of the underlying gravitational theory.

With these motivations in mind, there has been a significant effort in recent years to calculate~\cite{Wiseman:1991ss,Favata:2008yd,Favata:2009ii,Favata:2010zu,Pollney:2010hs,Favata:2011qi,Nichols:2017rqr,Talbot:2018sgr,Khera:2020mcz} and search for~\cite{Kennefick:1994nw,vanHaasteren:2009fy,Seto:2009nv,Cordes:2012zz,Madison:2014vca,Wang:2014zls,Arzoumanian:2015cxr,Lasky:2016knh,Yang:2018ceq,Johnson:2018xly,Islo:2019qht,Divakarla:2019zjj,Aggarwal:2019ypr,Hubner:2019sly,Boersma:2020gxx,Ebersold:2020zah,Burko:2020gse} memory waveforms associated with CBCs, as these are the primary target of GW observatories like LIGO/Virgo and pulsar timing arrays (PTAs), and are predicted to be abundant sources of nonlinear memory.

In this article, we focus instead on the nonlinear memory generated by another key GW source: \emph{cosmic strings}~\cite{Kibble:1976sj,Vilenkin:1984ib,Hindmarsh:1994re,Vilenkin:2000jqa}---linelike topological defects which may have formed in a cosmological phase transition in the early Universe due to a spontaneously-broken $U(1)$ symmetry, whose production is a generic prediction of many theories beyond the Standard Model~\cite{Jeannerot:2003qv}.
These strings self-intersect to form closed loops~\cite{Kibble:1982cb,Turok:1984cn}, which oscillate at relativistic speeds and emit strong GW bursts through sharp features called cusps and kinks~\cite{Damour:2001bk}, making them important targets for GW searches.
The amplitude of these GW signals is set by the dimensionless string tension $G\mu$, which is related to the symmetry breaking scale $\eta$ by $G\mu\sim(\eta/m_\Pl)^2$, with $m_\Pl$ the Planck mass.
Searches for cosmic strings with LIGO/Virgo~\cite{Abbott:2009rr,Abbott:2017mem,LIGOScientific:2019vic,Abbott:2021xxi,LIGOScientific:2021nrg} and pulsar timing arrays (PTAs)~\cite{Lasky:2015lej,Blanco-Pillado:2017rnf,Yonemaru:2020bmr} have so far returned only null results,\footnote{%
The NANOGrav Collaboration have recently found evidence for a stochastic process in the pulsar timing residuals of their 12.5-year dataset, with a common spectrum across all pulsars~\cite{Arzoumanian:2020vkk}.
Several authors have pointed out that this is consistent with a stochastic GW background from cosmic strings~\cite{Ellis:2020ena,Blasi:2020mfx,Buchmuller:2020lbh,Bian:2020urb,Blanco-Pillado:2021ygr}.
While this is very exciting, we note that the NANOGrav data do not yet provide conclusive evidence for the quadrupolar cross-correlations between pulsars that one would expect from a GW signal, and further data are thus needed to confidently identify the cause of the common stochastic process.
If the observed signal is indeed due to GWs, it is perfectly consistent with an astrophysical background from inspiralling supermassive black hole binaries, meaning that more exotic interpretations must be explored with caution.
} allowing us to place a conservative upper limit on the string tension of $G\mu\lesssim10^{-11}$ (there is some uncertainty due to the modelling of the cosmic string loop network---the most stringent constraints are at the level of $G\mu\lesssim10^{-15}$~\cite{LIGOScientific:2021nrg}).

There are at least two reasons to expect \emph{a priori} that cosmic strings could be an important source of nonlinear memory.\footnote{%
We note that cosmic string loops can also source significant amounts of \emph{linear} memory by emitting radiation in the underlying matter fields.
These effects are ignored by the Nambu-Goto approximation we adopt here, and can only be resolved by field-theory simulations.
In reality, we expect loops to generate a combination of linear and nonlinear memory, by radiating both matter and GWs.
This has recently been demonstrated for collapsing circular loops, using numerical-relativity/field-theory simulations~\cite{Aurrekoetxea:2020tuw}.}
    \begin{enumerate}
        \item GW memory waveforms tend to be associated with lower frequencies than the primary GW emission that sources them~\cite{Favata:2010zu}.
        This is because the memory grows monotonically over a period of order the duration of the primary signal, which is typically much longer than the oscillation period of the primary signal.
        This means that, e.g., massive stellar binary black holes which merge near the bottom end of the LIGO/Virgo frequency band produce memory signals which are shifted to lower frequencies, and are thus challenging to detect with LIGO/Virgo.
        Cosmic string signals, on the other hand, have durations which are comparable to their oscillation period (i.e. they have a `burst-like' morphology, which is very different to an inspiralling compact binary), meaning that we should expect the resulting memory signals to have power at similar frequencies to the original signal.
        What's more, the cosmic string signals we consider also have significant power at very high frequencies, and it is possible that the hereditary nature of the memory effect could transfer some of this power, enhancing the amplitude of the signal at observable frequencies.
        (A similar idea of `orphan' signals, where the memory emission is detectable even though the primary signal is not, was studied in reference~\cite{McNeill:2017uvq}.)
        \item The angular pattern of the GW memory signal on the sky is typically different to that of the primary GW emission that sources it.
        Cosmic string cusps and kinks emit GWs in narrow beams, meaning that only a very small fraction of all cusps and kinks are oriented such that their GWs can be observed.
        However, if the associated GW memory signal is more broadly distributed on the sphere, this might allow us to observe some of the many cusps and kinks whose beams are not oriented towards us.
    \end{enumerate}
We find that both expectations are borne out by our calculations below: the GW memory from cusps and kinks is indeed emitted in a much broader range of directions than the initial beam, and the memory signal does indeed have a similar frequency profile to the primary signal, with the high-frequency behaviour of the primary GWs playing an important r\^ole in determining the strength of the memory effect.
In fact, we show that these two ingredients lead to a \emph{divergence} in the memory signal from cusps on sufficiently large loops, potentially signifying a breakdown of the standard weak-field approach for calculating the GW signal from cusps.
We attempt to clarify the root cause of this breakdown, and suggest one possible resolution, based on the cusp-collapse scenario of reference~\cite{Jenkins:2020ctp}.
Other resolutions are possible however, and ultimately a fully general-relativistic treatment will be required to understand the true behaviour of cusps.

The remainder of this article is structured as follows.
In section~\ref{sec:nonlinear-gw-memory} we introduce the standard expression for the nonlinear GW memory, equation~\eqref{eq:memory-flux}, and develop from it some useful formulae for calculating the late-time memory (i.e. the total strain offset after the GW signal has passed), equation~\eqref{eq:late-time-memory}, and the frequency-domain memory waveform, equation~\eqref{eq:memory-frequency-domain}.
In section~\ref{sec:cosmic-strings} we briefly recap the standard cusp and kink waveforms as derived in reference~\cite{Damour:2001bk}, which are the other key ingredient of our analysis.
In section~\ref{sec:cusps}, we calculate the nonlinear GW memory signal associated with cusps, obtaining the simple frequency-domain waveform~\eqref{eq:cusp-result}; we show that the total GW energy radiated by this memory signal diverges for Nambu-Goto strings, and regularise this divergence by imposing a cutoff at the scale of the string width $\delta$; we then go on to consider higher-order memory effects (the nonlinear GW memory sourced by the memory GWs themselves) and show that accounting for all such contributions leads to a divergence for large loops, which persists even after applying the string-width regularisation; finally, we show that this divergence is cured if the cusp collapses to form a primordial black hole (PBH), as we recently proposed in reference~\cite{Jenkins:2020ctp}.
In section~\ref{sec:kinks} we repeat the memory calculation for kinks, obtaining the leading-order waveform~\eqref{eq:kink-result}; unlike in the cusp case, the memory signal is strongly suppressed at high frequencies due to interference effects, and no divergence occurs.
In section~\ref{sec:detection-prospects} we study the observable consequences of our memory waveforms for GW searches, under the assumption that the cusp memory divergence is cured by PBH formation; we find that in this scenario the memory is strongly suppressed, and is beyond the reach of current or planned GW searches.
Finally, we summarise our results in section~\ref{sec:summary}.
We discuss the cause of the higher-order memory divergence in appendix~\ref{sec:rhdot}, and give some technical details of the angular distribution of the memory GWs from cusps and kinks in appendices~\ref{sec:iota-polynomials} and~\ref{sec:K_n}.
We set $c=1$ throughout, but keep $G$ and $\hbar$ explicit.

\section{Nonlinear GW memory}
\label{sec:nonlinear-gw-memory}

We work in terms of the complex GW strain,
    \begin{equation}
        h(t,\vb*r)\equiv h_+-\rmi h_\times=\frac{1}{2}(e^+_{ij}-\rmi e^\times_{ij})h_{ij},
    \end{equation}
    where $t$ is the retarded time, $\vb*r$ is a 3-vector pointing from the source to the observer, and $e^A_{ij}$ are the transverse-traceless (TT) polarisation tensors.
We distinguish between the primary, oscillatory strain signal (sourced at linear order) and the additional strain due to the memory effect by writing these as $h^{(0)}$ and $h^{(1)}$ respectively (reserving $h^{(n)}$ with $n\ge2$ for the `memory of the memory' and other higher-order memory contributions, which we discuss in sections~\ref{sec:cusp-higher-order} and~\ref{sec:kink-higher-order}).
The leading nonlinear memory correction term can be written in gauge-invariant form as~\cite{Thorne:1992sdb,Favata:2010zu}
    \begin{equation}
    \label{eq:memory-flux}
        h^{(1)}(t,\vb*r)=\frac{2G}{r}\int_{-\infty}^t\dd{t'}\int_{S^2}\dd[2]{\vu*r'}\frac{(e^+_{ij}-\rmi e^\times_{ij})\hat{r}'_i\hat{r}'_j}{1-\vu*r\vdot\vu*r'}\frac{\dd[3]{E^{(0)}_\mathrm{gw}}}{\dd{t'}\dd[2]{\vu*r'}}.
    \end{equation}
We see immediately that since the integral is over the (non-negative) GW energy flux from the source, $\dd[3]{E^{(0)}_\mathrm{gw}}/\dd{t}\dd[2]{\vu*r}$, we generically obtain a nonzero memory correction which grows monotonically with time while the source is `on'.
By projecting onto the GW polarisation tensors we ensure that only the TT part of the angular integral contributes; this TT part vanishes if the emitted flux is exactly isotropic, but is generically non-vanishing for anisotropic emission.

Note that equation~\eqref{eq:memory-flux} corresponds to just one of many quadratic terms that appear on the right-hand side of the relaxed Einstein field equation~\cite{Favata:2008yd}.
While this term has received particular attention in the literature due to its hereditary nature and its pleasing intuitive interpretation as the nonlinear gravitational counterpart to the linear GW memory effect, the other quadratic terms will also give nonlinear corrections to the GWs emitted by cosmic strings.
Our results here therefore represent just one particular nonlinear contribution to these GW signals.
Indeed, since we argue below that nonlinear effects are important near cusps on large cosmic string loops, it would be interesting to explore these additional contributions further in future work.

Returning to equation~\eqref{eq:memory-flux}, we write the polarisation tensors as~\cite{Maggiore:1900zz}
    \begin{equation}
        e^+_{ij}=\hat{\theta}_i\hat{\theta}_j-\hat{\phi}_i\hat{\phi}_j,\qquad e^\times_{ij}=\hat{\theta}_i\hat{\phi}_j+\hat{\phi}_i\hat{\theta}_j,
    \end{equation}
    where $\vu*\theta$ and $\vu*\phi$ are the standard spherical polar unit vectors orthogonal to $\vu*r$.
It is then straightforward to show that
    \begin{equation}
        (e^+_{ij}-\rmi e^\times_{ij})\hat{r}'_i\hat{r}'_j=[(\vu*\theta-\rmi\vu*\phi)\vdot\vu*r']^2.
    \end{equation}
We can also rewrite the energy flux~\eqref{eq:memory-flux} in terms of the linear GW strain of the source, using the Isaacson formula~\cite{Isaacson:1967zz,Isaacson:1968zza}
    \begin{equation}
    \label{eq:flux}
        \frac{\dd[3]{E_\mathrm{gw}}}{\dd{t}\dd[2]{\vu*r}}=\frac{\ev*{|r\dot{h}|^2}}{16\uppi G},
    \end{equation}
    where the dot denotes a time derivative.
The leading GW memory term then becomes
    \begin{equation}
    \label{eq:memory-strain}
        h^{(1)}(t,\vb*r)=\int_{-\infty}^t\frac{\dd{t'}}{2r}\int_{\vu*r'}|r\dot{h}^{(0)}(t',\vb*r')|^2,
    \end{equation}
    where we have introduced the shorthand
    \begin{equation}
    \label{eq:integral-shorthand}
        \int_{\vu*r'}[\cdots]\equiv\int_{S^2}\frac{\dd[2]{\vu*r'}}{4\uppi}\frac{[(\vu*\theta-\rmi\vu*\phi)\vdot\vu*r']^2}{1-\vu*r\vdot\vu*r'}[\cdots],
    \end{equation}
    for brevity.
The temporal average $\ev{\cdots}$ is necessary in equation~\eqref{eq:flux} to give a gauge-invariant result, but is removed due to the time integral in equation~\eqref{eq:memory-strain}.
Throughout we refer to the oscillatory strain $h^{(0)}$ on the RHS that sources the memory as the `primary' GW emission.

\subsection{Late-time memory}

One drawback of equation~\eqref{eq:memory-strain} is that it is written in terms of the time-domain primary strain, whereas the cosmic string waveforms that we want to consider are much more naturally expressed in the frequency domain.
This problem disappears when we consider the \emph{late-time memory}, i.e. the total strain offset caused by the cusp,
    \begin{equation}
        \Updelta h^{(1)}\equiv\lim_{t\to\infty}h^{(1)}(t),
    \end{equation}
    as the time integral in equation~\eqref{eq:memory-strain} is then over the entire real line, and we can thus use Parseval's theorem to write
    \begin{equation}
    \label{eq:late-time-memory}
        \Updelta h^{(1)}(\vb*r)=\frac{2\uppi^2}{r}\int_\mathbb{R}\dd{f}\int_{\vu*r'}|rf\tilde{h}^{(0)}(f,\vb*r')|^2,
    \end{equation}
    where $\tilde{h}^{(0)}$ is the Fourier transform of the primary strain signal, and the extra factor of frequency comes from the time derivative on the strain.

\subsection{Frequency-domain memory waveforms}

The late-time memory~\eqref{eq:late-time-memory} is useful for giving a sense of the total size of the memory effect, but in many cases is not directly observable.
For example, the test masses in ground-based GW interferometers like LIGO and Virgo are not freely-falling in the plane of the interferometer arms; they are acted upon by feedback control systems at low frequencies to mitigate seismic noise~\cite{Saulson:1995zi}.
These low-frequency forces mean that the test masses cannot sustain a permanent displacement after the GW has passed.
However, the `ramping up' of the memory signal from zero at early times to $\Updelta h^{(1)}$ at late times can be measured if it contains power in the sensitive frequency band of the interferometer.

We are therefore interested in calculating the full memory signal in the frequency domain.
The simplest way of doing this is to use the result\footnote{%
One can show this by noting that $\int_{-\infty}^t\dd{t'}g(t')$ is just the convolution of $g(t)$ with the Heaviside step function $\Theta(t)$.
Since $\mathcal{F}[\Theta]=(1/2)\delta(f)-\rmi/(2\uppi f)$, the result follows from the convolution theorem, $\mathcal{F}[\Theta*g]=\mathcal{F}[\Theta]\mathcal{F}[g]$.}
    \begin{equation}
    \label{eq:step-function-fourier}
        \mathcal{F}\qty[\int_{-\infty}^t\dd{t'}g(t')]=\frac{1}{2}\tilde{g}(0)\delta(f)-\frac{\rmi}{2\uppi f}\tilde{g}(f)
    \end{equation}
    for a general function $g(t)$, where $\mathcal{F}[\cdots]$ denotes the Fourier transform, and $\tilde{g}=\mathcal{F}[g]$.
Applying this to equation~\eqref{eq:memory-strain}, we obtain
    \begin{equation}
        \tilde{h}^{(1)}(f)=\frac{1}{2}\Updelta h^{(1)}\delta(f)-\frac{\rmi}{2\uppi f}\int_\mathbb{R}\frac{\dd{t}}{2r}\int_{\vu*r'}\rme^{-2\uppi\rmi ft}|r\dot{h}^{(0)}|^2.
    \end{equation}
Note that we can neglect the term proportional to $\delta(f)$, as this only contributes at $f=0$, and we are interested here in frequencies which are accessible to GW experiments [this zero-frequency term is still captured in equation~\eqref{eq:late-time-memory}].
By replacing each factor of $\dot{h}^{(0)}$ with its Fourier transform and massaging the resulting expression, we find
    \begin{align}
    \begin{split}
    \label{eq:memory-frequency-domain}
        &\tilde{h}^{(1)}(f)=\\
        &-\frac{\rmi\uppi}{rf}\int_\mathbb{R}\dd{f'}\int_{\vu*r'}f'(f'-f)r^2\tilde{h}^{(0)}(f',\vu*r')\tilde{h}^{(0)*}(f'-f,\vu*r').
    \end{split}
    \end{align}
This is the simplest way of calculating the frequency-domain memory using only the primary frequency-domain signal $\tilde{h}^{(0)}$.

\section{Cosmic string burst waveforms}
\label{sec:cosmic-strings}

We consider cosmic string loops in the Nambu-Goto (i.e. zero-width) approximation (although at some points below it is necessary to reintroduce a finite string width $\delta$).
The Nambu-Goto equations of motion imply that these loops should generically develop sharp features called `cusps', where the tangent vector of a point on the loop instantaneously becomes null, generating a strong burst of GWs.
String intercommutations also generically lead to discontinuities in the loop's tangent vector called `kinks'.

Nambu-Goto loops oscillate at a frequency $2/\ell$ inversely related to their invariant length $\ell$, with their motion sourcing GWs at integer multiples of this base frequency.
Since this frequency is typically many orders of magnitude below the frequency band relevant to ground-based interferometers like LIGO/Virgo, we are typically interested in very high-order harmonics of the loops.
In this high-frequency regime, the GW emission of cosmic string loops is dominated by cusps and kinks.
(High-frequency GWs can also be produced through kink-kink collisions~\cite{Ringeval:2017eww}, but this emission mode is exactly isotropic and so does not contribute to the memory effect.)
The asymptotic high-frequency waveforms for cusps and kinks on a loop with dimensionless tension $G\mu$ and invariant length $\ell$ can be approximated by~\cite{Damour:2001bk}
    \begin{subequations}
    \begin{align}
    \label{eq:cusp-waveform}
        \tilde{h}^{(0)}_\rmc(f,\vu*r)&\simeq A_\rmc\frac{G\mu\ell^{2/3}}{r|f|^{4/3}}\Theta(\vu*r\vdot\vu*r_\rmc-\cos\theta_\rmb)\Theta(|f|-2/\ell),\\
    \label{eq:kink-waveform}
        \tilde{h}^{(0)}_\rmk(f,\vu*r)&\simeq A_\rmk\frac{G\mu\ell^{1/3}}{r|f|^{5/3}}\Theta(\vu*r\vdot\vu*r_\rmk-\cos\theta_\rmb)\Theta(|f|-2/\ell),
    \end{align}
    \end{subequations}
    where the dimensionless pre-factors are
    \begin{align}
    \begin{split}
        A_\rmc&\equiv\frac{8}{\Gamma^2(1/3)}\qty(\frac{2}{3})^{2/3}\approx0.8507,\\
        A_\rmk&\equiv\frac{\sqrt{A_\rmc}}{\uppi}\approx0.2936.
    \end{split}
    \end{align}
The fact that the frequency-domain waveforms~\eqref{eq:cusp-waveform} and~\eqref{eq:kink-waveform} are real and even around $f=0$ implies that the corresponding time-domain waveforms are real, and therefore that the cusps and kinks are linearly polarised.
The waveforms are nonzero only for frequencies above the base mode of the loop, $2/\ell$, and only if the observer lies inside a small beam with opening angle
    \begin{equation}
    \label{eq:beam-angle}
        \theta_\rmb(f)\simeq\frac{2^{2/3}}{3^{1/6}}(|f|\ell)^{-1/3}.
    \end{equation}
For cusps, there is a single well-defined beaming direction $\vu*r_\rmc$, whereas for kinks the beam is over a one-dimensional `fan' of directions on the sphere, and $\vu*r_\rmk$ represents the direction in this fan which is closest to the observer's line of sight $\vu*r$.

\begin{figure*}[t!]
    \includegraphics[width=\textwidth]{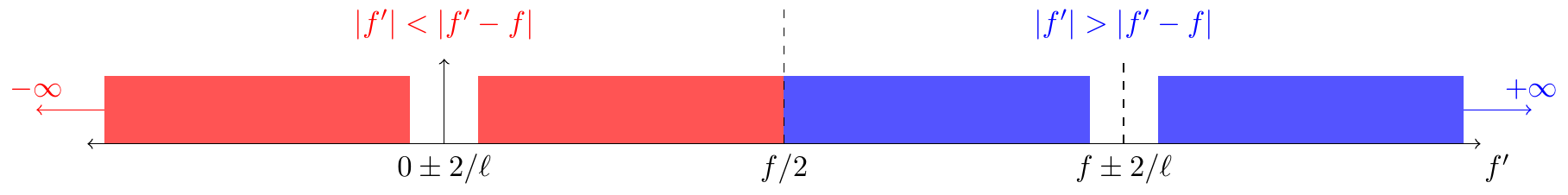}
    \caption{%
    Schematic illustration of the different contributions to $\tilde{h}^{(1)}_\rmc(f)$ from the integral over $f'$ in equation~\eqref{eq:memory-frequency-domain}.
    By introducing a dimensionless dummy variable $u\equiv f'/|f|$ we obtain the two integrals shown in equation~\eqref{eq:cusp-all-frequencies}, one corresponding to the finite interior region $2/\ell<f'<f-2/\ell$, and the other corresponding to the two semi-infinite exterior regions $f'<-2/\ell$ and $f'>f+2/\ell$.
    }
    \label{fig:fprime}
\end{figure*}

\section{Memory from cusps}
\label{sec:cusps}

We begin by calculating the nonlinear memory from cusps, inserting the primary waveform~\eqref{eq:cusp-waveform} into equations~\eqref{eq:late-time-memory} and~\eqref{eq:memory-frequency-domain} to obtain the late-time memory and frequency-domain waveform, and then iterating this process to obtain higher-order memory corrections.

\subsection{Beaming effects}

The anisotropic beaming of the GWs from cusps is what gives rise to a nonzero memory effect (due to its nonzero TT projection), and is captured in the spherical integral
    \begin{equation}
    \label{eq:spherical-integral-cusp}
        \int_{\vu*r'}\Theta(\vu*r_\rmc\vdot\vu*r'-\cos\theta_\rmb),
    \end{equation}
    where we are using the shorthand~\eqref{eq:integral-shorthand}.
To compute this integral, it is convenient to define polar coordinates $\vu*r'=(\theta',\phi')$ such that the North pole $\theta'=0$ coincides with the centre of the beam $\vu*r_\rmc$.
The integrand then only has support for $\theta'\in[0,\theta_\rmb]$, so that equation~\eqref{eq:spherical-integral-cusp} becomes
    \begin{align}
    \begin{split}
    \label{eq:spherical-integral-cusp-result}
        &\int_{\vu*r'}\Theta(\vu*r_\rmc\vdot\vu*r'-\cos\theta_\rmb)\\
        &=\int_0^{\theta_\rmb}\dd{\theta'}\frac{\sin\theta'}{2}\int_0^{2\uppi}\frac{\dd{\phi'}}{2\uppi}\frac{[(\vu*\theta-\rmi\vu*\phi)\vdot\vu*r']^2}{1-\vu*r\vdot\vu*r'}\\
        &=2\frac{\cos^4\frac{\iota}{2}}{\sin^2\iota}\qty[(1-\cos\theta_\rmb)\cos\theta_\rmb-\frac{1}{2}\cos\iota\sin^2\theta_\rmb],
    \end{split}
    \end{align}
    where $\iota\equiv\cos^{-1}\vu*r_\rmc\vdot\vu*r$ is the inclination of the beam to the observer's line of sight.
In the high-frequency regime where the primary cusp and kink waveforms are valid, the beam angle is very small, so we expand equation~\eqref{eq:spherical-integral-cusp-result} to leading order in $\theta_\rmb$ to obtain
    \begin{equation}
    \label{eq:spherical-integral-cusp-result-small-beam}
        \int_{\vu*r'}\Theta(\vu*r_\rmc\vdot\vu*r'-\cos\theta_\rmb)\simeq\frac{\theta_\rmb^2}{4}(1+\cos\iota).
    \end{equation}

There are a few remarks worth making about equation~\eqref{eq:spherical-integral-cusp-result-small-beam}.
First, we note that it assumes $\iota\gg\theta_\rmb$; when instead the inclination is much smaller than the beaming angle, $\iota\ll\theta_\rmb$, the integral~\eqref{eq:spherical-integral-cusp} drops to zero.
The fact that the result~\eqref{eq:spherical-integral-cusp-result-small-beam} is purely real, despite the integrand being complex, shows that the memory strain is linearly polarised, much like the primary cusp and kink waveforms.
Geometrically, the $\theta_\rmb^2/4$ factor represents the fraction of the sphere taken up by the beam, while the $(1+\cos\iota)$ factor shows how the strength of the memory effect varies with inclination.
In particular, we notice that the memory strain is nonzero when the observer lies outside of the beam, $\iota>\theta_\rmb$.
In fact, the observed memory strain vanishes only when the beam is face-on ($\iota\ll\theta_\rmb$) or face-off ($\iota=\uppi$), as in both cases the angular pattern of the primary GW emission is isotropic around the line of sight.

It is interesting to note that this angular pattern---a broad $\sim(1+\cos\iota)$ distribution, except at very small inclinations where the memory signal drops to zero---is exactly the same as that of the \emph{linear} GW memory generated by the ejection of an ultrarelativistic `blob' of matter from a massive object along a fixed axis~\cite{Segalis:2001ns}.
Intuitively, this makes complete sense: the setup here is essentially the same, except that the `blob' is replaced by a burst of GWs.

\subsection{Late-time memory}

Now that we have computed the angular integral~\eqref{eq:spherical-integral-cusp}, it is straightforward to obtain the late-time memory from the cusp.
Inserting equations~\eqref{eq:cusp-waveform} and~\eqref{eq:beam-angle} into equation~\eqref{eq:late-time-memory} and integrating over frequency, we find
    \begin{equation}
    \label{eq:cusp-late-time-memory-result}
        \Updelta h^{(1)}_\rmc=2\times3^{2/3}(\uppi A_\rmc G\mu)^2(1+\cos\iota)\frac{\ell}{r}.
    \end{equation}
Inserting the numerical factors, this has a maximum value of $\Updelta h^{(1)}_\rmc\approx59.43\times(G\mu)^2\ell/r$ for nearly-face-on cusps $\iota\gtrsim0$, and smoothly tapers to zero for face-off cusps $\iota=\uppi$.
We note that equation~\eqref{eq:cusp-late-time-memory-result} should be taken with a pinch of salt, as it is sensitive to the low-frequency regime where the primary waveform is less accurate; nonetheless, we expect this to give a reasonable estimate of the magnitude of the memory effect.

\subsection{Full waveform}
\label{sec:cusp-full-waveform}

We now calculate the full frequency-domain cusp memory waveform by inserting equations~\eqref{eq:cusp-waveform} and~\eqref{eq:beam-angle} into equation~\eqref{eq:memory-frequency-domain}.
In doing so, we must be careful to correctly account for the behaviour of the two frequency arguments $f'$ and $f'-f$; in particular, the integrand is only nonzero when both of these arguments have magnitude greater than $2/\ell$, and the size of the beam angle $\theta_\rmb$ must always be set by whichever of the two arguments has greater magnitude (as this corresponds to a smaller, more restrictive beam).
These different contributions are illustrated in figure~\ref{fig:fprime} for the case where $f$ is positive.
Assuming that $|f|>4/\ell$, we find
    \begin{align}
    \begin{split}
    \label{eq:cusp-all-frequencies}
        \tilde{h}^{(1)}_\rmc(f)&=-\frac{\rmi\Updelta h^{(1)}_\rmc}{2^{2/3}3\uppi\ell^{1/3}f|f|^{1/3}}\\
        &\times\qty[\int_{2/|f|\ell}^\infty\frac{\dd{u}}{u^{1/3}(1+u)}-\int_{2/|f|\ell}^{1/2}\frac{\dd{u}}{u^{1/3}(1-u)}],
    \end{split}
    \end{align}
    where $u$ is just a dimensionless dummy variable.
We can simplify this by taking $|f|\gg4/\ell$, as this is the regime where the primary waveforms are valid; in this high-frequency limit, both integrals can be evaluated analytically.
Inserting equation~\eqref{eq:cusp-late-time-memory-result}, the final result is
    \begin{equation}
    \label{eq:cusp-result}
        \tilde{h}^{(1)}_\rmc(f)\simeq-\rmi B_c\frac{(G\mu)^2\ell^{2/3}}{rf|f|^{1/3}}(1+\cos\iota)\Theta(|f|-2/\ell),
    \end{equation}
    with a numerical prefactor,
    \begin{align}
    \begin{split}
        B_\rmc&\equiv\uppi A_\rmc^2(3/2)^{2/3}\qty[\frac{4\uppi}{3\sqrt{3}}-{}_2F_1(\tfrac{2}{3},\tfrac{2}{3};\tfrac{5}{3};-1)]\\
        &\approx4.764,
    \end{split}
    \end{align}
    where ${}_2F_1$ is a hypergeometric function.
While we assumed $|f|>4/\ell$ in order to obtain equation~\eqref{eq:cusp-all-frequencies}, we have checked that our final expression~\eqref{eq:cusp-result} underestimates the true memory signal in the region $2/\ell<|f|<4/\ell$ (assuming the primary signal is accurate at these frequencies, which we expect to be the case to within an order of magnitude if one ignores the low-frequency motion not associated with the cusp), so we can safely leave the low-frequency cutoff at $2/\ell$ as in the primary waveform.
The simple expression~\eqref{eq:cusp-result} thus gives a conservative but generally accurate model of the time-varying part of the cusp memory waveform at all frequencies greater that the fundamental mode of the loop, while the zero-frequency offset is described by equation~\eqref{eq:cusp-late-time-memory-result}.

\begin{figure}[t!]
    \includegraphics[width=0.48\textwidth]{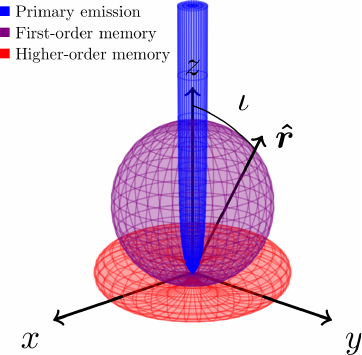}
    \caption{%
    A cartoon illustration of the angular distribution of the energy radiated by a cusp.
    The primary emission (blue) is concentrated in a narrow beam of width $\theta_\rmb\sim(f\ell)^{-1/3}$, while the first-order memory emission (violet) is proportional to $(1+\cos\iota)^2$, and the second-order memory emission (red) is proportional to $\sin^4\iota$.
    }
    \label{fig:cusp-memory-angular-patterns}
\end{figure}

Comparing equation~\eqref{eq:cusp-result} with the primary cusp waveform~\eqref{eq:cusp-waveform}, we see that they are remarkably similar to each other, with both being given by the same simple frequency power law $\sim f^{-4/3}$ and the same dependence on the loop length $\ell$ (the latter being required by dimensional arguments).
There are, however, some important differences:
    \begin{enumerate}
        \item The memory GW emission has broad support on the sphere, while the primary waveform only has support inside a narrow beam (see figure~\ref{fig:cusp-memory-angular-patterns}).
        \item The memory waveform is suppressed by an additional power of $G\mu$ (this makes intuitive sense, given that it is a nonlinear effect sourced by the primary GW emission).
        \item The numerical constant in front of the memory waveform, $B_\rmc$, is an order of magnitude larger than that in front of the primary waveform, $A_\rmc$.
    \end{enumerate}
The first point is particularly crucial, as it lies at the heart of the divergent behaviour that we investigate in sections~\ref{sec:energy-divergence} and~\ref{sec:cusp-higher-order}.

\subsection{Time-domain waveform near the arrival time}

\begin{figure}[t!]
    \includegraphics[width=0.48\textwidth]{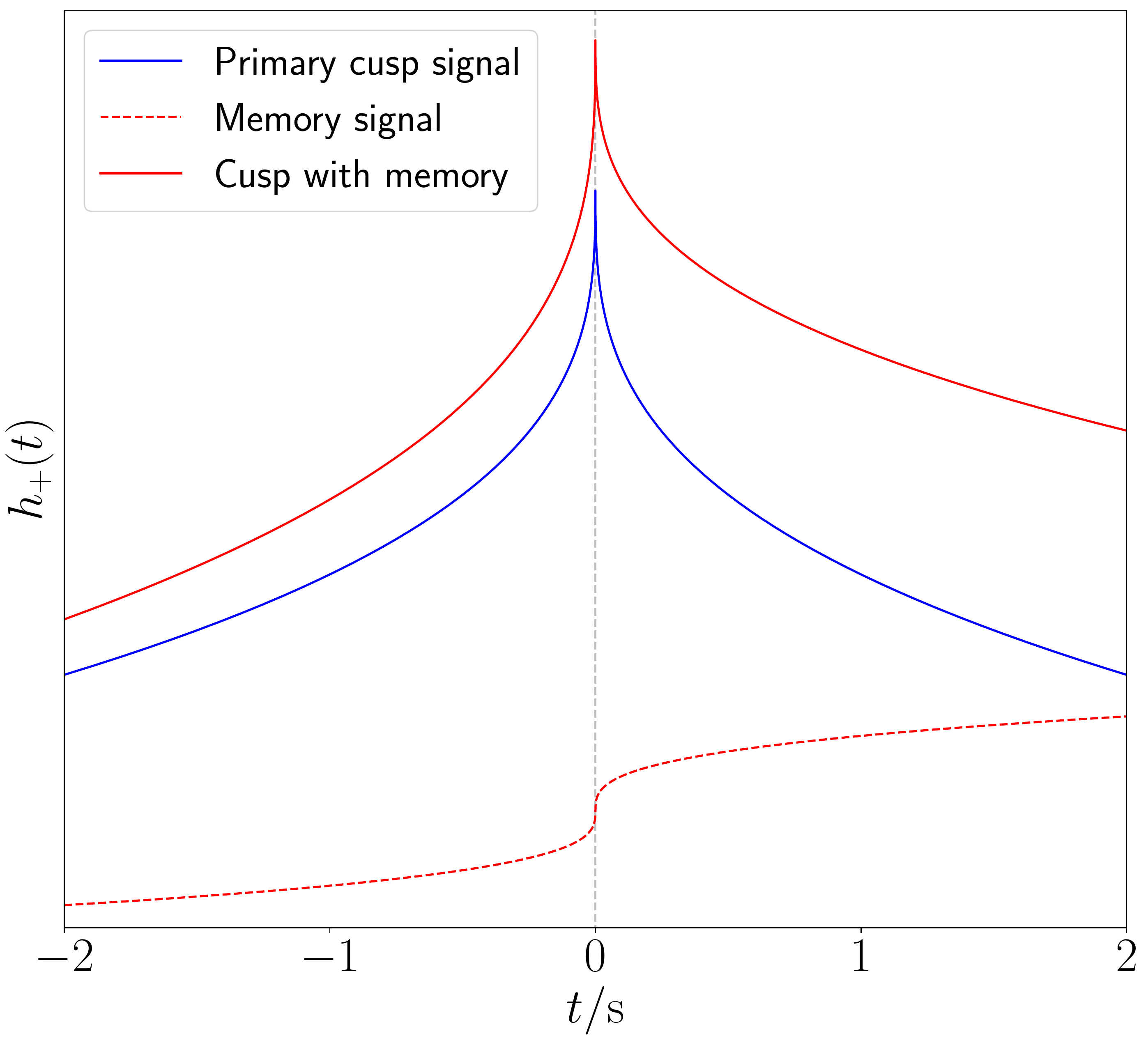}
    \caption{%
    The time-domain GW strain $h_\rmc(t)$ from a cusp, with and without the leading-order memory contribution~\eqref{eq:cusp-time-domain-approx}.
    The memory is exaggerated by a factor of $\sim1/G\mu$ here to make it visible.
    For small inclinations $\iota<\theta_\rmb$ the observer lies within the cusp's beam, and sees the memory superimposed on the primary cusp signal (solid red line).
    If the inclination is very small, $\iota\ll\theta_\rmb$, then the memory vanishes and only the primary cusp signal is observable (solid blue line).
    In most cases however, the observer lies outside of the beam, and only the memory is observable (red dashed line).
    Note that the higher-order memory contributions (order $n\ge2$) are \emph{not} shown here; these would look like step functions in the time domain, with height that either diverges rapidly with $n$ [`large' loops, $\ell\gtrsim\delta/(G\mu)^3$] or converges so rapidly that the contribution to the total signal is negligible [`small' loops, $\ell\lesssim\delta/(G\mu)^3$].
    }
    \label{fig:cusp-td-waveform}
\end{figure}

While we have focused on deriving the memory signal in the frequency domain, we can inverse-Fourier-transform equation~\eqref{eq:cusp-result} to obtain a simple closed-form expression in the time domain,
    \begin{align}
    \begin{split}
        \label{eq:cusp-time-domain-approx}
        h^{(1)}_\rmc(t)-h^{(1)}_\rmc(t_0)\simeq&-2^{5/3}3\uppi B_\rmc\frac{(G\mu)^2}{r}(1+\cos\iota)\\
        &\times(t-t_0)\qty[1-\frac{\Gamma(2/3)}{(4\uppi|t-t_0|/\ell)^{2/3}}],
    \end{split}
    \end{align}
    where $\Gamma(z)$ is Euler's Gamma function, and $\Gamma(2/3)\approx1.354$.
(Here we have re-introduced the time of arrival of the primary cusp signal, $t_0$, rather than setting it to zero.)
This time-domain waveform is shown in figure~\ref{fig:cusp-td-waveform}.
Note that this is real, which means that the signal is linearly $+$-polarised, just like the primary waveform $h^{(0)}$.

It is important to note that since equation~\eqref{eq:cusp-result} is only valid for high frequencies $f\gg2/\ell$, equation~\eqref{eq:cusp-time-domain-approx} must only be valid for a short duration $|t-t_0|\ll\ell$ around the arrival time.
However, for typical loop sizes this `short duration' is actually much longer than the relevant observational timescale (typically a few seconds), so the approximation in equation~\eqref{eq:cusp-time-domain-approx} may be a useful one.

\subsection{Radiated energy}
\label{sec:radiated-energy}

Given a frequency-domain waveform $\tilde{h}(f)$, we can calculate the corresponding (one-sided, logarithmic) GW energy flux spectrum,
    \begin{equation}
        \frac{\dd[3]{E_\mathrm{gw}}}{\dd{(\ln f)}\dd[2]{\vu*r}}=\frac{\uppi r^2f^3}{4G}\qty(|\tilde{h}(f)|^2+|\tilde{h}(-f)|^2).
    \end{equation}
In the context of radiation from a cosmic string loop, it is helpful to normalise this with respect to the total energy of the loop, $\mu\ell$.
We therefore define a dimensionless energy spectrum,
    \begin{equation}
    \label{eq:dimensionless-energy}
        \epsilon(f,\vu*r)\equiv\frac{1}{\mu\ell}\frac{\dd[3]{E_\mathrm{gw}}}{\dd{(\ln f)}\dd[2]{\vu*r}}.
    \end{equation}
For the primary cusp waveform~\eqref{eq:cusp-waveform}, this is given by
    \begin{equation}
        \epsilon^{(0)}_\rmc(f,\vu*r)\simeq\frac{\uppi}{2}A_\rmc^2G\mu(f\ell)^{1/3}\Theta(\theta_\rmb-\iota)\Theta(f-2/\ell),
    \end{equation}
    while for the cusp memory~\eqref{eq:cusp-result}, we find
    \begin{equation}
        \epsilon^{(1)}_\rmc(f,\vu*r)\simeq\frac{\uppi}{2}B_\rmc^2(G\mu)^3(f\ell)^{1/3}(1+\cos\iota)^2\Theta(f-2/\ell).
    \end{equation}
The fact that $\tilde{h}^{(0)}_\rmc$ is purely real while $\tilde{h}^{(1)}_\rmc$ is purely imaginary means that there is no coherent cross-energy between the two contributions, so the total energy is just $\epsilon^{(0)}_\rmc+\epsilon^{(1)}_\rmc$.

For observers in the beaming direction the energy spectra of the primary and memory signals scale in the exact same way with frequency, but with a much smaller coefficient for the memory signal.
However, the picture changes drastically when integrating the spectra over the sphere to compute the total emission, as the primary waveform is then suppressed by a factor of $\theta_\rmb^2\sim(f\ell)^{-2/3}$.
We write these isotropically-averaged spectra as
    \begin{equation}
    \label{eq:dimensionless-energy-isotropic}
        \bar{\epsilon}(f)\equiv\int_{S^2}\dd[2]{\vu*r}\epsilon(f,\vu*r),
    \end{equation}
    such that
    \begin{align}
    \begin{split}
    \label{eq:cusp-isotropic-energy}
        \bar{\epsilon}^{(0)}_\rmc(f)&\simeq\qty(\frac{2}{3})^{1/3}(\uppi A_\rmc)^2G\mu(f\ell)^{-1/3}\Theta(f-2/\ell),\\
        \bar{\epsilon}^{(1)}_\rmc(f)&\simeq\frac{8}{3}(\uppi B_\rmc)^2(G\mu)^3(f\ell)^{1/3}\Theta(f-2/\ell).
    \end{split}
    \end{align}
We see that the isotropic energy spectrum due to the memory emission is blue-tilted (i.e. grows with frequency), while the primary spectrum is red-tilted.
This means that the memory emission dominates at very high frequencies,
    \begin{align}
    \begin{split}
        f&>\frac{3}{16\ell}\qty(\frac{A_\rmc}{B_\rmc G\mu})^3\\
        &\approx10^{22}\,\mathrm{Hz}\times\qty(\frac{\ell}{\mathrm{pc}})^{-1}\qty(\frac{G\mu}{10^{-11}})^{-3}.
    \end{split}
    \end{align}

\subsection{Ultraviolet divergence of the radiated energy}
\label{sec:energy-divergence}

The total fraction of the loop's energy radiated by the cusp can be found by integrating equation~\eqref{eq:cusp-isotropic-energy},
    \begin{equation}
    \label{eq:cusp-total-energy}
        \mathcal{E}\equiv\int_0^\infty\frac{\dd{f}}{f}\bar{\epsilon}(f).
    \end{equation}
For the primary cusp signal, this gives
    \begin{equation}
    \label{eq:cusp-total-energy-primary}
        \mathcal{E}^{(0)}_\rmc\simeq3^{2/3}(\uppi A_\rmc)^2G\mu\ll1.
    \end{equation}
For the memory signal, however, the integral~\eqref{eq:cusp-total-energy} diverges.
This is clearly unphysical, and shows a breakdown in the validity of equation~\eqref{eq:cusp-result}.
Note however that this breakdown is \emph{not} in the low-frequency regime where we know that equation~\eqref{eq:cusp-result} is inaccurate; rather, the integral has an ultraviolet divergence that goes like $\sim f^{1/3}$ at high frequencies $f\to\infty$.
Instead, we can understand this divergence as a breakdown of the Nambu-Goto approximation for the loop dynamics.

We recall here the justification for the Nambu-Goto action, following section~6.1 of reference~\cite{Vilenkin:2000jqa}, so that we can clarify how this fails for cusps at high frequencies.
Assuming there are no non-gravitational long-range interactions between widely-separated string segments, a generic cosmic string can be described by a local worldsheet Lagrangian of the schematic form
    \begin{equation}
    \label{eq:lagrangian}
        \mathcal{L}=-\mu+\alpha\hbar\kappa+\beta\frac{\hbar^2\kappa^2}{\mu}+\cdots,
    \end{equation}
    where $\mu$ is the string tension as before, $\kappa$ represents spacetime curvature (with indices suppressed), and $\alpha,\beta$ are numerical constants.
In most situations, the curvature associated with a loop is of the order $\kappa\sim\ell^{-2}$, where $\ell$ is the length of the loop.
This is many orders of magnitude larger than the width of the loop,
    \begin{equation}
        \delta\sim(\mu/\hbar)^{-1/2}=\ell_\Pl/\sqrt{G\mu},
    \end{equation}
    where $\ell_\Pl$ is the Planck length.
This implies that
    \begin{equation}
    \label{eq:nambu-goto-approx}
        \hbar\kappa/\mu\sim(\delta/\ell)^2\ll1,
    \end{equation}
    so that we can drop all but the first term in equation~\eqref{eq:lagrangian}, leaving the Nambu-Goto action.
This `zero-width' approximation for the loop is equivalent to taking $\hbar\to0$ in equation~\eqref{eq:lagrangian}, effacing the microphysics of the underlying field-theory model and keeping only the classical part of the action.
This is valid for loops larger than a critical scale,
    \begin{equation}
        \ell\gtrsim\delta/(G\mu)\sim\ell_\Pl/(G\mu)^{3/2},
    \end{equation}
    as field-theory calculations show that loops smaller than this lose much of their energy through matter radiation~\cite{Hindmarsh:1994re,Srednicki:1986xg} or through topological unwinding and dispersion~\cite{Helfer:2018qgv,Aurrekoetxea:2020tuw}, such that the Nambu-Goto approximation breaks down.
Note, however, that these effects cannot resolve the divergence identified above, as this occurs for much larger loops $\ell\gg\delta/(G\mu)$, for which matter effects should be negligible.

\begin{figure}[t!]
    \includegraphics[width=0.48\textwidth]{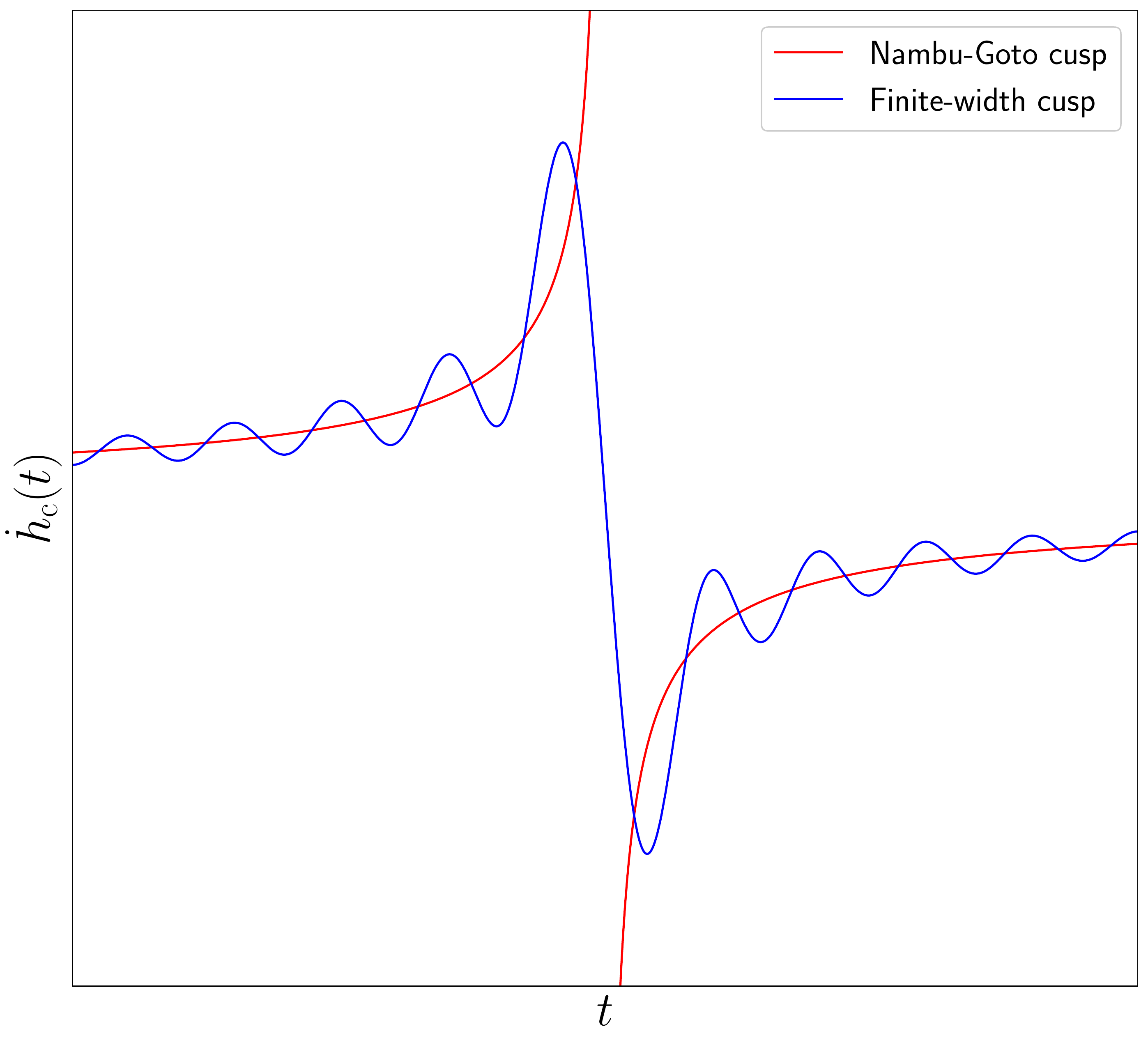}
    \caption{%
    The time derivative of the cusp strain signal close to the peak, $|t-t_0|\ll\ell$, for an observer in the beaming direction, $\iota=0$.
    This diverges for the standard Nambu-Goto waveform~\eqref{eq:cusp-waveform}, ruining the validity of the Nambu-Goto approximation, and causing a divergence in the energy radiated by the first-order memory.
    By introducing a frequency cutoff due to the finite string width, $f<1/\delta$, we see that the derivative becomes finite and continuous, and the first-order memory divergence is regularised.
    }
    \label{fig:hdot-cusp}
\end{figure}

The Nambu-Goto dynamics of loops generically predicts the formation of cusps, where the GW strain looks locally like $h(t)\sim|t-t_0|^{1/3}$.
As was already recognised in reference~\cite{Damour:2001bk}, the curvature associated with this scales as
    \begin{equation}
        \kappa\sim\ddot{h}\sim|t-t_0|^{-5/3},
    \end{equation}
    which diverges at the peak, clearly ruining the validity of the Nambu-Goto approximation~\eqref{eq:nambu-goto-approx}.
This problem does not manifest itself in the primary cusp waveform, since the beaming angle $\theta_\rmb$ decreases fast enough with frequency to ensure the total radiated energy is finite, cf. equation~\eqref{eq:cusp-total-energy-primary}.
However, the energy flux in the centre of the beam is still divergent, and as we have shown here, this sources further divergences due to the nonlinear nature of gravity.
Since the memory effect is not beamed, there is nothing suppressing it at high frequencies.

The simplest way to regularise this divergence is to impose an ultraviolet cutoff at the string width scale.
By truncating the frequency-domain cusp waveform at $f\sim1/\delta$, the cusp is smoothed out on timescales $|t-t_0|\sim\delta$, and the curvature reaches a finite maximum value that scales like $\kappa\sim\delta^{-5/3}$.
This smoothing is shown explicitly in figure~\ref{fig:hdot-cusp}, where we see that the time derivative of the strain diverges in the Nambu-Goto case, but is finite and continuous if a finite width is introduced.
[In this heuristic setup, it is not immediately clear whether or not the smoothing afforded by a finite string width is strong enough to prevent higher-curvature terms in the Lagrangian~\eqref{eq:lagrangian} from becoming important near the cusp; we return to this point later.]

We therefore consider only GWs with frequency $f<1/\delta$.
This has no impact on the primary waveform, since
    \begin{equation}
        1/\delta\approx10^{38}\,\mathrm{Hz}\times\qty(\frac{G\mu}{10^{-11}})^{1/2},
    \end{equation}
    which is beyond the reach of any current or planned GW experiments.
However, the cutoff \emph{does} impact the GW memory.
For example, setting an upper limit of $f=1/\delta$ in the integral~\eqref{eq:cusp-total-energy}, we find a finite value for the energy radiated by the cusp memory,
    \begin{equation}
    \begin{split}
    \label{eq:cusp-total-energy-memory}
        \mathcal{E}^{(1)}_\rmc&\simeq8(\uppi B_\rmc)^2(G\mu)^3(\ell/\delta)^{1/3}\\
        &\sim8(\uppi B_\rmc)^2(G\mu)^{19/6}(\ell/\ell_\Pl)^{1/3}.
    \end{split}
    \end{equation}
Numerically, this gives
    \begin{equation}
        \mathcal{E}^{(1)}_\rmc\approx3\times10^{-15}\times\qty(\frac{G\mu}{10^{-11}})^{19/6}\qty(\frac{\ell}{\mathrm{pc}})^{1/3},
    \end{equation}
    which shows that the energy radiated due to the first-order memory effect is smaller than that from the primary emission for observationally-allowed values of the string tension, $G\mu\lesssim10^{-11}$.
However, the factor of $(\ell/\delta)^{1/3}\gg1$ in equation~\eqref{eq:cusp-total-energy-memory} is concerning.
Based on equation~\eqref{eq:cusp-total-energy-memory}, cusps on GUT-scale strings ($G\mu=10^{-6}$) would radiate far more energy through the memory effect than through the primary GWs; for $\ell\gtrsim10^{-4}\,\mathrm{pc}$ they would radiate more than the entire energy of the loop.
Such a situation would be unphysical, and would indicate the breakdown of the validity of the primary cusp waveform.
One might argue that this is not an issue, since GUT-scale Nambu-Goto strings are already ruled out by observations; however, we show below that accounting for higher-order memory effects exacerbates the problem, leading to similar unphysical results even for much lower string tensions.

We should note that the cutoff imposed here is rather ad-hoc, and is done in a way that is agnostic to the underlying microphysics of the string.
In reality, the divergence will be resolved in a way which may depend on the microphysics, and the GW memory observables may be sensitive to this; indeed, this is implied by the fact that equation~\eqref{eq:cusp-total-energy-primary} depends directly on the string width $\delta$.

\subsection{Second-order memory}
\label{sec:cusp-higher-order}

The memory waveform~\eqref{eq:cusp-result} describes the spacetime curvature generated by the energy-momentum of the primary GWs from the cusp.
However, these memory GWs themselves carry energy-momentum, and will in turn act as a source of their own GW memory (see reference~\cite{Talbot:2018sgr} for a discussion of this effect in the context of binary black hole coalescences).
We refer to this `memory of the memory' here as the \emph{second-order memory effect}.

The calculation of this second-order memory contribution is straightforward for cusps, simply substituting equation~\eqref{eq:cusp-result} into the RHS of equation~\eqref{eq:memory-frequency-domain} and following all of the same steps as before.
The key difference is in the angular integral,
    \begin{equation}
        \int_{\vu*r'}(1+\cos\iota')^2=\frac{1}{6}\sin^2\iota,
    \end{equation}
    which, due to the broader emission of the first-order memory signal, is not suppressed by a factor of $\theta_\rmb^2$; cf. equation~\eqref{eq:spherical-integral-cusp-result-small-beam}.
The resulting expressions for the late-time memory and frequency-domain waveform are
    \begin{align}
    \begin{split}
        \Updelta h^{(2)}_\rmc&=\frac{2(\uppi B_\rmc)^2}{3r}(G\mu)^4\ell^{4/3}\sin^2\iota\int_{2/\ell}^\infty\frac{\dd{f}}{f^{2/3}},\\
        \tilde{h}^{(2)}_\rmc(f)&=-\frac{\rmi\uppi B_\rmc^2}{3rf}(G\mu)^4\ell^{4/3}\sin^2\iota\bigg[\int_{2/\ell}^\infty\frac{\dd{f'}}{f'^{1/3}(f'+|f|)^{1/3}}\\
        &\qquad\qquad\qquad\qquad\quad+\int_{2/\ell}^{|f|/2}\frac{\dd{f'}}{f'^{1/3}(|f|-f')^{1/3}}\bigg],
    \end{split}
    \end{align}
    both of which contain integrals which diverge due to high-frequency contributions from the first-order memory.
Introducing the $f<1/\delta$ cutoff from section~\ref{sec:energy-divergence} once again, we find to leading order in $\delta$,
    \begin{align}
    \begin{split}
    \label{eq:cusp-2nd-order-mem}
        \Updelta h^{(2)}_\rmc&\simeq\frac{2\uppi^2\ell}{r}B_\rmc^2(G\mu)^4(\ell/\delta)^{1/3}\sin^2\iota,\\
        \tilde{h}^{(2)}_\rmc(f)&\simeq-\frac{\rmi\Updelta h^{(2)}_\rmc}{2\uppi f}\Theta(1/\delta-|f|)\Theta(|f|-2/\ell).
    \end{split}
    \end{align}
This is interesting in that it departs from the the $\sim f^{-4/3}$ scaling of both the primary waveform~\eqref{eq:cusp-waveform} and the first-order memory waveform~\eqref{eq:cusp-result}; the second-order memory instead has a slower decay with frequency, due to the lack of beaming in the first-order memory.
In fact, for intermediate frequencies $2/\ell<|f|<1/\delta$ the second-order memory is identical to the Fourier transform of a Heaviside step function of height $\Updelta h^{(2)}_\rmc$.
This means that on timescales much shorter than the loop oscillation period $\ell/2$ and much longer than the light-crossing time of the loop width $\delta$, the second-order memory signal looks like a step function in the time domain; this is because the signal is dominated by high frequencies $f\lesssim1/\delta$, and therefore `switches on' in a very short time interval $|t-t_0|\lesssim\delta$.\footnote{%
This picture is closely related to the zero-frequency limit (ZFL) method for calculating radiation from high-energy gravitational scattering~\cite{Smarr:1977fy,Turner:1978jj,Bontz:1979zfl,Wagoner:1979dd}, which leverages the fact that the scattering is effectively instantaneous compared to the GW frequencies considered.}

The waveform~\eqref{eq:cusp-2nd-order-mem} can be used to calculate a second-order correction to the energy radiated by the cusp,
    \begin{align}
    \begin{split}
    \label{eq:cusp-energy-memory-2nd-order}
        \epsilon^{(2)}_\rmc&=\frac{\uppi r^2f^3}{2G\mu\ell}\qty(|\tilde{h}^{(0)}_\rmc+\tilde{h}^{(1)}_\rmc+\tilde{h}^{(2)}_\rmc|^2-|\tilde{h}^{(0)}_\rmc+\tilde{h}^{(1)}_\rmc|^2)\\
        &=\frac{\uppi r^2f^3}{2G\mu\ell}\qty(|\tilde{h}^{(2)}_\rmc|^2+2\tilde{h}^{(1)}_\rmc\tilde{h}^{(2)*}_\rmc)\\
        &=\bigg[\frac{\uppi^3}{2}B_\rmc^4(G\mu)^7(\ell/\delta)^{2/3}f\ell\sin^4\iota\\
        &\qquad+\uppi^2B_\rmc^3(G\mu)^5(\ell/\delta)^{1/3}(f\ell)^{2/3}\sin^2\iota(1+\cos\iota)\bigg]\\
        &\qquad\times\Theta(1/\delta-|f|)\Theta(|f|-2/\ell).
    \end{split}
    \end{align}
We see that, unlike for the first-order correction, there are now nonzero cross-terms due to $\tilde{h}^{(1)}_\rmc$ and $\tilde{h}^{(2)}_\rmc$ being exactly in phase with each other, resulting in two new contributions to the energy.
Integrating over frequency and emission direction, the total energy is given by
    \begin{equation}
    \label{eq:cusp-total-energy-memory-2nd-order}
        \mathcal{E}^{(2)}_\rmc=\frac{16}{15}(\uppi B_\rmc)^4(G\mu)^7(\ell/\delta)^{5/3}+4(\uppi B_\rmc)^3(G\mu)^5(\ell/\delta).
    \end{equation}
Comparing with the corresponding first-order result~\eqref{eq:cusp-total-energy-memory} we see a clear pattern emerging, where the second term in equation~\eqref{eq:cusp-total-energy-memory-2nd-order} is multiplied by a factor of $\sim\uppi B_\rmc(G\mu)^2(\ell/\delta)^{2/3}$ compared to the first-order energy, and multiplying by the same factor again gives the first term in equation~\eqref{eq:cusp-total-energy-memory-2nd-order}.
This factor is greater than unity so long as the loop length $\ell$ is larger than
    \begin{align}
    \begin{split}
    \label{eq:ell-star}
        \ell_*&\equiv\frac{\delta}{(\uppi B_\rmc)^{3/2}(G\mu)^{3}}\\
        &\approx90\,\mathrm{m}\times\qty(\frac{G\mu}{10^{-11}})^{-7/2},
    \end{split}
    \end{align}
    which applies to all macroscopically-large loops.

One would naively expect successive memory corrections for a generic GW source to become less and less important at higher order; the fact that they become \emph{more} important for such a large class of cosmologically-relevant cosmic string loops is surprising, and suggests that something unphysical is happening.
Indeed, if we plug numerical values into equation~\eqref{eq:cusp-total-energy-memory-2nd-order}, we find that loops larger than about $10^{14}\,\mathrm{m}\times(G\mu/10^{-11})^{-47/10}$ (i.e. not much larger than the solar system) have $\mathcal{E}^{(2)}_\rmc$ greater than unity, meaning that they radiate away more than the total energy of the loop.
This represents a significant worsening of the issue we identified at the end of section~\ref{sec:energy-divergence}.
The problem only gets worse when we go beyond second-order corrections, as we demonstrate below.

\subsection{Higher-order memory, and another divergence}
\label{sec:higher-order-divergence}

Iterating the procedure described above to calculate the third-order GW memory, it is straightforward to show that it obeys the same step-function-like relation that we found at second order,
    \begin{equation}
        \tilde{h}^{(3)}_\rmc(f)=-\frac{\rmi\Updelta h^{(3)}_\rmc}{2\uppi f}\Theta(1/\delta-|f|)\Theta(|f|-2/\ell),
    \end{equation}
    with the late-time memory given by
    \begin{equation}
        \Updelta h^{(3)}_\rmc=\frac{2\uppi^2}{r}\int_\mathbb{R}\dd{f}\int_{\vu*r'}r^2f^2\qty(|\tilde{h}^{(2)}_\rmc|^2+2\tilde{h}^{(1)}_\rmc\tilde{h}^{(2)*}_\rmc),
    \end{equation}
    where we have made sure to include both of the second-order energy contributions as source terms for the third-order memory.
Upon integration, this becomes
    \begin{align}
    \begin{split}
        \Updelta h^{(3)}_\rmc=&-\frac{4\delta}{5r}(\ell/\ell_*)^{8/3}\sin^2\iota\qty(1-\frac{1}{3}\cos^2\iota)\\
        &-\frac{2\delta}{r}(\ell/\ell_*)^2\sin^2\iota\qty(1+\frac{3}{5}\cos\iota),
    \end{split}
    \end{align}
    where $\ell_*$ is the $G\mu$-dependent lengthscale defined in equation~\eqref{eq:ell-star}.

The third-order memory clearly scales differently for loops with $\ell\gg\ell_*$ (which we call `large loops') compared to those with $\ell\ll\ell_*$ (which we call `small loops').
When going to fourth order and beyond, we obtain an increasing number of cross-terms in the energy at each order (starting with $|\tilde{h}^{(n)}_\rmc|^2$, then $2\tilde{h}^{(n-1)}_\rmc\tilde{h}^{(n)*}_\rmc$, $2\tilde{h}^{(n-2)}_\rmc\tilde{h}^{(n)*}_\rmc$, and so on, down to $2\tilde{h}^{(1)}_\rmc\tilde{h}^{(n)*}_\rmc$), resulting in a proliferation of terms in the resulting memory expressions, each with a different power of $\ell/\ell_*$.
It is therefore much simpler to treat small and large loops separately, and focus on the leading power of $\ell/\ell_*$ in each case.
This results in a very economical formula for iterating the memory calculation, valid for all $n\ge2$,
    \begin{equation}
    \label{eq:iterate-memory}
        \Updelta h^{(n)}_\rmc=
        \begin{cases}
            \displaystyle\frac{r}{\delta}\int_{\vu*r'}|\Updelta h^{(n-1)}_\rmc|^2, & \text{for}\;\ell\gg\ell_*\\[8pt]
            \displaystyle\frac{6\uppi r}{\delta^2}\int_{\vu*r'}\Updelta h^{(n-1)}_\rmc|\tilde{h}^{(1)}_\rmc(1/\delta)|, & \text{for}\;\ell\ll\ell_*
        \end{cases}
    \end{equation}
    where we have evaluated the frequency integral in each case, leaving just the angular integral.
The frequency-domain waveform is then given by the same quasi-step-function form as before,
    \begin{equation}
        \tilde{h}^{(n)}_\rmc=-\frac{\rmi\Updelta h^{(n)}_\rmc}{2\uppi f}\Theta(1/\delta-|f|)\Theta(|f|-2/\ell),
    \end{equation}
    and the corresponding energy spectra are given by
    \begin{equation}
        \epsilon^{(n)}_\rmc=
        \begin{cases}
            \displaystyle\frac{r^2f}{8\uppi G\mu\ell}|\Updelta h^{(n)}_\rmc|^2, & \text{for}\;\ell\gg\ell_*\\[8pt]
            \displaystyle\frac{r^2f^2}{2G\mu\ell}\Updelta h^{(n)}_\rmc|\tilde{h}^{(1)}_\rmc|, & \text{for}\;\ell\ll\ell_*
        \end{cases}
    \end{equation}

Solving equation~\eqref{eq:iterate-memory} iteratively with equation~\eqref{eq:cusp-2nd-order-mem} as an input, we find for all $n\ge2$,
    \begin{equation}
    \label{eq:cusp-higher-order-final}
        \Updelta h^{(n)}_\rmc=
        \begin{cases}
            \displaystyle\frac{\delta}{r}(\ell/\ell_*)^{2^n/3}L_n(\iota), & \text{for}\;\ell\gg\ell_*\\[8pt]
            \displaystyle\frac{\delta}{r}(\ell/\ell_*)^{2n/3}S_n(\iota), & \text{for}\;\ell\ll\ell_*
        \end{cases}
    \end{equation}
    where $L_n(\iota)$ and $S_n(\iota)$ are polynomials in $\cos\iota$ which describe the angular pattern of the memory for large and small loops, respectively---these are described in detail in appendix~\ref{sec:iota-polynomials}.
For small loops, all memory effects are subdominant compared to the primary emission, and equation~\eqref{eq:cusp-higher-order-final} gives a convergent geometric series.
For large loops, on the other hand, the memory emission becomes stronger at each order, and equation~\eqref{eq:cusp-higher-order-final} gives a lacunary series which diverges extremely quickly.
One might hope that the polynomials $L_n(\iota)$ decrease in magnitude fast enough to counteract the divergence, but we find empirically in appendix~\ref{sec:iota-polynomials} that
    \begin{equation}
        |L_n(\iota)|\approx5(2/5)^{2^{n-2}}\sin^2\iota,
    \end{equation}
    so the series diverges as long as $\ell\gtrsim(5/2)^3\ell_*\approx1\,\text{km}\times(G\mu/10^{-11})^{-7/2}$, as shown in figure~\ref{fig:cusp-energy}.
We discuss the cause of this divergence in appendix~\ref{sec:rhdot}, and argue that it is caused by a super-Planckian GW energy flux from the cusp.

\subsection{Memory from cusp collapse}
\label{sec:cusp-collapse}

\begin{figure*}[p!]
    \includegraphics[width=0.66\textwidth]{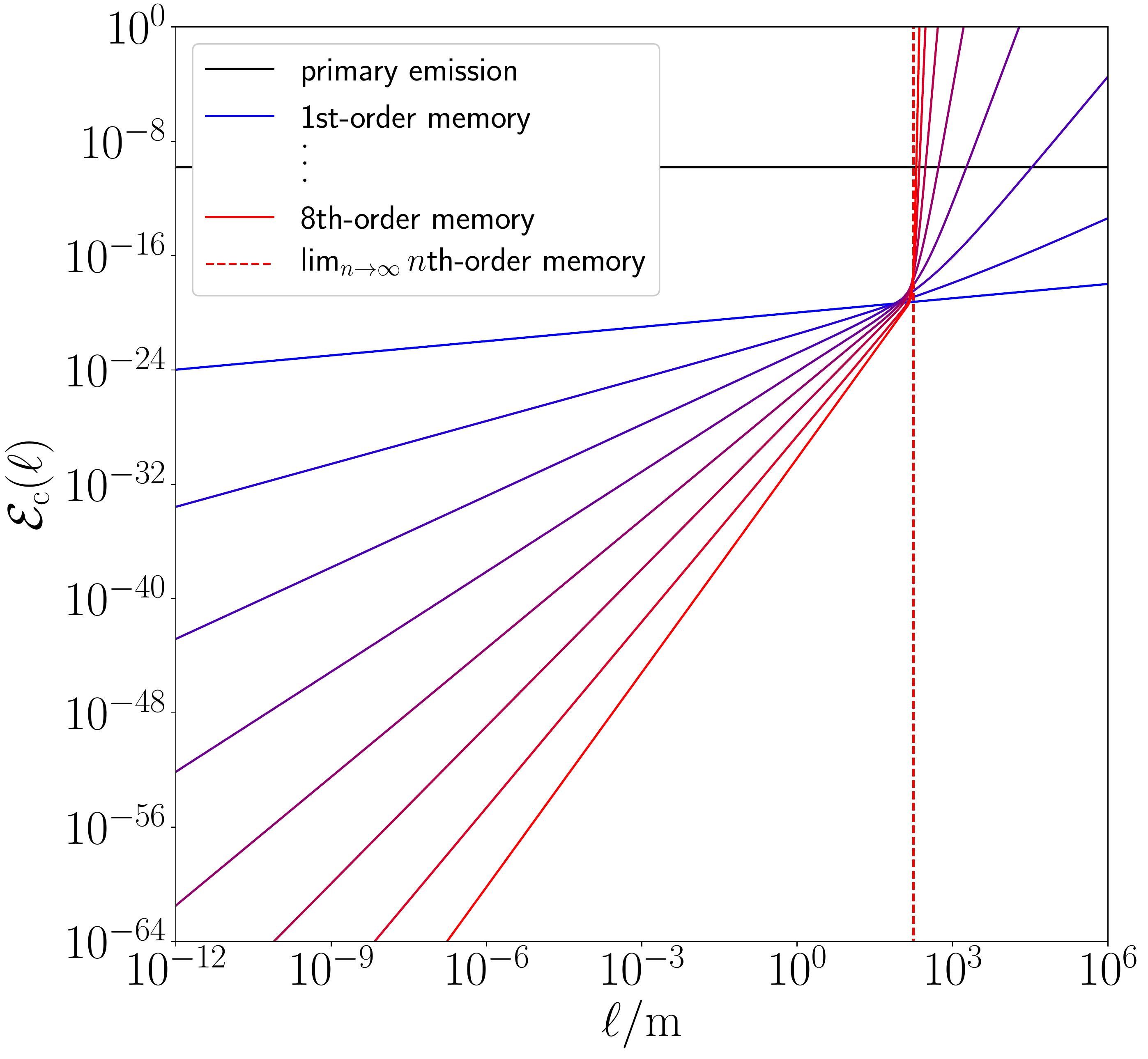}
    \includegraphics[width=0.66\textwidth]{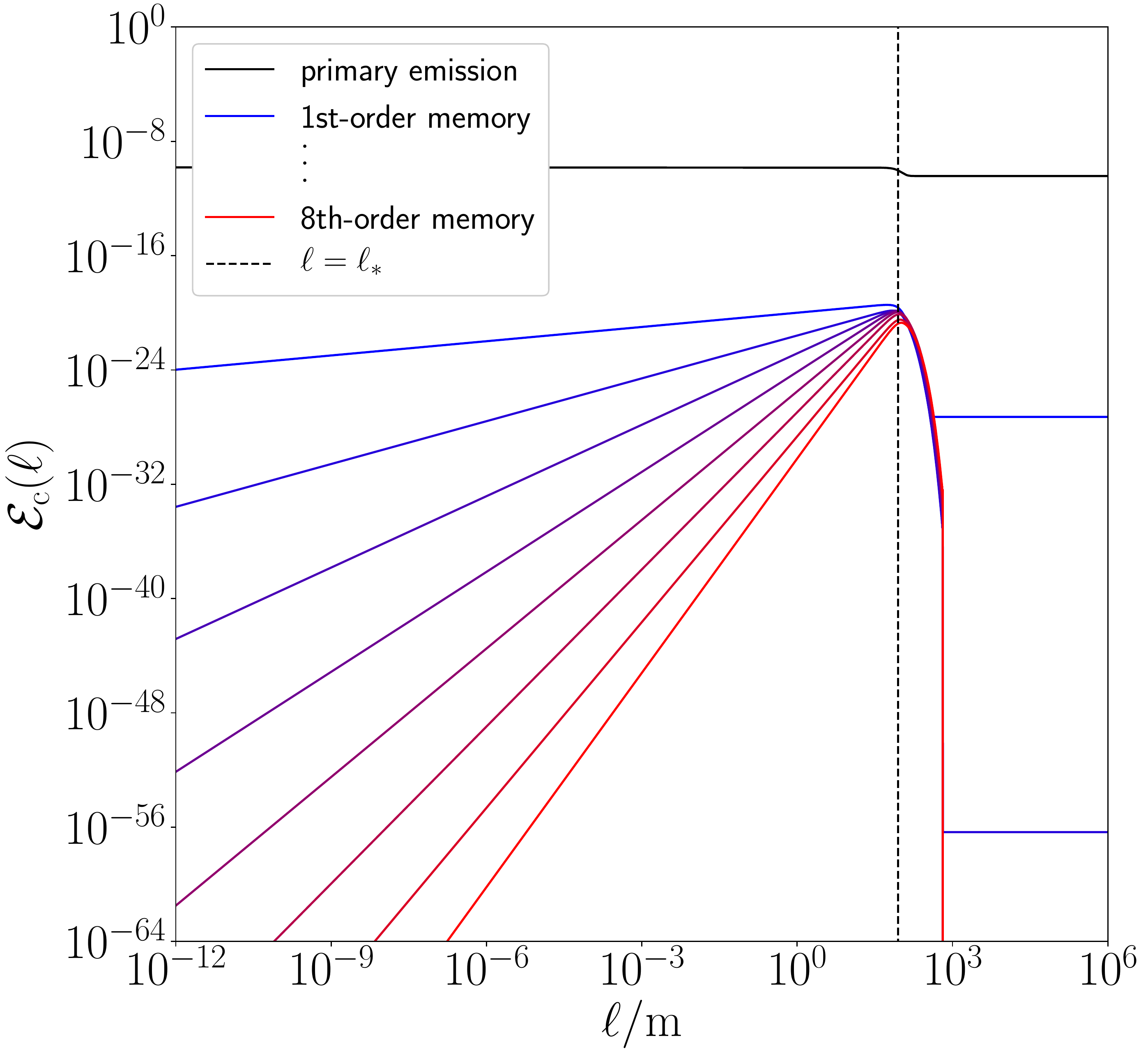}
    \caption{%
    The fractional energy radiated by a cusp at different orders in the memory expansion as a function of loop length, with $G\mu=10^{-11}$.
    The top panel shows the standard cusp case, clearly illustrating the divergence at $\ell\gtrsim\ell_*\approx90\,\mathrm{m}$.
    The bottom panel shows the cusp collapse case, for which the radiated energy at each order drops to a small, $\ell$-independent value for $\ell\gtrsim\ell_*$, curing the divergence.
    }
    \label{fig:cusp-energy}
\end{figure*}

The results of the previous section show that, even with an ultraviolet cutoff in place at the scale of the string width, the standard cusp waveform~\eqref{eq:cusp-waveform} leads to a divergence for all `large' loops with length $\ell\gtrsim\delta/(G\mu)^3$.
The fact that the divergence appears in observable, gauge-independent quantities (the memory strain and, therefore, the radiated energy) means that something unphysical must be going on.
Since the only inputs to our calculation are the cusp waveform~\eqref{eq:cusp-waveform} and the GW memory formula~\eqref{eq:memory-flux}, at least one of these two ingredients must break down for cusps on loops of this size.

In order to track down the cause of the divergence, let us list the assumptions that go into equations~\eqref{eq:memory-flux} and~\eqref{eq:cusp-waveform}:
    \begin{enumerate}
        \item The GW frequency is assumed to be much greater than the fundamental mode of the loop, $f\gg2/\ell$.
        This is because the waveform~\eqref{eq:cusp-waveform} is derived using the universal behaviour of the loop on scales $\ll\ell$ near the cusp.
        \item The loop's dynamics are assumed to follow the Nambu-Goto action~\eqref{eq:lagrangian} on lengthscales larger than the loop width $\delta$.
        (We have imposed a cutoff that effaces scales below this, in order to regularise the first divergence we encountered in section~\ref{sec:energy-divergence}.)
        \item The loop is assumed to evolve according to the flat-space equations of motion derived from~\eqref{eq:lagrangian}; i.e. gravitational backreaction is assumed to be negligible.
        \item The GWs generated by the loop are assumed to be well-described by linear perturbations on a flat background.
    \end{enumerate}

As mentioned earlier, the first assumption cannot be the source of the problem, as the divergence is associated with very high frequencies near the string width scale.

The second assumption is robust so long as ($i$) the higher-order curvature terms in the worldsheet Lagrangian~\eqref{eq:lagrangian} are negligible, and ($ii$) the strings are created through the breaking of a local gauge symmetry, so that the underlying field theory does not give rise to long-range interactions (i.e. we are not considering global strings, which would instead be described by the Kalb-Ramond action~\cite{Vilenkin:2000jqa}).
As mentioned in section~\ref{sec:energy-divergence}, while introducing a finite string width prevents the curvature from diverging, it does not necessarily guarantee that the higher-order curvature terms are negligible.
In principle the memory divergence identified here could be cured by departures from the Nambu-Goto action near the cusp.
However, these departures would have to take place on scales much greater than the string width, which seems very difficult to achieve.

By elimination, it seems that the problem is mostly likely due to assumptions 3 and 4: i.e., that the flat-space description of the cusp's dynamics and GW generation is inconsistent.
Indeed, we can trace the divergence back to the fact that the integrated GW energy flux diverges in the centre of the cusp's beam, $\int\dd{(\ln f)}\epsilon^{(0)}_\rmc(f,\vu*r_\rmc)\to\infty$, which already suggests that a flat-space description is insufficient.
This divergent flux is hidden somewhat by the narrowness of the beam, $\theta_\rmb\sim(f\ell)^{-1/3}$, which ensures that $\mathcal{E}^{(0)}_\rmc$ is finite, but we have shown that the GW memory inherits and amplifies the divergence.

One possible resolution is that the gravitational backreaction of the loop on itself could smooth out the cusp on scales much larger than $\delta$.
However, there is large body of literature on cosmic string backreaction~\cite{Thompson:1988yj,Quashnock:1990wv,Copeland:1990qu,Battye:1994qa,Buonanno:1998is,Carter:1998ix,Wachter:2016hgi,Wachter:2016rwc,Blanco-Pillado:2018ael,Chernoff:2018evo,Blanco-Pillado:2019nto} which indicates that the backreaction timescale is $\sim\ell/(G\mu)$, much longer than the loop's oscillation period.
While it is true that these studies all assume linearised gravity, and are thus likely to underestimate the magnitude of the effect near cusps, it is still hard to see how the loop could backreact fast enough to smooth out the cusp on relatively large scales before the peak of the signal.

All of this suggests that we need some strong-gravity mechanism which acts on a very short timescale while the cusp is forming, and suppresses the cusp's GW emission at frequencies far below the cutoff, $f\ll1/\delta$.
In reference~\cite{Jenkins:2020ctp} we proposed exactly such a mechanism.
We argued that when cusps form on sufficiently large cosmic string loops, they source such extreme spacetime curvature that a small portion of the loop could collapse to form a black hole at a time $\sim G\mu\ell$ before the peak of the cusp emission.
Remarkably, the loops for which this `cusp collapse' process is predicted to occur are those with length $\ell\gtrsim\delta/(G\mu)^3$---exactly the same loops for which the higher-order memory divergence occurs.
We suggest that this is no coincidence, and that one plausible solution to the divergence we have identified here is given by the results of reference~\cite{Jenkins:2020ctp}.

It is difficult to calculate the precise GW signal associated with cusp collapse, but there are two main qualitative differences it introduces compared to the standard cusp waveform: ($i$) the Fourier transform of the primary strain signal, $\tilde{h}^{(0)}_\rmc$, is reduced by a factor $\approx1/2$, as it only receives contributions from the half of the signal at times before the peak; ($ii$) more importantly, there is a loss of power at frequencies $f\gtrsim1/(G\mu\ell)$, due to the truncation immediately before the peak.
Both of these effects influence the corresponding GW memory signal.
We can account for ($i$) by multiplying the strain at each order in the memory expansion by the appropriate power of $1/2$, and can approximate the effect of ($ii$) by introducing a sharp cutoff at frequency $f=1/(G\mu\ell)$, which is equivalent to replacing $\delta\to G\mu\ell$ in all previous expressions.
The critical lengthscale~\eqref{eq:ell-star} which previously marked the onset of the divergence then becomes
    \begin{equation}
        \ell_*\to\frac{\ell}{(\uppi B_\rmc)^{3/2}(G\mu)^2}\gg\ell,
    \end{equation}
    which means that we are always in the `small loop' regime, $\ell\ll\ell_*$; higher-order memory corrections are suppressed by powers of $G\mu$, and the divergence is avoided completely, as shown in figure~\ref{fig:cusp-energy}.
Note that the cusp collapse process is only predicted to take place for loops with $\ell\gtrsim\delta/(G\mu)^3$, and that the GW memory from smaller loops is still described by the results given above, with $\ell_*$ given by equation~\eqref{eq:ell-star}.

More explicitly, if the divergence is indeed cured by invoking cusp collapse, then the GW observables from the cusp are as follows: the primary waveform is
    \begin{align}
    \begin{split}
        \tilde{h}^{(0)}_\rmc&\simeq A_\rmc\frac{G\mu\ell^{2/3}}{r|f|^{4/3}}\Theta(\vu*r\vdot\vu*r_\rmc-\cos\theta_\rmb)\Theta(|f|-2/\ell)\\
        &\qquad\times
        \begin{cases}
            (1/2)\Theta(1/G\mu\ell-|f|), & \text{for}\;\ell\gg\ell_*\\
            \Theta(1/\delta-|f|), & \text{for}\;\ell\ll\ell_*
        \end{cases}
    \end{split}
    \end{align}
    with $\ell_*$ given by~\eqref{eq:ell-star}; the first-order memory waveform is
    \begin{align}
    \begin{split}
        \tilde{h}^{(1)}_\rmc&\simeq-\rmi B_\rmc\frac{(G\mu)^2\ell^{2/3}}{rf|f|^{1/3}}(1+\cos\iota)\Theta(|f|-2/\ell)\\
        &\qquad\times
        \begin{cases}
            (1/4)\Theta(1/G\mu\ell-|f|), & \text{for}\;\ell\gg\ell_*\\
            \Theta(1/\delta-|f|), & \text{for}\;\ell\ll\ell_*
        \end{cases}
    \end{split}
    \end{align}
    with the corresponding late-time memory given by
    \begin{align}
    \begin{split}
        \Updelta h^{(1)}_\rmc&\approx2\times3^{2/3}(\uppi A_\rmc G\mu)^2(1+\cos\iota)\frac{\ell}{r}\\
        &\qquad\times
        \begin{cases}
            1/4, & \text{for}\;\ell\gg\ell_*\\
            1, & \text{for}\;\ell\ll\ell_*
        \end{cases}
    \end{split}
    \end{align}
    and the $n$th-order memory waveforms for $n\ge2$ are all step-function-like,
    \begin{align}
    \begin{split}
        \tilde{h}^{(n)}_\rmc&\approx-\frac{\rmi\Updelta h^{(n)}_\rmc}{2\uppi f}\Theta(|f|-2/\ell)\\
        &\qquad\times
        \begin{cases}
            \Theta(1/G\mu\ell-|f|), & \text{for}\;\ell\gg\ell_*\\
            \Theta(1/\delta-|f|), & \text{for}\;\ell\ll\ell_*
        \end{cases}
    \end{split}
    \end{align}
    with height given by
    \begin{align}
    \begin{split}
        \Updelta h^{(n)}_\rmc&\approx
        \begin{cases}
            \displaystyle\frac{\ell}{r}\qty(\frac{\uppi B_\rmc}{4})^{2^{n-1}}(G\mu)^{1+\frac{2^{n+1}}{3}}L_n(\iota), & \text{for}\;\ell\gg\ell_*\\
            \displaystyle\frac{\delta}{r}(\ell/\ell_*)^{2n/3}S_n(\iota), & \text{for}\;\ell\ll\ell_*
        \end{cases}
    \end{split}
    \end{align}

It is interesting to note that loops with length just below the cusp-collapse threshold, $\ell\lesssim\ell_*$, emit more energy through GW memory than those above the threshold.
We can see this already in the primary emission, where the total energy emission is
    \begin{equation}
    \label{eq:cusp-collapse-total-energy-0th-order}
        \mathcal{E}^{(0)}_\rmc\approx3^{2/3}(\uppi A_\rmc)^2G\mu\times
        \begin{cases}
            1/4, & \text{for}\;\ell\gg\ell_*\\
            1, & \text{for}\;\ell\ll\ell_*
        \end{cases}
    \end{equation}
    with only cusps above the collapse threshold being subject to the $1/4$ reduction in GW power due to the truncation of the signal.
However, the difference becomes more significant in the memory emission,
    \begin{equation}
    \label{eq:cusp-collapse-total-energy-1st-order}
        \mathcal{E}^{(1)}_\rmc\approx
        \begin{cases}
            \displaystyle\frac{1}{2}(\uppi B_\rmc)^2(G\mu)^{8/3}, & \text{for}\;\ell\gg\ell_*\\[8pt]
            \displaystyle8(\uppi B_\rmc)^{3/2}(G\mu)^2(\ell/\ell_*), & \text{for}\;\ell\ll\ell_*
        \end{cases}
    \end{equation}
    where there is an extra power of $(G\mu)^{2/3}$ for cusps above the collapse threshold, compared to those just below it.
This pattern continues for higher-order memory, where the total radiated energy for all $n\ge2$ is given by
\begin{widetext}
    \begin{equation}
    \label{eq:cusp-collapse-total-energy-nth-order}
        \mathcal{E}^{(n)}_\rmc\approx
        \begin{cases}
            \displaystyle\frac{1}{2}\qty(\frac{\uppi B_\rmc}{4})^{2^n}(G\mu)^{2^{n+2}/3}\int_{S^2}\frac{\dd[2]{\vu*r}}{4\uppi}|L_n(\iota)|^2, & \text{for}\;\ell\gg\ell_*\\[8pt]
            \displaystyle\frac{1}{2}(\uppi B_\rmc)^{3/2}(G\mu)^2(\ell/\ell_*)^{(2n-1)/3}\int_{S^2}\frac{\dd[2]{\vu*r}}{4\uppi}6(1+\cos\iota)S_n(\iota), & \text{for}\;\ell\ll\ell_*
        \end{cases}
    \end{equation}
\end{widetext}

We note briefly that reference~\cite{Jenkins:2020ctp} also discussed a further GW signature associated with cusp collapse: the high-frequency quasi-normal ringing of the newly-formed black hole after the collapse.
We neglect this effect here however, as it is hard to say anything concrete about the phase evolution and angular pattern of this additional GW emission, and this prevents us from calculating the associated memory signal.
It would be interesting to revisit this contribution to the memory if and when more detailed phase-coherent cusp collapse waveform models become available.

\section{Memory from kinks}
\label{sec:kinks}

Unlike cusps (which are transient), kinks are persistent features of cosmic string loops: they propagate around the loop at the speed of light, continuously emitting GWs in a beam which traces out a one-dimensional `fan' of directions, like a lighthouse.
This fact is usually unimportant when computing the primary GW emission from kinks, since the beam only overlaps with the observer's line of sight for a small fraction of the loop oscillation time, meaning that kinks are effectively transient sources for a given observer.
However, in order to calculate a GW memory signal, we need to know the primary GW flux in all directions over the entire history of the source, which for kinks means specifying the beaming direction as a function of time.

We consider here the simplest case, where the beam of the kink traces out a great circle on the sphere at a constant rate.
We choose our polar coordinates such that this circle lies in the equatorial plane, $\theta_\rmk=\uppi/2$.
The azimuthal direction of the kink's beam is then given by $\phi_\rmk=4\uppi\sigma t/\ell$ with $\sigma=\pm1$, where the two different signs correspond to left- and right-moving kinks respectively.\footnote{%
The definition of whether a given kink is left- or right-moving is somewhat arbitrary here.
For concreteness, we call kinks with $\sigma=+1$ `left-moving'; these move anti-clockwise around the loop when viewed from the North pole $\theta=0$.
Conversely, kinks with $\sigma=-1$ are called `right-moving'; these move clockwise when viewed from $\theta=0$.}
From equation~\eqref{eq:kink-waveform} we then have
    \begin{equation}
    \label{eq:kink-phase}
        \tilde{h}^{(0)}_\rmk(f,\vu*r)\simeq A_\rmk\frac{G\mu\ell^{1/3}}{r|f|^{5/3}}\rme^{-\rmi\sigma\phi f\ell/2}\Theta(\theta_\rmb-|\iota|)\Theta(|f|-2/\ell),
    \end{equation}
    where the inclination $\iota\equiv\uppi/2-\theta$ describes the angle between the observer's line of sight and the closest point on the equatorial plane, and takes values $\iota\in[-\uppi/2,\uppi/2]$ (with positive/negative values corresponding to the observer being above/below the plane).
Note that we have picked up a phase factor $\rme^{-\rmi\sigma\phi f\ell/2}$ to account for the time at which the kink passes closest to the line of sight.
As we show below, this direction-dependent phase ultimately leads to a strong suppression of the kink memory signal compared to the cusp case.

\subsection{Beaming effects}

As in the cusp case, the first step in calculating the GW memory signal is to compute the angular integral that captures the beaming effects of the kink,
    \begin{equation}
        \int_{\vu*r'}\rme^{-\rmi\sigma\phi'f\ell/2}\Theta(\theta_\rmb-|\iota|).
    \end{equation}
Here the small circle of radius $\theta_\rmb$ around the North pole that we considered for cusps has been replaced with a narrow band of half-width $\theta_\rmb$ around the equator, and we have included the $\phi$-dependent phase factor from equation~\eqref{eq:kink-phase}.
It is straightforward to integrate out the zenith angle $\theta'$ if we keep only the leading-order term in $\theta_\rmb$; this gives
    \begin{equation}
    \label{eq:kink-angular-integral}
        \int_{\vu*r'}\rme^{-\rmi\sigma\phi'f\ell/2}\Theta(\theta_\rmb-|\iota|)\simeq\theta_\rmb K_{f\ell/2}(-\sigma\iota),
    \end{equation}
    where we have defined a family of azimuthal integrals,
    \begin{equation}
    \label{eq:K_n-definition}
        K_n(\iota)=\int_0^{2\uppi}\frac{\dd{\phi}}{2\uppi}\rme^{\rmi n\phi}\frac{(\sin\iota\cos\phi-\rmi\sin\phi)^2}{1-\cos\iota\cos\phi}.
    \end{equation}
Notice that the argument $n$ is always an integer, as the GW frequency is always an integer multiple of the loop's fundamental mode $2/\ell$ (although we often ignore this when taking the continuum limit at high frequencies).
Computing the integral equation~\eqref{eq:K_n-definition} for general $n$ therefore corresponds to finding the Fourier spectrum of a complicated nonlinear function of $\phi$; we perform this calculation explicitly in appendix~\ref{sec:K_n}.

\subsection{Late-time memory}

Using equation~\eqref{eq:late-time-memory}, along with the expression for the angular integral $K_0(\iota)$ from equation~\eqref{eq:K_n-final}, we find that the late-time memory from the kink is given by
    \begin{equation}
        \Updelta h^{(1)}_\rmk=\frac{3^{5/6}\ell}{r}(\uppi A_\rmk G\mu)^2\frac{4\sin|\iota|+\cos2\iota-3}{\cos^2\iota}.
    \end{equation}
Inserting numerical values, the strongest effect is $\Updelta h^{(1)}_\rmk\approx-4.250\times(G\mu)^2\ell/r$ when the observer lies in the plane of the kink $\iota=0$, an order of magnitude smaller than the maximum cusp memory.
The late-time kink memory decreases smoothly to zero as one approaches the poles $\iota\to\pm\uppi/2$.

\subsection{Full waveform}

The calculation here is very similar to the cusp case in section~\ref{sec:cusp-full-waveform}.
Inserting equation~\eqref{eq:kink-phase} into equation~\eqref{eq:memory-frequency-domain}, and taking care to enforce the frequency cutoffs and to account for the two competing beam angles $\theta_\rmb(f')$ and $\theta_\rmb(f'-f)$, we obtain
    \begin{align}
    \begin{split}
        \tilde{h}_\rmk^{(1)}(f)&\simeq-\frac{\rmi2^{5/3}\uppi(A_\rmk G\mu)^2\ell^{1/3}}{3^{1/6}rf|f|^{2/3}}K_{f\ell/2}(-\sigma\iota)\\
        &\times\qty[\int_{2/|f|\ell}^\infty\frac{\dd{u}}{u^{2/3}(1+u)}-\int_{2/|f|\ell}^{1/2}\frac{\dd{u}}{u^{2/3}(1-u)}]
    \end{split}
    \end{align}
    for $|f|>4/\ell$.
(This is analogous to equation~\eqref{eq:cusp-all-frequencies} from the cusp case.)
Taking the limit $|f|\gg2/\ell$, we can evaluate the integrals analytically to find
    \begin{equation}
    \label{eq:kink-result}
        \tilde{h}^{(1)}_\rmk(f)\simeq-\rmi B_\rmk\frac{(G\mu)^2\ell^{1/3}}{rf|f|^{2/3}}K_{f\ell/2}(-\sigma\iota)\Theta(|f|-4/\ell),
    \end{equation}
    where the numerical constant is
    \begin{align}
    \begin{split}
        B_\rmk&\equiv\frac{\uppi A_\rmk^2}{\sqrt{3}}\qty(\frac{2}{3})^{5/3}\qty[\frac{2\uppi}{3\sqrt{3}}-{}_2F_1(\tfrac{1}{3},\tfrac{1}{3};\tfrac{4}{3};-1)]\\
        &\approx0.5830,
    \end{split}
    \end{align}
    and where we have set the frequency cutoff at double the fundamental mode $f_0=2/\ell$ due to the different behaviour of the angular integral $K_n(\iota)$ for $n=1$ compared to $n\ge2$.

This waveform~\eqref{eq:kink-result} shares many features with both the cusp memory waveform~\eqref{eq:cusp-result} and the primary kink waveform~\eqref{eq:kink-phase}.
The most important difference from both of those waveforms is the dependence on inclination, which here is a function of frequency.
Using the results derived in appendix~\ref{sec:K_n}, we can rewrite equation~\eqref{eq:kink-result} as
    \begin{align}
    \begin{split}
    \label{eq:kink-result-cases}
        \tilde{h}^{(1)}_\rmk(f)&\simeq-\rmi B_\rmk\frac{(G\mu)^2\ell^{1/3}}{rf|f|^{2/3}}\frac{4\sin|\iota|}{\cos^2\iota}\\
        &\times
        \begin{cases}
            \displaystyle\qty(\frac{\cos\iota}{1+\sin|\iota|})^{f\ell/2}\Theta(f-4/\ell), & \text{for}\:\sigma\iota<0\\
            \displaystyle\qty(\frac{\cos\iota}{1+\sin|\iota|})^{-f\ell/2}\Theta(-f-4/\ell), & \text{for}\:\sigma\iota>0
        \end{cases}
    \end{split}
    \end{align}
    meaning that the memory signal from left-moving kinks contains only negative frequencies above the equatorial plane and only positive frequencies below the plane, and vice versa for right-moving kinks.

\begin{figure}[t!]
    \includegraphics[width=0.48\textwidth]{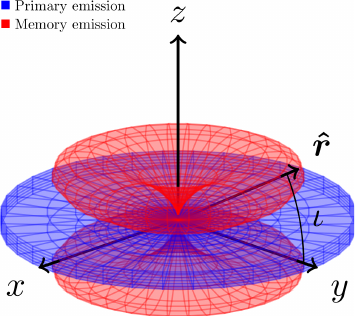}
    \caption{%
    A cartoon illustration of the angular distribution of the energy radiated by a kink.
    The primary emission (blue) is concentrated in a narrow fan of half-width $\theta_\rmb\sim(f\ell)^{-1/3}$ around the plane of the kink, while the memory emission (red) is concentrated in two lobes either side of this plane, which are exponentially suppressed as one approaches either of the directions normal to the plane.
    }
    \label{fig:kink-memory-angular-patterns}
\end{figure}

For high frequencies $|n|\gg1$ the angular integral $K_n(\iota)$ has a maximum value of $\simeq4/(\rme|n|)$ at inclination $\iota\simeq1/n$.
This means that the kink memory signal is only observable very close to the plane of the kink (but \emph{not} in the plane, where it vanishes; see figure~\ref{fig:kink-memory-angular-patterns}), and is suppressed by an extra power of frequency compared to the primary signal, $\tilde{h}^{(1)}_\rmk\sim f^{-8/3}$.

\subsection{Time-domain waveform near the arrival time}

As with the cusp case, we can reverse-Fourier-transform equation~\eqref{eq:kink-result-cases} to find the time-domain memory strain around time of arrival of the primary kink signal, $|t-t_0|\ll\ell$.
Unlike the cusp case, we obtain a signal which is not linearly polarised, but contains both $+$ and $\times$ polarisation content,
    \begin{align}
    \begin{split}
    \label{eq:kink-memory-time-domain}
        h^{(1)}_{\rmk,+}(t)-h^{(1)}_{\rmk,+}(t_0)&\simeq\frac{16\uppi B_\rmk(G\mu)^2}{2^{1/3}r}(t-t_0)\\
        &\times\frac{\sin|\iota|}{\cos^2\iota}E_{2/3}\qty(2\ln\frac{1+\sin|\iota|}{\cos\iota}),\\
        h^{(1)}_{\rmk,\times}(t)-h^{(1)}_{\rmk,\times}(t_0)&\simeq\frac{64\uppi^2B_\rmk(G\mu)^2}{2^{1/3}\ell r}(t-t_0)^2\\
        &\times\frac{\sin\sigma\iota}{\cos^2\iota}E_{-1/3}\qty(2\ln\frac{1+\sin|\iota|}{\cos\iota}),
    \end{split}
    \end{align}
    where we have used the generalised exponential integral function, $E_n(z)\equiv\int_1^\infty\dd{x}\rme^{-zx}x^{-n}$.
The $\times$-polarised component is suppressed by an additional factor of $(t-t_0)/\ell\ll1$, meaning that the signal is still approximately $+$-polarised.

An important difference with respect to the cusp case is that since kinks are long-lived rather than transient, their memory signal is not concentrated around a particular arrival time, but can in principle be observed at \emph{all} times.
One can easily obtain expressions analogous to equation~\eqref{eq:kink-memory-time-domain} at any point in the kink's periodic motion by substituting in the appropriate time when evaluating the reverse Fourier transform; in general this leads to a mixing between the $+$- and $\times$-polarisation modes.

\subsection{Radiated energy}

As in the cusp case, we are interested in the total energy radiated by the kink, and how the memory emission adds to this.
Using equation~\eqref{eq:dimensionless-energy}, we find that the dimensionless energy spectra for the primary emission and the first-order memory are
    \begin{align}
    \begin{split}
        \epsilon^{(0)}_\rmk(f,\vu*r)&\simeq\frac{\uppi}{2}A_\rmk^2G\mu(f\ell)^{-1/3}\Theta(\theta_\rmb-|\iota|)\Theta(f-2/\ell),\\
        \epsilon^{(1)}_\rmk(f,\vu*r)&\simeq4\uppi B_\rmk^2\frac{(G\mu)^3}{(f\ell)^{1/3}}\frac{\sin^2\iota\cos^{f\ell-4}\iota}{(1+\sin|\iota|)^{f\ell}}\Theta(f-4/\ell).
    \end{split}
    \end{align}
Integrating over the sphere, the primary emission is suppressed by a factor of the beaming angle $\theta_\rmb\sim(f\ell)^{-1/3}$, giving
    \begin{equation}
        \bar{\epsilon}^{(0)}_\rmk(f)\simeq\frac{2^{5/3}}{3^{1/6}}(\uppi A_\rmk)^2G\mu(f\ell)^{-2/3}\Theta(f-2/\ell).
    \end{equation}
The spherical integral for the memory contribution can be evaluated explicitly to give
    \begin{equation}
        \int_{S^2}\dd[2]{\vu*r}\frac{\sin^2\iota\cos^{f\ell-4}\iota}{(1+\sin|\iota|)^{f\ell}}=\frac{8\uppi}{f\ell(f\ell-2)(f\ell+2)},
    \end{equation}
    so that at high frequencies, the isotropic energy spectrum is approximately
    \begin{equation}
        \bar{\epsilon}^{(1)}_\rmk(f)\simeq32(\uppi B_\rmk)^2\frac{(G\mu)^3}{(f\ell)^{10/3}}\Theta(f-4/\ell).
    \end{equation}
We see that the angular pattern of the kink memory strongly suppresses the isotropic spectrum at high frequencies, meaning that unlike in the cusp case, there is no frequency range where the memory contribution dominates, and the total radiated energy converges without imposing an ultraviolet cutoff,
    \begin{align}
    \begin{split}
    \label{eq:kink-total-energy}
        \mathcal{E}^{(0)}_\rmk&\simeq3^{5/6}(\uppi A_\rmk)^2G\mu,\\
        \mathcal{E}^{(1)}_\rmk&\simeq\frac{3(\uppi B_\rmk)^2}{2^{8/3}\times5}(G\mu)^3.
    \end{split}
    \end{align}

\subsection{Higher-order memory}
\label{sec:kink-higher-order}

As with the cusp case, we can iterate the memory calculation with our 1st-order memory waveform~\eqref{eq:kink-result} as an input to calculate the 2nd-order memory effect (the `memory of the memory').
This involves calculating the angular integral
    \begin{equation}
        \int_{\vu*r'}\rme^{-\rmi\sigma\phi'f\ell/2}K_{f'\ell/2}(-\sigma\iota')K_{(f'-f)\ell/2}(-\sigma\iota'),
    \end{equation}
    which in the high-frequency regime $|f'|,|f'-f|\gg2/\ell$ is well-approximated by
    \begin{align}
    \begin{split}
        \simeq\frac{16K_{f\ell/2}(-\sigma\iota)}{(f'\ell-f\ell/2)^3}\bigg[&\Theta(f'-4/\ell)\Theta(f'-f-4/\ell)\\
        &-\Theta(-f'-4/\ell)\Theta(-f'+f-4/\ell)\bigg].
    \end{split}
    \end{align}
With this result to hand, the remaining steps are very similar to the calculations for the 1st-order memory described above, yielding
    \begin{align}
    \begin{split}
        \tilde{h}^{(2)}_\rmk(f)&\simeq-\rmi\frac{512\uppi B_\rmk^2(G\mu)^4}{rf|f|^{10/3}\ell^{7/3}}K_{f\ell/2}(-\sigma\iota)\Theta(|f|-4/\ell)\\
        &\times\int_{4/|f|\ell}^\infty\frac{\dd{u}}{u^{2/3}(u+1)^{2/3}(u+1/2)^3}\\
        &\approx-\rmi\frac{6664\times(G\mu)^4}{rf|f|^{10/3}\ell^{7/3}}K_{f\ell/2}(-\sigma\iota)\Theta(|f|-4/\ell).
    \end{split}
    \end{align}
The fact that this has the exact same dependence on the inclination $\iota$ as the 1st-order memory makes it straightforward to iterate the process to higher orders.
Doing this, we find that for all $n\ge2$, the kink GW memory is given schematically by
    \begin{equation}
        \tilde{h}^{(n)}_\rmk(f)\sim\frac{-\rmi(G\mu)^{2n}f\ell^3}{r(|f|\ell)^{8n/3}}K_{f\ell/2}(-\sigma\iota)\Theta(|f|-4/\ell),
    \end{equation}
    multiplied by some numerical constant (for which there does not seem to be a simple expression for all $n$).

The situation here is drastically different from the cusp case.
For cusps we saw that each successive order in the memory was suppressed by larger powers of $G\mu\ll1$, but also enhanced by larger powers of $\ell/\delta\gg1$, and that in certain situations the latter would dominate, causing a divergence.
For kinks, we see instead that each order in the memory is not only suppressed by powers of $G\mu$, but is further suppressed by a factor of $(|f|\ell)^{-8/3}\ll1$ each time.
The signal falls off quickly enough with frequency that higher-order memory contributions are not sensitive to the string-width scale, and no factors of $\ell/\delta$ appear.
This means that there is no situation where the kink memory diverges, and that the higher-order contributions are negligible in any observational scenario.
This lack of divergence is in agreement with the cusp-collapse mechanism we have invoked as a possible resolution for the cusp divergence, as kinks are not predicted to form PBHs~\cite{Jenkins:2020ctp}.

\subsection{Caveats of our approach}

We have only considered the simplest case where kink's beam traverses a fixed plane at a constant rate, in order to make detailed analytical calculations feasible.
This situation is highly idealised, and is not representative of the loops one would find in a cosmological loop network, which would typically contain structure on scales smaller than the loop length $\ell$, causing the path of the beam to vary on those scales.
We expect that such structure is only likely to make a significant \emph{qualitative} difference to our results if it is on a scale corresponding to the GW frequency of interest.
We note that small-scale structure on loops is expected to be damped over time through gravitational backreaction~\cite{Quashnock:1990wv}, so our simple treatment here is not likely to be too unreasonable.
In any case, we do not expect such considerations to change the main conclusion of this section: that memory from kinks is highly suppressed compared to the cusp case.

We have also neglected the fact that kinks always appear in pairs on loops, with one left-mover for every right-mover.
A realistic loop is likely to have several pairs of kinks, with each of these sourcing a GW memory signal like the one calculated here.
Since the kinks travel around the loop at the same average rate, it is possible that the superposition of their memory signals could give rise to interesting coherent effects.
However, regardless of whether the kink memory signals are coherent or not, the GW energy flux will still be of the same order of magnitude, so our main conclusions are unaffected.

\section{Detection prospects}
\label{sec:detection-prospects}

\begin{figure}[t!]
    \includegraphics[width=0.48\textwidth]{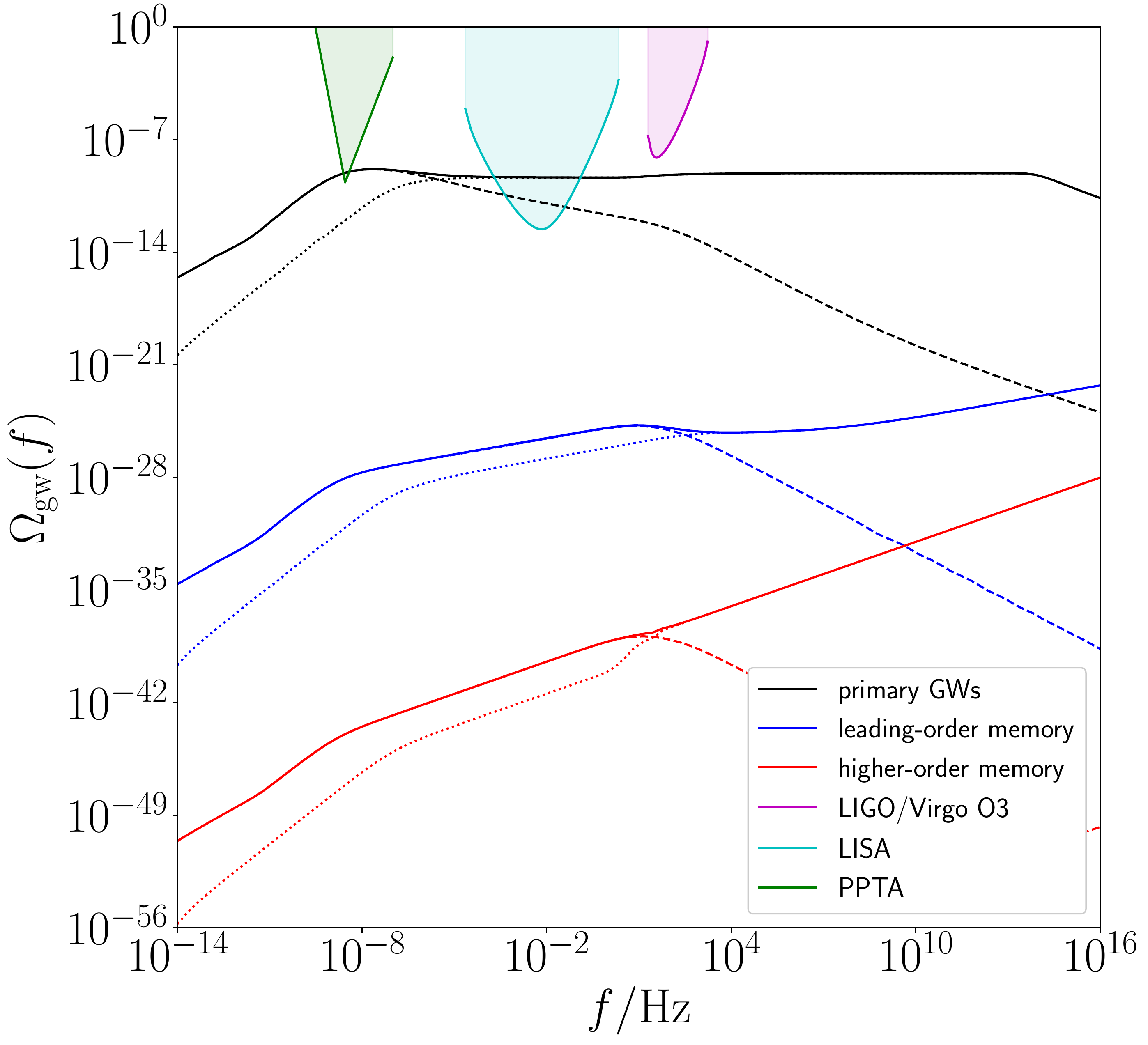}
    \caption{%
    Contributions to the SGWB energy spectrum $\Omega_\mathrm{gw}(f)$ from cosmic strings at different orders in the memory expansion, assuming that the memory divergence is resolved through the cusp-collapse scenario described in section~\ref{sec:cusp-collapse}.
    We see that the memory effect is negligible compared to the primary emission.
    Here we assume `model 2'~\cite{Blanco-Pillado:2013qja,Blanco-Pillado:2017rnf} of the string network, and a string tension of $G\mu=8.1\times10^{-11}$ (this is the largest string tension allowed by current constraints in the cusp-collapse scenario~\cite{Jenkins:2020ctp}; smaller tensions suppress the memory effect even further).
    The dashed/dotted curves show the contributions from the matter/radiation era, respectively, with the solid curves showing the combined spectra.
    A distinct change in all three of the radiation-era spectra is visible at around $f\approx10\,\mathrm{Hz}$; at frequencies above this, the spectra are increasingly dominated by small loops which do not undergo cusp-collapse, and this is why the memory effect becomes more prominent (though still undetectable).
    The magenta curve shows the power-law-integrated (PI) sensitivity curve~\cite{Thrane:2013oya} from the LIGO/Virgo O3 isotropic stochastic search~\cite{Abbott:2021xxi}, which is publicly available at reference~\cite{LIGOScientific:2021PI}.
    The green curve shows the Parkes Pulsar Timing Array (PPTA) PI curve~\cite{Shannon:2015ect,Verbiest:2016vem}, calculating using the code from reference~\cite{Thrane:2013oya}, which is publicly available at reference~\cite{Thrane:2013PI}.
    The cyan curve shows the projected LISA~\cite{Audley:2017drz} power-law-integrated sensitivity curve, as described in references~\cite{Caprini:2019pxz,Smith:2019wny}.
    }
    \label{fig:Omega_gw}
\end{figure}

Having calculated the nonlinear GW memory waveforms associated with cusps and kinks, it is natural to ask whether these signals are detectable with current or future GW observatories.
Clearly the divergent behaviour diagnosed for cusps in section~\ref{sec:cusp-higher-order} could, in principle, have important observational implications, depending on how the divergence is regulated.
For the purposes of this section we assume the divergence is resolved along the lines of the cusp-collapse scenario described in section~\ref{sec:cusp-collapse}.
We show below that, under this assumption, the cusp and kink memory signals are suppressed so strongly that they are well beyond the reach of GW observatories, even futuristic third-generation interferometers like Einstein Telescope (ET)~\cite{Punturo:2010zz} and Cosmic Explorer (CE)~\cite{Reitze:2019iox}.

\subsection{Burst searches}

We start by calculating the expected detection horizons for individual bursts of GW memory from cusps and kinks---i.e. the maximum distance at which a burst can be detected, on average.
We assume a matched-filter search, such that the optimal root-mean-square SNR (averaging over sky location and polarisation angle) for a frequency-domain waveform $\tilde{h}(f)$ which is isotropically averaged over the source inclination is given by~\cite{Maggiore:1900zz}
    \begin{equation}
        \rho_\mathrm{rms}=\qty[\sum_I\frac{4}{5}\sin^2\alpha_I\int_0^\infty\dd{f}\frac{|\tilde{h}(f)|^2}{P_I(f)}]^{1/2}.
    \end{equation}
The sum here is over different GW detectors, with $P_I(f)$ representing the noise power spectral density (PSD) of detector $I$, and $\alpha_I$ the opening angle between the two interferometer arms (this angle enters through the detector's response function; LIGO, Virgo, and CE have an opening angle of $\uppi/2$, while each of ET's three interferometers has an opening angle of $\uppi/3$).
It is convenient to rewrite this in terms of the fractional energy spectrum, using equation~\eqref{eq:dimensionless-energy} to give
    \begin{equation}
    \label{eq:snr-rms}
        \rho_\mathrm{rms}=\qty[\sum_I\frac{2G\mu\ell}{5\uppi^2r^2}\sin^2\alpha_I\int_0^\infty\dd{f}\frac{\bar{\epsilon}(f_\mathrm{s})}{f_\mathrm{s}^3P_I(f)}]^{1/2},
    \end{equation}
    where $r(z)$ is now the comoving distance to the source, and $f_\mathrm{s}=(1+z)f$ is the source-frame frequency, with $z$ the redshift.
We assume that any cosmic string signal with $\rho_\mathrm{rms}\ge12$ can be confidently detected; in reality, this threshold depends on the distribution of non-Gaussian noise transients (`glitches') in the network, and how closely these are able to mimic the waveforms of interest, but we ignore these details here.

Even assuming an optimistic third-generation GW detector network consisting of Einstein Telescope plus two Cosmic Explorers,\footnote{%
For ET we use the `ET-D' noise PSD developed in reference~\cite{Hild:2010id}, while for CE we use the `CE-2' noise PSD developed in reference~\cite{Hall:2020dps}.%
} we find that given current constraints on the string tension\footnote{%
Stochastic GW background constraints from PTAs~\cite{Lasky:2015lej,Blanco-Pillado:2017rnf} imply that $G\mu\lesssim8.1\times10^{-11}$ in the cusp-collapse scenario~\cite{Jenkins:2020ctp} for `model 2'~\cite{Blanco-Pillado:2013qja} of the loop network.
The other network model considered most frequently in the literature (`model 3'~\cite{Ringeval:2005kr,Lorenz:2010sm}) gives an even more stringent constraint of $G\mu\lesssim4.0\times10^{-15}$~\cite{LIGOScientific:2021nrg}.
(See references~\cite{Abbott:2017mem,Auclair:2019wcv,LIGOScientific:2021nrg} for an overview of these models.)%
} the detection horizon for a cusp memory signal is at most $\approx2\,\mathrm{pc}$.
This corresponds to a negligibly small detection rate, as very few cosmic strings are expected within a volume of this size.
For kink memory the result is even more pessimistic, with a detection horizon of $\approx0.008\,\mathrm{AU}$ (a few times larger than the Earth-Moon distance).
Larger values of the string tension $G\mu$ would boost the detectability of these memory bursts, but would be in conflict with existing observational results.

\subsection{Stochastic background searches}

The combined GW emission from many loops throughout cosmic history gives rise to a stochastic GW background (SGWB)~\cite{Vilenkin:1981bx,Vilenkin:1984ib,Hindmarsh:1994re,Allen:1996vm,Maggiore:1999vm,Vilenkin:2000jqa,Damour:2001bk,Siemens:2006yp,Abbott:2017mem,Caprini:2018mtu,Jenkins:2018nty,Auclair:2019wcv,LIGOScientific:2021nrg}.
The intensity of this SGWB as a function of frequency is usually expressed as a fraction of the cosmological critical density $\rho_\mathrm{crit}=3H_0^2/(8\uppi G)$ in terms of the density parameter,
    \begin{equation}
        \Omega_\mathrm{gw}(f)\equiv\frac{1}{\rho_\mathrm{crit}}\dv{\rho_\mathrm{gw}}{(\ln f)},
    \end{equation}
    which for cosmic string loops can be written as
    \begin{equation}
        \Omega_\mathrm{gw}(f)=\frac{16\uppi G\mu}{3H_0^2}\sum_iN_i\int\dd{t}\int\dd{\ell}a^4n(\ell,t)\bar{\epsilon}_i(f/a,\ell),
    \end{equation}
    where $t$ is cosmic time, $a(t)$ is the FLRW scale factor, and $n(\ell,t)\dd{\ell}$ is the comoving number density of loops with length between $\ell$ and $\ell+\dd{\ell}$.
The sum is over different GW signals (cusps, kinks, and kink-kink collisions), with $N_i$ the mean number of signals per loop per oscillation period, and $\bar{\epsilon}_i$ the isotropic energy spectrum of each signal, as defined in equation~\eqref{eq:dimensionless-energy-isotropic}.

In figure~\ref{fig:Omega_gw} we show the SGWB spectrum from a particular model of the loop network, including the contributions from first-order and higher-order GW memory from cusps and kinks.
We see that while the primary SGWB reaches a plateau at high frequencies, the memory contributions to the SGWB grow with frequency above $\approx10\,\mathrm{Hz}$; this makes sense given the slower fall-off of the cusp memory emission at high frequency compared to the primary cusp and kink signals.
However, each order in memory is suppressed by an additional factor of $(G\mu)^2$, and this suppression is strong enough to render the memory contribution unobservable at all frequencies.

\subsection{Consequences for the loop distribution function}

Additional energy loss due to GW memory emission means that loops decay more rapidly, modifying the distribution of loop sizes.
This could in principle leave distinguishable imprints on the two observational probes mentioned in the previous sections.
However, we show below that such imprints are suppressed by powers of $G\mu$ in the cusp-collapse scenario, rendering them unobservable.

Following reference~\cite{Auclair:2019jip}, we write the loop decay rate as
    \begin{equation}
        \dv{\ell}{t}=-\Gamma G\mu\mathcal{J}(\ell),
    \end{equation}
    where $\Gamma\approx50$ is a numerical constant, and $\mathcal{J}(\ell)$ is an arbitrary function of the loop length which is equal to unity in the standard Nambu-Goto case.
If $\mathcal{J}$ differs from unity then the faster/slower loop decay rate leads to a smaller/greater number of loops once the equilibrium `scaling' distribution is reached; reference~\cite{Auclair:2019jip} shows that the modified loop distribution function $n(\ell,t)$ can then be obtained from the standard Nambu-Goto distribution by inserting appropriate factors of $\mathcal{J}$.

We account for enhanced energy loss through GW memory radiation by writing
    \begin{equation}
        \mathcal{J}(\ell)=\sum_{n=0}^\infty\frac{\mathcal{E}^{(n)}}{\mathcal{E}^{(0)}}\simeq1+\frac{\mathcal{E}^{(1)}}{\mathcal{E}^{(0)}},
    \end{equation}
    where $\mathcal{E}^{(n)}$ is the dimensionless GW energy radiated through $n$th-order memory, as given by equations~\eqref{eq:cusp-collapse-total-energy-0th-order},~\eqref{eq:cusp-collapse-total-energy-1st-order}, and~\eqref{eq:cusp-collapse-total-energy-nth-order} for cusps and equation~\eqref{eq:kink-total-energy} for kinks.
We find that $\mathcal{J}(\ell)-1$ is at most $\sim G\mu$ for both cusps and kinks, meaning that any modifications to the loop distribution function are suppressed by powers of $G\mu$, and are thus unobservable.
(This is assuming the memory divergence is resolved through cusp collapse; of course, it is still possible that the memory emission could be much larger, and thereby lead to significant imprints on the loop distribution, and indeed on the other observational probes we have discussed.)

\section{Summary and conclusion}
\label{sec:summary}

We have thoroughly explored the nonlinear GW memory associated with cusps and kinks on Nambu-Goto cosmic strings, deriving detailed analytical waveform models for the memory GWs, including the `memory of the memory' and other higher-order memory effects.
These are among the first memory observables computed for a cosmological source of GWs, with previous literature having focused almost entirely on astrophysical sources.

The leading-order cusp memory waveform~\eqref{eq:cusp-result} that we have found is strikingly similar to the primary GW signal from the cusp~\eqref{eq:cusp-waveform}, with the same characteristic $\sim f^{-4/3}$ frequency power-law.
However, one very important difference is that this memory signal is emitted in all directions, unlike the primary signal which is confined to a narrow beam of width $\theta_\rmb\sim f^{-1/3}$.
As a result, we find that the total GW energy radiated by the cusp memory \emph{diverges} for a Nambu-Goto (i.e., zero-width) loop.
This divergence can be regularised by introducing a cutoff at the scale of the string width $\delta\sim\ell_\Pl/\sqrt{G\mu}$, but this then introduces powers of $\ell/\delta\gg1$ to the higher-order memory terms, causing the sum of all memory contributions to diverge for loops of length $\ell\gtrsim\delta/(G\mu)^3$.

In section~\ref{sec:cusp-collapse} we have argued that the most plausible explanation for this unphysical memory divergence is the assumption that the spacetime containing the cosmic string loop is well-described by a flat background with linear perturbations.
We have suggested that some strong-gravity mechanism must kick in on loops of length $\ell\gtrsim\delta/(G\mu)^3$ to suppress the high-frequency GW emission from cusps, thereby curing the divergence.
Our earlier work reference~\cite{Jenkins:2020ctp} provides exactly such a mechanism: the collapse of a small cosmic string segment near the cusp to form a PBH.
Remarkably, this cusp-collapse process is predicted to occur for all loops of length $\ell\gtrsim\delta/(G\mu)^3$---exactly the same loops for which we have diagnosed the memory divergence.
We have shown explicitly that if cusp collapse does indeed occur for these loops, the corresponding GW memory is strongly suppressed, and the divergence is cured.

We have also calculated the memory emission associated with kinks, and have shown that this is suppressed due to interference between GWs emitted by the kink at different points in its history.
There is thus no situation in which the kink memory signal diverges; this accords with the cusp-collapse description, as kinks are not predicted to form PBHs in that scenario.

In section~\ref{sec:detection-prospects} we have investigated the detection prospects for these cusp and kink memory signals, calculating the expected detection rate of individual memory signals for matched-filter searches with third-generation GW observatories like Einstein Telescope and Cosmic Explorer, as well as the contribution of memory GWs to the SGWB spectrum, and possible imprints in the cosmic string loop distribution function.
We find that by requiring the string tension to agree with existing observational bounds ($G\mu\lesssim10^{-10}$) and invoking cusp collapse to prevent the memory from diverging, the resulting memory signal is very strongly suppressed, and is not likely to be detected by any current or upcoming GW observatories.
Of course, if cusp collapse does not occur, then it is possible that large loops could source much stronger memory signals; however, one would then need an alternative means of resolving the cusp divergence.

Our work demonstrates the importance of considering the nonlinear memory associated with a broader class of GW sources than just compact binaries; we have shown that the memory effect is interesting not just from an observational point of view, but also as a tool for sharpening our theoretical understanding and modelling of said GW sources.
For cosmic strings in particular, as an unexpected by-product of our analysis we have shown that the standard Nambu-Goto description of cusps is unphysical, and that strong gravity effects (possibly including PBH formation) could play an important r\^ole in a more complete understanding of their dynamics.
This motivates further work to better understand cusps in full GR, including nonlinear gravitational effects beyond those considered here, and thereby compute reliable waveform predictions for GW observatories.
It would also be very interesting to study the \emph{linear} memory associated with cusps, and to understand how this contributes to the total GW emission; however, such a study would require us to go beyond the Nambu-Goto approach adopted here.
More generally, our results motivate a broader examination of the nonlinear memory effect in GW astronomy and cosmology, to see what other surprises may be in store.

\begin{acknowledgments}
    We thank Josu Aurrekoetxea and Eric Thrane for their valuable feedback on this work.
    A.~C.~J. is supported by King's College London through a Graduate Teaching Scholarship.
    M.~S. is supported in part by the Science and Technology Facility Council (STFC), United Kingdom, under research grant No. ST/P000258/1.
    This is LIGO document number P2100040.
\end{acknowledgments}

\appendix
\section{Understanding the origin of the higher-order memory divergence}
\label{sec:rhdot}

Here we derive a condition~\eqref{eq:divergence} on a generic GW signal $h(t)$ which, we argue heuristically, is necessary for the associated nonlinear memory to diverge.
We find this condition by considering a simple toy model in which we can tune the `sharpness' of the signal to find where the divergence sets in.
We then show that this condition predicts the divergence for cusps on `large' cosmic string loops, while also providing an explanation for why the memory from compact binaries never diverges.

\subsection{Gaussian pulse as a toy model}
Our goal here is to derive a condition on how `sharp' a putative GW signal must be to give rise to a divergent nonlinear memory expansion.
To investigate this, we use a toy model in which the primary GW is a Gaussian pulse which reaches our detector at retarded time $t=0$,
    \begin{equation}
    \label{eq:pulse}
        h^{(0)}(t)=\frac{A}{r}\exp(-\frac{t^2}{2\sigma^2}).
    \end{equation}
This has amplitude $A$ and width $\sigma$, both of which have dimensions of length.
We ignore the polarisation and angular pattern of the pulse, though these can play an important role (e.g., for the case of kinks, as discussed above), and focus on the pulse's behaviour as a function of time.
The Fourier transform of equation~\eqref{eq:pulse} is
    \begin{equation}
        \tilde{h}^{(0)}(f)=\frac{A}{r}\sqrt{2\uppi}\sigma\exp(-\frac{(2\uppi f\sigma)^2}{2}),
    \end{equation}
    so we see that the pulse has a characteristic frequency of $\sim1/\sigma$.
The pulse is smooth and infinitely differentiable for all finite $\sigma$; however, by making $\sigma$ small, we can make the pulse arbitrarily `sharp' (equivalently, we can make the characteristic frequency arbitrarily large).
This sharpness can be quantified by the maximum value of the time derivative of the strain,
    \begin{align}
    \begin{split}
        \max_t|r\dot{h}^{(0)}(t)|&=\rme^{-1/2}A/\sigma\\
        &\sim[\text{amplitude}]\times[\text{characteristic frequency}],
    \end{split}
    \end{align}
    i.e. if the dimensionless ratio $A/\sigma$ is much greater than unity the pulse is `sharp', and if $A/\sigma$ is much less than unity the pulse is `soft'.

The pulse's first-order memory can be written as
    \begin{equation}
        r\dot{h}^{(1)}=C|r\dot{h}^{(0)}|^2=C\qty(\frac{At}{\sigma^2})^2\exp(-\frac{t^2}{\sigma^2}),
    \end{equation}
    where $C$ is some constant arising from the angular integral, which we assume to be $\order{1}$ for all orders in the memory expansion.
(This assumption can easily be violated: e.g., if the integrand has vanishing TT component then $C=0$.)
We immediately see that the memory signal is larger than the primary GW signal near the arrival time ($|t|\sim\sigma$) if and only if $A/\sigma\gg1$, or equivalently,
    \begin{equation}
    \label{eq:divergence}
        \max_t|r\dot{h}(t)|\gg1.
    \end{equation}
Using the Isaacson formula~\eqref{eq:flux}, we see that this is equivalent to
    \begin{equation}
        \max_t\qty|\dv{E_\mathrm{gw}}{t}|\gg\frac{1}{G}=\frac{m_\Pl}{t_\Pl},
    \end{equation}
    i.e. if the GW energy flux is super-Planckian.

Moving on to the second-order memory, there are two contributions: the self-energy of the first-order memory, and its cross-energy with the primary signal,
    \begin{align}
    \begin{split}
        r\dot{h}^{(2)}&=Cr^2(|\dot{h}^{(1)}+\dot{h}^{(0)}|^2-|\dot{h}^{(0)}|^2)\\
        &=C|r\dot{h}^{(1)}|^2+2Cr^2\dot{h}^{(1)}\dot{h}^{(0)}\\
        &=C^3\qty(\frac{At}{\sigma^2})^4\exp(-\frac{2t^2}{\sigma^2})\\
        &\qquad-2C^2\qty(\frac{At}{\sigma^2})^3\exp(-\frac{3t^2}{2\sigma^2}).
    \end{split}
    \end{align}
The general pattern is straightforward from here.
Treating the `sharp' regime $A/\sigma\gg1$ and the `soft' regime $A/\sigma\ll1$ separately (analogous to the separation between the `large loop' and `small loop' regimes for cusps), we find for all $n\ge1$,
    \begin{equation}
        r\dot{h}^{(n)}\simeq
        \begin{cases}
            \displaystyle\frac{1}{C}\qty[\frac{CAt}{\sigma^2}\exp(-\frac{t^2}{2\sigma^2})]^{2^n}, & A/\sigma\gg1\\[10pt]
            \displaystyle\frac{1}{4C}\qty[-\frac{2CAt}{\sigma^2}\exp(-\frac{t^2}{2\sigma^2})]^{n+1}, & A/\sigma\ll1
        \end{cases}
    \end{equation}
    which shows that the memory expansion diverges if equation~\eqref{eq:divergence} holds.

Of course, there are many ways in which this argument could fail, some of which we have already mentioned (in particular, if $C\ll1$).
However, the general takeaway is that if the time derivative of a GW strain signal is large, then a memory divergence \emph{might} occur.
For $\max_t|r\dot{h}|\ll1$ on the other hand, it seems very likely that the memory expansion must always converge.
If this is indeed the case, then we could interpret equation~\eqref{eq:divergence} as a necessary, but not sufficient, condition for the memory divergence.

\begin{figure*}[t!]
    \includegraphics[width=0.49\textwidth]{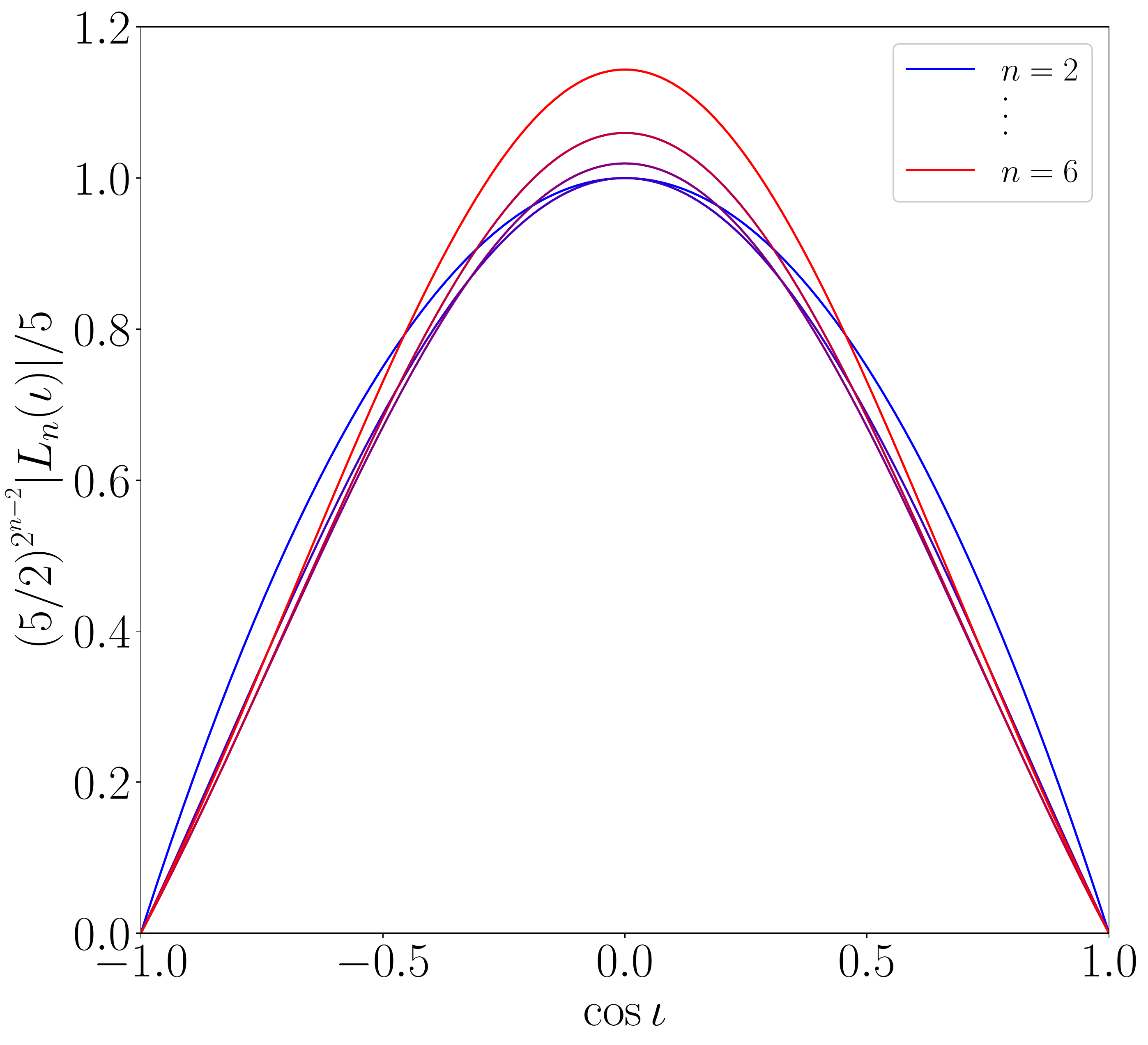}
    \includegraphics[width=0.49\textwidth]{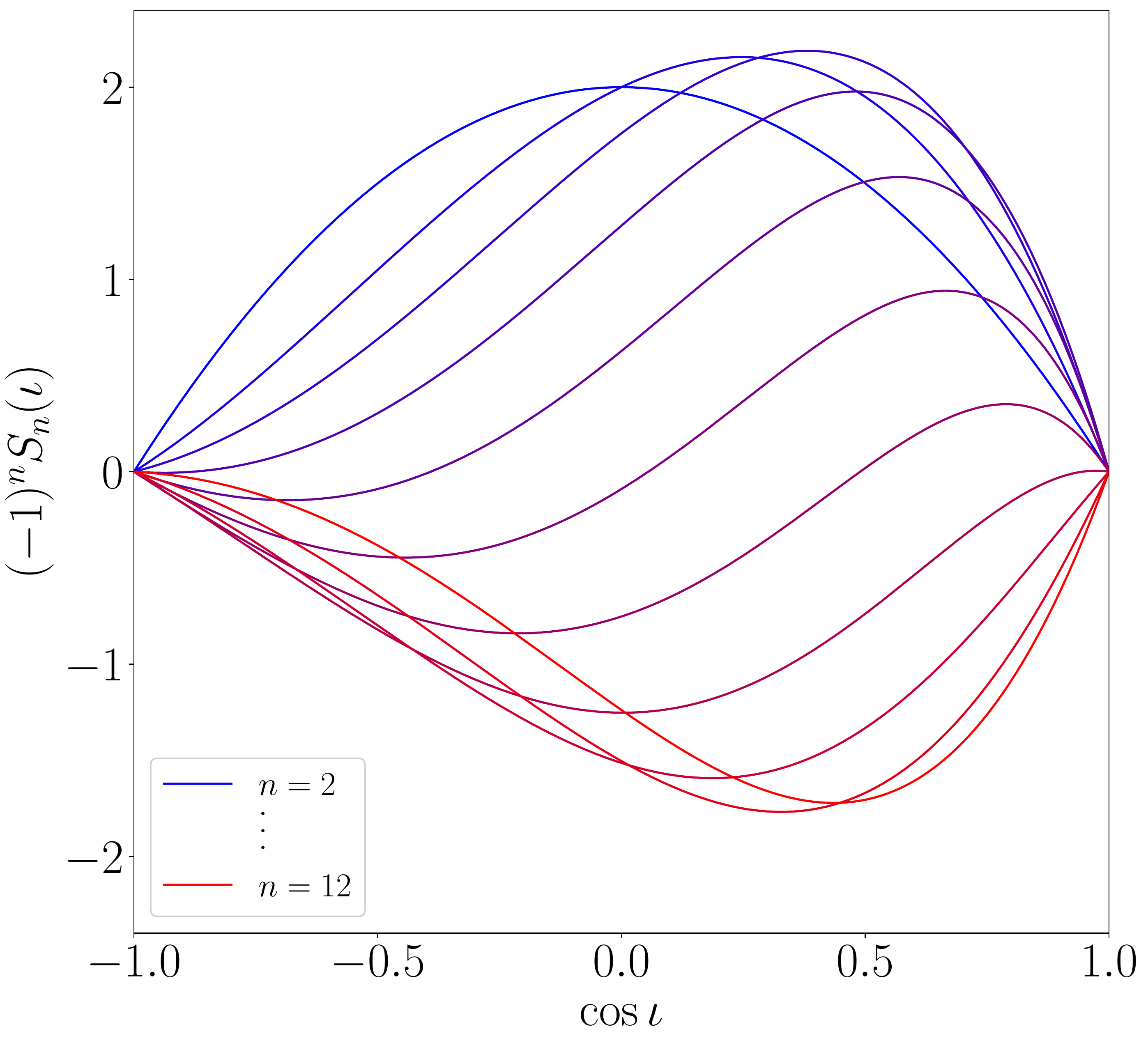}
    \caption{%
    Functions describing the strength of the $n$th-order cusp memory signal as a function of inclination, with $L_n(\iota)$ representing `large' loops $\ell\gg\ell_*$, and $S_n(\iota)$ representing `small' loops $\ell\ll\ell_*$.
    }
    \label{fig:L_n-S_n}
\end{figure*}

\subsection{Application to cosmic strings}
How does this fit into our results for cosmic string cusps?
Neglecting numerical constants, the time-domain cusp waveform looks like
    \begin{equation}
        h_\rmc^{(0)}(t)\sim-\frac{G\mu}{r}\ell^{2/3}|t|^{1/3}+\text{constant},
    \end{equation}
    so the time derivative \emph{diverges} at $t=0$ due to the absolute value function.
If we introduce a cutoff at the string width scale $\delta$, we find that
    \begin{equation}
        \max_t|r\dot{h}_\rmc^{(0)}|\sim G\mu(\ell/\delta)^{2/3}.
    \end{equation}
Naively applying the condition~\eqref{eq:divergence}, we would thus expect the cusp memory signal to diverge for loops of length
    \begin{equation}
        \ell\gtrsim\delta/(G\mu)^{3/2}.
    \end{equation}
However, this does \emph{not} happen, for the simple reason that there is a small factor $C\ll1$ arising from the angular integral, and as mentioned above this causes the condition~\eqref{eq:divergence} to fail.

We can circumvent this issue by considering the first-order memory signal from the cusp as the source of the divergence: this should give $C=\order{1}$ as it has a broad emission pattern, and is not concentrated into a narrow beam.
Indeed, using equation~\eqref{eq:cusp-time-domain-approx} we find that the maximum time derivative for the first-order cusp memory signal is
    \begin{equation}
        \max_t|r\dot{h}_\rmc^{(1)}|\sim(G\mu)^2(\ell/\delta)^{2/3}.
    \end{equation}
Applying the condition~\eqref{eq:divergence} again, we find that the cusp memory signal should diverge if
    \begin{equation}
        \ell\gtrsim\delta/(G\mu)^3,
    \end{equation}
    which is exactly what we find from the careful analysis in section~\ref{sec:higher-order-divergence}.
This supports the idea that equation~\eqref{eq:divergence} gives a necessary condition for the memory divergence, which is generally applicable if the factor $C$ arising from the angular integral is not too small.

\subsection{Application to compact binaries}
Here we use a very simple heuristic analysis to show that $|r\dot{h}(t)|$ is at most $\order{1}$ for compact binary coalescences, so that the memory divergence condition~\eqref{eq:divergence} never holds.
We neglect numerical factors throughout.

The GW signal from a CBC can be written as
    \begin{equation}
    \label{eq:cbc-strain}
        h(t)=\mathcal{A}(t)\rme^{\rmi\phi(t)},
    \end{equation}
    where the leading-order Newtonian contributions to the amplitude and phase are~\cite{Maggiore:1900zz}
    \begin{equation}
        \mathcal{A}\sim\frac{(G\mathcal{M})^{5/4}}{r\tau^{1/4}},\qquad\phi\sim\qty(\frac{\tau}{G\mathcal{M}})^{5/8},
    \end{equation}
    with $\tau$ the time until coalescence and $\mathcal{M}\equiv\eta^{3/5}M$ the chirp mass, where $\eta\le1/4$ is the dimensionless mass ratio.
The time derivative of the strain~\eqref{eq:cbc-strain} is therefore
    \begin{equation}
        |r\dot{h}(t)|=\sqrt{|r\dot{\mathcal{A}}|^2+|r\dot{\phi}\mathcal{A}|^2}\sim\sqrt{\qty(\frac{G\mathcal{M}}{\tau})^{5/2}+\qty(\frac{G\mathcal{M}}{\tau})^{5/4}}.
    \end{equation}
Formally, this diverges in the Newtonian analysis as $\tau\to0$.
However, introducing a cutoff at the ISCO radius truncates the signal at a finite value of $\tau$,
    \begin{equation}
        \tau_\mathrm{min}\sim\eta^{-8/5}G\mathcal{M},
    \end{equation}
    so that
    \begin{equation}
    \label{eq:cbc-max-hdot}
        \max_t|r\dot{h}(t)|\sim\sqrt{\eta^2+\eta^4}.
    \end{equation}
Since $\eta$ is at most $\order{1}$, we see that the condition~\eqref{eq:divergence} is never met.
This makes complete sense, given that we know that the memory from CBC signals cannot diverge.
As a by-product, equation~\eqref{eq:cbc-max-hdot} leads us to conjecture that equal-mass binaries ($\eta=1/4$) should give rise to stronger higher-order memory effects than extreme mass-ratio inspirals ($\eta\ll1$).

\section{Angular integrals for higher-order cusp memory}
\label{sec:iota-polynomials}

Here we derive the angular patterns of the higher-order cusp memory in the `large-loop' ($\ell\gg\ell_*$) and `small-loop' ($\ell\ll\ell_*$) limits, which are described by the functions $L_n(\iota)$ and $S_n(\iota)$ respectively, with $\iota$ the inclination between the cusp beaming direction and the observer's line of sight.
These are defined by inserting the $n$th-order memory formula~\eqref{eq:cusp-higher-order-final} into the iterative relation~\eqref{eq:iterate-memory}, which gives
    \begin{align}
    \begin{split}
    \label{eq:iota-polynomials-iterative}
        L_{n+1}(\iota)&=\int_{\vu*r'}|L_n(\iota')|^2,\\
        S_{n+1}(\iota)&=\int_{\vu*r'}6(1+\cos\iota')S_n(\iota'),
    \end{split}
    \end{align}
    for all $n\ge2$, with
    \begin{equation}
    \label{eq:iota-polynomials-iterative-n2}
        L_2(\iota)=S_2(\iota)=2\sin^2\iota.
    \end{equation}
(The integral symbol $\int_{\vu*r'}$ is defined in equation~\eqref{eq:integral-shorthand}.)

We find that $L_n$ and $S_n$ can be written as polynomials in $\cos\iota$ (of order $2^{n-1}$ and $n$, respectively).
The iterative process~\eqref{eq:iota-polynomials-iterative} can therefore be carried out by evaluating a simpler family of integrals,
    \begin{equation}
        C_n(\iota)\equiv\int_{\vu*r'}\cos^n\iota'.
    \end{equation}
To do so, we choose our $\vu*r'=(\theta',\phi')$ coordinates such that $\theta'$ is zero along the line of sight, so that $\cos\theta'\equiv\vu*r\vdot\vu*r'$, with the cusp beaming direction defined relative to the line of sight by $\vu*r_\rmc\vdot\vu*r=\cos\iota$.
We then have
    \begin{equation}
        \cos\iota'\equiv\vu*r_\rmc\vdot\vu*r'=\cos\theta'\cos\iota-\cos\phi'\sin\theta'\sin\iota,
    \end{equation}
    and equation~\eqref{eq:integral-shorthand} becomes
    \begin{equation}
        \int_{\vu*r'}=\int_{S^2}\frac{\dd[2]{\vu*r'}}{4\uppi}(1+\cos\theta')\rme^{-2\rmi\phi'},
    \end{equation}
    so that we can use a binomial expansion of $\cos^n\iota'=(\cos\theta'\cos\iota-\cos\phi'\sin\theta'\sin\iota)^n$ to write
    \begin{align}
    \begin{split}
        C_n(\iota)&=\sum_{k=0}^n
        \begin{pmatrix}
            n \\ k
        \end{pmatrix}
        (-\sin\iota)^k\cos^{n-k}\iota\\
        &\times\int_{-1}^{+1}\frac{\dd{x}}{2}(1+x)x^{n-k}(1-x^2)^{k/2}\\
        &\times\int_0^{2\uppi}\frac{\dd{\phi'}}{2\uppi}\rme^{-2\rmi\phi'}\cos^k\phi',
    \end{split}
    \end{align}
    where we have set $x=\cos\theta'$.
Using the Beta function identity
    \begin{equation}
        \mathrm{B}(a,b)\equiv\int_0^1\dd{x}x^{a-1}(1-x)^{b-1}=\frac{\Gamma(a)\Gamma(b)}{\Gamma(a+b)},
    \end{equation}
    we then obtain, for all $n\ge0$,
    \begin{align}
    \begin{split}
        C_{2n}(\iota)&=\sum_{k=0}^n\frac{\sqrt{\uppi}(n-k)(2n)!\cos^{2k}\iota\sin^{2(n-k)}\iota}{2^{2n+1}k!(n-k+1)!\Gamma(n+\tfrac{3}{2})},\\
        C_{2n+1}(\iota)&=\sum_{k=0}^n\frac{\sqrt{\uppi}(n-k)(2n+1)!\cos^{2k+1}\iota\sin^{2(n-k)}\iota}{2^{2n+2}k!(n-k+1)!\Gamma(n+\tfrac{5}{2})}.
    \end{split}
    \end{align}
Calculating each iteration of equation~\eqref{eq:iota-polynomials-iterative} is then reduced to writing down the appropriate linear combination of $C_n(\iota)$ for different $n$.

The resulting expressions for $S_n(\iota)$ and $L_n(\iota)$ for the first few $n\ge2$ are shown in Tables~\ref{tab:S_n} and~\ref{tab:L_n} respectively.
We find empirically that the large-loop angular functions can be approximated by
    \begin{equation}
    \label{eq:L_n-approx}
        |L_n(\iota)|\approx5(2/5)^{2^{n-2}}\sin^2\iota
    \end{equation}
    with an accuracy of $\sim10\%$, as is illustrated in the left panel of figure~\ref{fig:L_n-S_n}.
The small-loop angular functions $S_n(\iota)$ do not seem to follow such a simple pattern.

\begin{table*}[p!]
    \caption{\label{tab:L_n}%
    The first few large-loop angular functions, as defined by equations~\eqref{eq:iota-polynomials-iterative} and~\eqref{eq:iota-polynomials-iterative-n2}.
    Note that $L_n(\iota)$ is a polynomial in $\cos\iota$ of order $2^{n-1}$, such that the number of terms grows exponentially with $n$.
    Despite this apparent complexity, we find that these formulae can be approximated by the simple expression~\eqref{eq:L_n-approx}.}
    \begin{ruledtabular}
    \begin{tabular}{c | c}
        $n$ & $L_n(\iota)$ \\
        \hline
        $2$ & $2\sin^2\iota$ \\
        $3$ & $-\dfrac{2\sin^2\iota}{15}(5-\cos2\iota)$ \\
        $4$ & $\dfrac{\sin^2\iota}{141750}[3737\cos2\iota-262\cos4\iota+7(-2070+\cos6\iota)]$ \\
        $5$ & $\dfrac{\sin^2\iota}{123092512042500000}\bigg[91170878299511\cos2\iota-8844505203863\cos4\iota+627703754313\cos6\iota-31676232034\cos8\iota$ \\
        & $+1106354755\cos10\iota+1001(-326445712830-24241\cos12\iota+245\cos14\iota)\bigg]$ \\
        $6$ & $\dfrac{\sin^2\iota}{4375937836474054406594550954572130000000000000000}\bigg[2337436144549410554623113364688535605014856\cos2\iota$ \\
        & $-252383762462012251730735377642370477964415\cos4\iota+22676810284676710047197528844734128748260\cos6\iota$ \\
        & $-1739064548768558706903275538411759890046\cos8\iota+114707816611411474747133898122922244740\cos10\iota$ \\
        & $-6538676998608577029462451915641765109\cos12\iota+322110616165743776447552901776475040\cos14\iota$ \\
        & $-13638911209793535304662793111510580\cos16\iota+490882191972236519729403447127344\cos18\iota$ \\
        & $+1463(-10081167364753857595829662265\cos20\iota+ 246246861424248720285801892\cos22\iota$ \\
        & $+1495(-3145080481324598891074\cos24\iota+44031911264492955804\cos26\iota$ \\
        & $+6525(-569906751862925826959924173050076-61436527080929\cos28\iota+271199768840\cos30\iota)))\bigg]$ \\
    \end{tabular}
    \end{ruledtabular}
\end{table*}

\begin{table*}[p!]
    \caption{\label{tab:S_n}%
    The first few small-loop angular functions, as defined by equations~\eqref{eq:iota-polynomials-iterative} and~\eqref{eq:iota-polynomials-iterative-n2}.
    Note that $S_n(\iota)$ is a polynomial in $\cos\iota$ of order $n$.}
    \begin{ruledtabular}
    \begin{tabular}{c | c}
        $n$ & $S_n(\iota)$ \\
        \hline
        $2$ & $2\sin^2\iota$ \\
        $3$ & $-\dfrac{2\sin^2\iota}{5}(5+3\cos\iota)$ \\
        $4$ & $\dfrac{2\sin^2\iota}{25}(25+24\cos\iota+3\cos2\iota)$ \\
        $5$ & $-\dfrac{2\sin^2\iota}{175}(154+201\cos\iota+42\cos2\iota+3\cos3\iota)$ \\
        $6$ & $\dfrac{\sin^2\iota}{12250}(15575+28032\cos\iota+7932\cos2\iota+960\cos3\iota+45\cos4\iota)$ \\
        $7$ & $-\dfrac{\sin^2\iota}{245000}[466422\cos\iota+172824\cos2\iota+5(29702+5619\cos3\iota+450\cos4\iota+15\cos5\iota)]$ \\
        $8$ & $\dfrac{\sin^2\iota}{245000}(-30194+299024\cos\iota+157771\cos2\iota+32552\cos3\iota+3490\cos4\iota+200\cos5\iota+5\cos6\iota)$ \\
        $9$ & $-\dfrac{\sin^2\iota}{13475000}[4934669\cos\iota+6432382\cos2\iota+5(-2137388+347047\cos3\iota+46508\cos4\iota+3565\cos5\iota+154\cos6\iota+3\cos7\iota)]$ \\
        $10$ & $\dfrac{\sin^2\iota}{10375750000}\bigg[-5257311104\cos\iota+2472686192\cos2\iota+5(214599584\cos3\iota+36187180\cos4\iota$ \\
        & $+7(-383512899+491360\cos5\iota+28400\cos6\iota+960\cos7\iota+15\cos8\iota))\bigg]$ \\
        $11$ & $-\dfrac{\sin^2\iota}{1888386500000}\bigg[-2359095296230\cos\iota-55089058544\cos2\iota+116827655852\cos3\iota+27832319320\cos4\iota$ \\
        & $+3261899900\cos5\iota+49(-59685302066+4749680\cos6\iota+214995\cos7\iota+5850\cos8\iota+75\cos9\iota)\bigg]$ \\
        $12$ & $\dfrac{\sin^2\iota}{122745122500000}\bigg[-213800187213104\cos\iota-34153675430026\cos2\iota+5(-37491558286234+292405153376\cos3\iota$ \\
        & $+238091184664\cos4\iota+37145938400\cos5\iota+3229010267\cos6\iota$ \\
        & $+179962104\cos7\iota+6563970\cos8\iota+147000\cos9\iota+1575\cos10\iota)\bigg]$
    \end{tabular}
    \end{ruledtabular}
\end{table*}

\begin{figure}[t!]
    \includegraphics[width=0.48\textwidth]{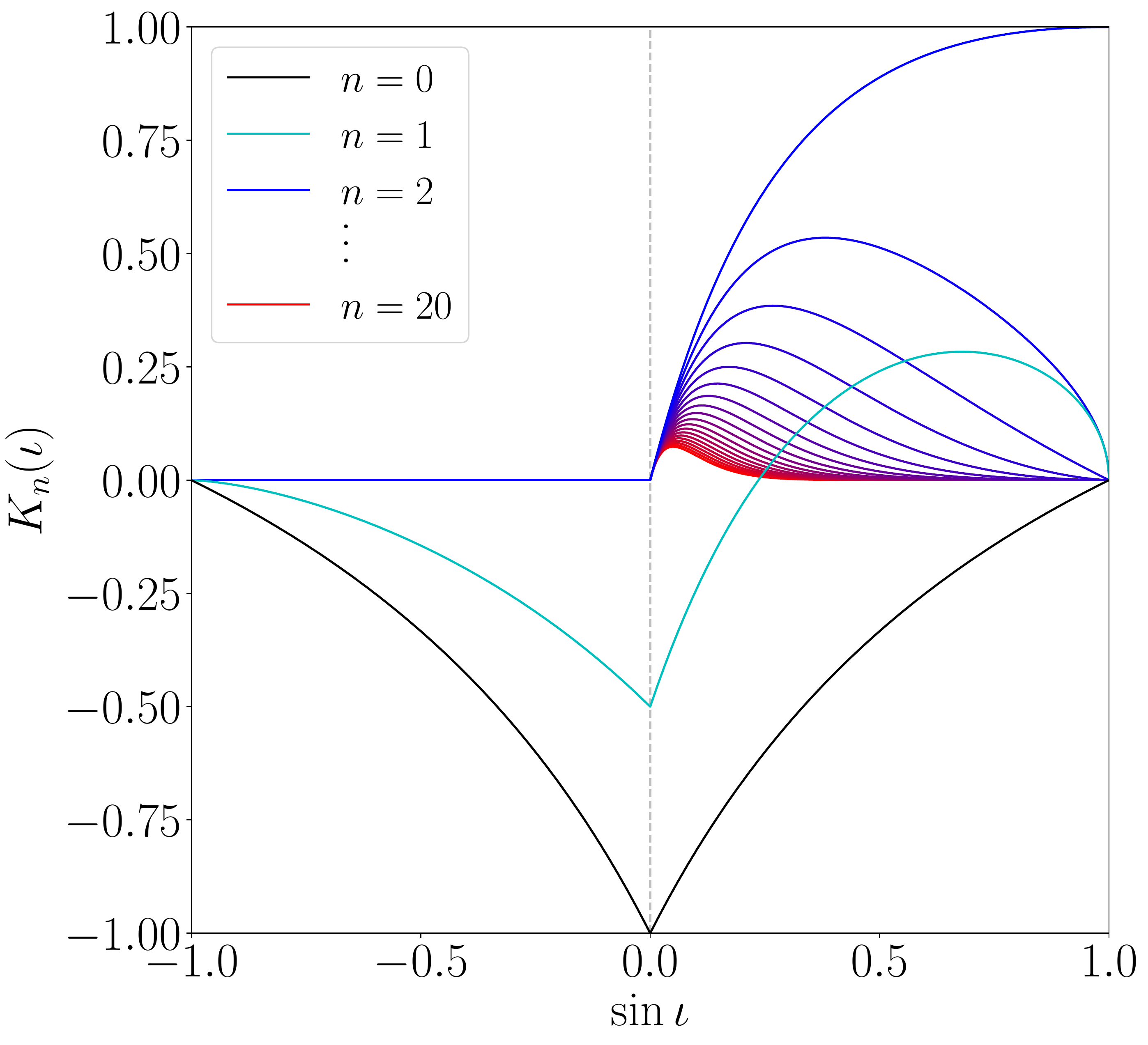}
    \caption{%
    The angular integral $K_n(\iota)$ for non-negative $n$, as given by equation~\eqref{eq:K_n-final}.
    The corresponding curves for negative $n$ are obtained by reflecting around $\iota=0$.
    }
    \label{fig:kink-angular-integral}
\end{figure}

\section{Angular integrals for kink memory}
\label{sec:K_n}

Here we compute the integral $K_n(\iota)$ defined in equation~\eqref{eq:K_n-definition} for all angular modes $n\in\mathbb{Z}$.
Note that while the integrand is complex, the integral itself is always real, as the imaginary part of the integrand is an odd function of $\phi$.
Note also that we can focus on non-negative $n\ge0$ by exploiting the symmetry property
    \begin{equation}
        K_{-n}(\iota)=K_n(-\iota).
    \end{equation}

For some special values of $\iota$ we can obtain the full spectrum immediately from equation~\eqref{eq:K_n-definition},
    \begin{align}
    \begin{split}
    \label{eq:K_n-special-values}
        K_n(0)&=-\delta_{n,0}-\frac{1}{2}(\delta_{n,1}+\delta_{n,-1}),\\
        K_n(\pm\uppi/2)&=\delta_{n,\pm2}.
    \end{split}
    \end{align}
For general values $\iota\in[-\uppi/2,+\uppi/2]$, we proceed by expanding the denominator of equation~\eqref{eq:K_n-definition} using the geometric series and the binomial theorem,
    \begin{align}
    \begin{split}
    \label{eq:K_n-denominator-series}
        \frac{1}{1-\cos\iota\cos\phi}&=\sum_{k=0}^\infty\cos^k\iota\cos^k\phi\\
        &=\sum_{k=0}^\infty\sum_{m=0}^k
        \begin{pmatrix}
            k \\ m
        \end{pmatrix}
        \frac{\cos^k\iota}{2^k}\rme^{-\rmi(k-2m)\phi}.
    \end{split}
    \end{align}
This converges everywhere on $\iota\in[-\uppi/2,+\uppi/2]$ except for $\iota=0$, in which case we use equation~\eqref{eq:K_n-special-values} instead.
We can then integrate term-by-term to extract the contribution from each order in the series.
The result, valid for all $n\ge0$, is
    \begin{align}
    \begin{split}
        &K_n(\iota)=\\
        &\sum_{k=\max(0,2-n)}^\infty
        \begin{pmatrix}
            2k+n-2 \\ k
        \end{pmatrix}
        \qty(\frac{\cos\iota}{2})^{2k+n-2}\qty(\frac{1+\sin\iota}{2})^2\\
        &-2\sum_{k=0}^\infty
        \begin{pmatrix}
            2k+n \\ k
        \end{pmatrix}
        \qty(\frac{\cos\iota}{2})^{2k+n+2}\\
        &+\sum_{k=0}^\infty
        \begin{pmatrix}
            2k+n+2 \\ k
        \end{pmatrix}
        \qty(\frac{\cos\iota}{2})^{2k+n+2}\qty(\frac{1-\sin\iota}{2})^2.
    \end{split}
    \end{align}
The first sum must be carried out separately for the three cases $n=0$, $n=1$, and $n>1$, yielding the following expressions:
    \begin{align}
    \begin{split}
    \label{eq:K_n-final}
        K_0(\iota)&=\frac{4\sin|\iota|+\cos2\iota-3}{2\cos^2\iota},\\
        K_1(\iota)&=\frac{\sin|\iota|-1}{2\sin|\iota|}\bigg[\cos\iota\\
        &+(\sin\iota-1)\frac{1-\sin|\iota|+\sin\iota(1+2\sin\iota+\sin|\iota|)}{\cos^3\iota}\bigg],\\
        K_n(\iota)&=
        \begin{cases}
            \dfrac{4\sin\iota}{\cos^2\iota}\qty(\dfrac{\cos\iota}{1+\sin\iota})^n, & \iota>0\\
            0, & \iota\le0
        \end{cases},\qquad\text{for }n>1.
    \end{split}
    \end{align}
These clearly agree with equation~\eqref{eq:K_n-special-values} for $\iota=\pm\uppi/2$.
Note that for all $n$, the integral $K_n(\iota)$ is not differentiable at $\iota=0$; this is related to the fact that the series in equation~\eqref{eq:K_n-denominator-series} diverges at $\iota=0$.
However, despite this formal divergence, we see that equation~\eqref{eq:K_n-final} agrees with equation~\eqref{eq:K_n-special-values} in the limit $\iota\to0$, whether this limit is taken from above or from below.
We therefore use equation~\eqref{eq:K_n-final} over the full domain $\iota\in[-\uppi/2,+\uppi/2]$.
The resulting curves for the first few non-negative $n$ are illustrated in figure~\ref{fig:kink-angular-integral}.

For $n\ge2$, the integral only has support for $\iota>0$, and peaks at an inclination $\iota_*$ given by
    \begin{equation}
        \sin\iota_*=\frac{1}{2}(n-\sqrt{n^2-4}).
    \end{equation}
When calculating the kink memory signal we are interested in the high-frequency regime, which corresponds to large $n$.
In the limit $n\to\infty$, we have
    \begin{equation}
        \iota_*\simeq1/n,\qquad K_n(\iota_*)\simeq4/(\rme n),
    \end{equation}
    meaning that the memory signal is strongly suppressed, and is only observable very close to (but strictly outside of) the plane of the kink.

\bibliography{memory}

\begin{thebibliography}{135}%
\makeatletter
\providecommand \@ifxundefined [1]{%
 \@ifx{#1\undefined}
}%
\providecommand \@ifnum [1]{%
 \ifnum #1\expandafter \@firstoftwo
 \else \expandafter \@secondoftwo
 \fi
}%
\providecommand \@ifx [1]{%
 \ifx #1\expandafter \@firstoftwo
 \else \expandafter \@secondoftwo
 \fi
}%
\providecommand \natexlab [1]{#1}%
\providecommand \enquote  [1]{``#1''}%
\providecommand \bibnamefont  [1]{#1}%
\providecommand \bibfnamefont [1]{#1}%
\providecommand \citenamefont [1]{#1}%
\providecommand \href@noop [0]{\@secondoftwo}%
\providecommand \href [0]{\begingroup \@sanitize@url \@href}%
\providecommand \@href[1]{\@@startlink{#1}\@@href}%
\providecommand \@@href[1]{\endgroup#1\@@endlink}%
\providecommand \@sanitize@url [0]{\catcode `\\12\catcode `\$12\catcode
  `\&12\catcode `\#12\catcode `\^12\catcode `\_12\catcode `\%12\relax}%
\providecommand \@@startlink[1]{}%
\providecommand \@@endlink[0]{}%
\providecommand \url  [0]{\begingroup\@sanitize@url \@url }%
\providecommand \@url [1]{\endgroup\@href {#1}{\urlprefix }}%
\providecommand \urlprefix  [0]{URL }%
\providecommand \Eprint [0]{\href }%
\providecommand \doibase [0]{http://dx.doi.org/}%
\providecommand \selectlanguage [0]{\@gobble}%
\providecommand \bibinfo  [0]{\@secondoftwo}%
\providecommand \bibfield  [0]{\@secondoftwo}%
\providecommand \translation [1]{[#1]}%
\providecommand \BibitemOpen [0]{}%
\providecommand \bibitemStop [0]{}%
\providecommand \bibitemNoStop [0]{.\EOS\space}%
\providecommand \EOS [0]{\spacefactor3000\relax}%
\providecommand \BibitemShut  [1]{\csname bibitem#1\endcsname}%
\let\auto@bib@innerbib\@empty
\bibitem [{\citenamefont {Abbott}\ \emph {et~al.}(2016)\citenamefont {Abbott}
  \emph {et~al.}}]{TheLIGOScientific:2016src}%
  \BibitemOpen
  \bibfield  {author} {\bibinfo {author} {\bibfnamefont {B.~P.}\ \bibnamefont
  {Abbott}} \emph {et~al.} (\bibinfo {collaboration} {LIGO Scientific
  Collaboration, Virgo Collaboration}),\ }\bibfield  {title} {\enquote
  {\bibinfo {title} {{Tests of general relativity with GW150914}},}\ }\href
  {\doibase 10.1103/PhysRevLett.116.221101} {\bibfield  {journal} {\bibinfo
  {journal} {Phys. Rev. Lett.}\ }\textbf {\bibinfo {volume} {116}},\ \bibinfo
  {pages} {221101} (\bibinfo {year} {2016})},\ \bibinfo {note} {[Erratum: Phys.
  Rev. Lett. 121, 129902 (2018)]},\ \Eprint {http://arxiv.org/abs/1602.03841}
  {arXiv:1602.03841 [gr-qc]} \BibitemShut {NoStop}%
\bibitem [{\citenamefont {Yunes}\ \emph {et~al.}(2016)\citenamefont {Yunes},
  \citenamefont {Yagi},\ and\ \citenamefont {Pretorius}}]{Yunes:2016jcc}%
  \BibitemOpen
  \bibfield  {author} {\bibinfo {author} {\bibfnamefont {Nicolas}\ \bibnamefont
  {Yunes}}, \bibinfo {author} {\bibfnamefont {Kent}\ \bibnamefont {Yagi}}, \
  and\ \bibinfo {author} {\bibfnamefont {Frans}\ \bibnamefont {Pretorius}},\
  }\bibfield  {title} {\enquote {\bibinfo {title} {{Theoretical Physics
  Implications of the Binary Black-Hole Mergers GW150914 and GW151226}},}\
  }\href {\doibase 10.1103/PhysRevD.94.084002} {\bibfield  {journal} {\bibinfo
  {journal} {Phys. Rev. D}\ }\textbf {\bibinfo {volume} {94}},\ \bibinfo
  {pages} {084002} (\bibinfo {year} {2016})},\ \Eprint
  {http://arxiv.org/abs/1603.08955} {arXiv:1603.08955 [gr-qc]} \BibitemShut
  {NoStop}%
\bibitem [{\citenamefont {Abbott}\ \emph
  {et~al.}(2021{\natexlab{a}})\citenamefont {Abbott} \emph
  {et~al.}}]{LIGOScientific:2020tif}%
  \BibitemOpen
  \bibfield  {author} {\bibinfo {author} {\bibfnamefont {R.}~\bibnamefont
  {Abbott}} \emph {et~al.} (\bibinfo {collaboration} {LIGO Scientific
  Collaboration, Virgo Collaboration}),\ }\bibfield  {title} {\enquote
  {\bibinfo {title} {{Tests of general relativity with binary black holes from
  the second LIGO-Virgo gravitational-wave transient catalog}},}\ }\href
  {\doibase 10.1103/PhysRevD.103.122002} {\bibfield  {journal} {\bibinfo
  {journal} {Phys. Rev. D}\ }\textbf {\bibinfo {volume} {103}},\ \bibinfo
  {pages} {122002} (\bibinfo {year} {2021}{\natexlab{a}})},\ \Eprint
  {http://arxiv.org/abs/2010.14529} {arXiv:2010.14529 [gr-qc]} \BibitemShut
  {NoStop}%
\bibitem [{\citenamefont {Braginsky}\ and\ \citenamefont
  {Grishchuk}(1985)}]{Braginsky:1986ia}%
  \BibitemOpen
  \bibfield  {author} {\bibinfo {author} {\bibfnamefont {V.~B.}\ \bibnamefont
  {Braginsky}}\ and\ \bibinfo {author} {\bibfnamefont {L.~P.}\ \bibnamefont
  {Grishchuk}},\ }\bibfield  {title} {\enquote {\bibinfo {title} {{Kinematic
  Resonance and Memory Effect in Free Mass Gravitational Antennas}},}\ }\href
  {http://www.jetp.ac.ru/cgi-bin/e/index/e/62/3/p427?a=list} {\bibfield
  {journal} {\bibinfo  {journal} {Sov. Phys. JETP}\ }\textbf {\bibinfo {volume}
  {62}},\ \bibinfo {pages} {427--430} (\bibinfo {year} {1985})}\BibitemShut
  {NoStop}%
\bibitem [{\citenamefont {Braginsky}\ and\ \citenamefont
  {Thorne}(1987)}]{Braginsky:1987gw}%
  \BibitemOpen
  \bibfield  {author} {\bibinfo {author} {\bibfnamefont {Vladimir~B.}\
  \bibnamefont {Braginsky}}\ and\ \bibinfo {author} {\bibfnamefont {Kip~S.}\
  \bibnamefont {Thorne}},\ }\bibfield  {title} {\enquote {\bibinfo {title}
  {{Gravitational-wave bursts with memory and experimental prospects}},}\
  }\href {\doibase 10.1038/327123a0} {\bibfield  {journal} {\bibinfo  {journal}
  {Nature}\ }\textbf {\bibinfo {volume} {327}},\ \bibinfo {pages} {123--125}
  (\bibinfo {year} {1987})}\BibitemShut {NoStop}%
\bibitem [{\citenamefont {Favata}(2010)}]{Favata:2010zu}%
  \BibitemOpen
  \bibfield  {author} {\bibinfo {author} {\bibfnamefont {Marc}\ \bibnamefont
  {Favata}},\ }\bibfield  {title} {\enquote {\bibinfo {title} {{The
  gravitational-wave memory effect}},}\ }\href {\doibase
  10.1088/0264-9381/27/8/084036} {\bibfield  {journal} {\bibinfo  {journal}
  {Class. Quant. Grav.}\ }\textbf {\bibinfo {volume} {27}},\ \bibinfo {pages}
  {084036} (\bibinfo {year} {2010})},\ \Eprint {http://arxiv.org/abs/1003.3486}
  {arXiv:1003.3486 [gr-qc]} \BibitemShut {NoStop}%
\bibitem [{\citenamefont {Zel'dovich}\ and\ \citenamefont
  {Polnarev}(1974)}]{Zeldovich:1974gvh}%
  \BibitemOpen
  \bibfield  {author} {\bibinfo {author} {\bibfnamefont {Y.~B.}\ \bibnamefont
  {Zel'dovich}}\ and\ \bibinfo {author} {\bibfnamefont {A.~G.}\ \bibnamefont
  {Polnarev}},\ }\bibfield  {title} {\enquote {\bibinfo {title} {{Radiation of
  gravitational waves by a cluster of superdense stars}},}\ }\href
  {https://ui.adsabs.harvard.edu/abs/1974SvA....18...17Z/abstract} {\bibfield
  {journal} {\bibinfo  {journal} {Sov. Astron.}\ }\textbf {\bibinfo {volume}
  {18}},\ \bibinfo {pages} {17} (\bibinfo {year} {1974})}\BibitemShut {NoStop}%
\bibitem [{\citenamefont {Smarr}(1977)}]{Smarr:1977fy}%
  \BibitemOpen
  \bibfield  {author} {\bibinfo {author} {\bibfnamefont {Larry}\ \bibnamefont
  {Smarr}},\ }\bibfield  {title} {\enquote {\bibinfo {title} {{Gravitational
  Radiation from Distant Encounters and from Headon Collisions of Black Holes:
  The Zero Frequency Limit}},}\ }\href {\doibase 10.1103/PhysRevD.15.2069}
  {\bibfield  {journal} {\bibinfo  {journal} {Phys. Rev. D}\ }\textbf {\bibinfo
  {volume} {15}},\ \bibinfo {pages} {2069--2077} (\bibinfo {year}
  {1977})}\BibitemShut {NoStop}%
\bibitem [{\citenamefont {Turner}(1977)}]{Turner:1977hyp}%
  \BibitemOpen
  \bibfield  {author} {\bibinfo {author} {\bibfnamefont {Michael}\ \bibnamefont
  {Turner}},\ }\bibfield  {title} {\enquote {\bibinfo {title} {{Gravitational
  radiation from point-masses in unbound orbits: Newtonian results}},}\ }\href
  {\doibase 10.1086/155501} {\bibfield  {journal} {\bibinfo  {journal}
  {Astrophys. J.}\ }\textbf {\bibinfo {volume} {216}},\ \bibinfo {pages}
  {610--619} (\bibinfo {year} {1977})}\BibitemShut {NoStop}%
\bibitem [{\citenamefont {Turner}\ and\ \citenamefont
  {Will}(1978)}]{Turner:1978zz}%
  \BibitemOpen
  \bibfield  {author} {\bibinfo {author} {\bibfnamefont {Michael}\ \bibnamefont
  {Turner}}\ and\ \bibinfo {author} {\bibfnamefont {Clifford~M.}\ \bibnamefont
  {Will}},\ }\bibfield  {title} {\enquote {\bibinfo {title} {{Post-Newtonian
  gravitational bremsstrahlung}},}\ }\href {\doibase 10.1086/155996} {\bibfield
   {journal} {\bibinfo  {journal} {Astrophys. J.}\ }\textbf {\bibinfo {volume}
  {220}},\ \bibinfo {pages} {1107--1124} (\bibinfo {year} {1978})}\BibitemShut
  {NoStop}%
\bibitem [{\citenamefont {Kov\'acs}\ and\ \citenamefont
  {Thorne}(1978)}]{Kovacs:1978eu}%
  \BibitemOpen
  \bibfield  {author} {\bibinfo {author} {\bibfnamefont {S\'andor~J.}\
  \bibnamefont {Kov\'acs}}\ and\ \bibinfo {author} {\bibfnamefont {Kip~S.}\
  \bibnamefont {Thorne}},\ }\bibfield  {title} {\enquote {\bibinfo {title}
  {{The Generation of Gravitational Waves. 4. Bremsstrahlung}},}\ }\href
  {\doibase 10.1086/156350} {\bibfield  {journal} {\bibinfo  {journal}
  {Astrophys. J.}\ }\textbf {\bibinfo {volume} {224}},\ \bibinfo {pages}
  {62--85} (\bibinfo {year} {1978})}\BibitemShut {NoStop}%
\bibitem [{\citenamefont {Bontz}\ and\ \citenamefont
  {Price}(1979)}]{Bontz:1979zfl}%
  \BibitemOpen
  \bibfield  {author} {\bibinfo {author} {\bibfnamefont {Robert~J.}\
  \bibnamefont {Bontz}}\ and\ \bibinfo {author} {\bibfnamefont {Richard~H.}\
  \bibnamefont {Price}},\ }\bibfield  {title} {\enquote {\bibinfo {title} {{The
  spectrum of radiation at low frequencies}},}\ }\href {\doibase
  10.1086/156880} {\bibfield  {journal} {\bibinfo  {journal} {Astrophys. J.}\
  }\textbf {\bibinfo {volume} {228}},\ \bibinfo {pages} {560--575} (\bibinfo
  {year} {1979})}\BibitemShut {NoStop}%
\bibitem [{\citenamefont {Epstein}(1978)}]{Epstein:1978dv}%
  \BibitemOpen
  \bibfield  {author} {\bibinfo {author} {\bibfnamefont {Reuben}\ \bibnamefont
  {Epstein}},\ }\bibfield  {title} {\enquote {\bibinfo {title} {{The Generation
  of Gravitational Radiation by Escaping Supernova Neutrinos}},}\ }\href
  {\doibase 10.1086/156337} {\bibfield  {journal} {\bibinfo  {journal}
  {Astrophys. J.}\ }\textbf {\bibinfo {volume} {223}},\ \bibinfo {pages}
  {1037--1045} (\bibinfo {year} {1978})}\BibitemShut {NoStop}%
\bibitem [{\citenamefont {Turner}(1978)}]{Turner:1978jj}%
  \BibitemOpen
  \bibfield  {author} {\bibinfo {author} {\bibfnamefont {Michael~S.}\
  \bibnamefont {Turner}},\ }\bibfield  {title} {\enquote {\bibinfo {title}
  {{Gravitational Radiation from Supernova Neutrino Bursts}},}\ }\href
  {\doibase 10.1038/274565a0} {\bibfield  {journal} {\bibinfo  {journal}
  {Nature}\ }\textbf {\bibinfo {volume} {274}},\ \bibinfo {pages} {565--566}
  (\bibinfo {year} {1978})}\BibitemShut {NoStop}%
\bibitem [{\citenamefont {Burrows}\ and\ \citenamefont
  {Hayes}(1996)}]{Burrows:1995bb}%
  \BibitemOpen
  \bibfield  {author} {\bibinfo {author} {\bibfnamefont {Adam}\ \bibnamefont
  {Burrows}}\ and\ \bibinfo {author} {\bibfnamefont {John}\ \bibnamefont
  {Hayes}},\ }\bibfield  {title} {\enquote {\bibinfo {title} {{Pulsar recoil
  and gravitational radiation due to asymmetrical stellar collapse and
  explosion}},}\ }\href {\doibase 10.1103/PhysRevLett.76.352} {\bibfield
  {journal} {\bibinfo  {journal} {Phys. Rev. Lett.}\ }\textbf {\bibinfo
  {volume} {76}},\ \bibinfo {pages} {352--355} (\bibinfo {year} {1996})},\
  \Eprint {http://arxiv.org/abs/astro-ph/9511106} {arXiv:astro-ph/9511106}
  \BibitemShut {NoStop}%
\bibitem [{\citenamefont {Kotake}\ \emph {et~al.}(2006)\citenamefont {Kotake},
  \citenamefont {Sato},\ and\ \citenamefont {Takahashi}}]{Kotake:2005zn}%
  \BibitemOpen
  \bibfield  {author} {\bibinfo {author} {\bibfnamefont {Kei}\ \bibnamefont
  {Kotake}}, \bibinfo {author} {\bibfnamefont {Katsuhiko}\ \bibnamefont
  {Sato}}, \ and\ \bibinfo {author} {\bibfnamefont {Keitaro}\ \bibnamefont
  {Takahashi}},\ }\bibfield  {title} {\enquote {\bibinfo {title} {{Explosion
  mechanism, neutrino burst, and gravitational wave in core-collapse
  supernovae}},}\ }\href {\doibase 10.1088/0034-4885/69/4/R03} {\bibfield
  {journal} {\bibinfo  {journal} {Rept. Prog. Phys.}\ }\textbf {\bibinfo
  {volume} {69}},\ \bibinfo {pages} {971--1144} (\bibinfo {year} {2006})},\
  \Eprint {http://arxiv.org/abs/astro-ph/0509456} {arXiv:astro-ph/0509456}
  \BibitemShut {NoStop}%
\bibitem [{\citenamefont {Ott}(2009)}]{Ott:2008wt}%
  \BibitemOpen
  \bibfield  {author} {\bibinfo {author} {\bibfnamefont {Christian~D.}\
  \bibnamefont {Ott}},\ }\bibfield  {title} {\enquote {\bibinfo {title} {{The
  Gravitational Wave Signature of Core-Collapse Supernovae}},}\ }\href
  {\doibase 10.1088/0264-9381/26/6/063001} {\bibfield  {journal} {\bibinfo
  {journal} {Class. Quant. Grav.}\ }\textbf {\bibinfo {volume} {26}},\ \bibinfo
  {pages} {063001} (\bibinfo {year} {2009})},\ \Eprint
  {http://arxiv.org/abs/0809.0695} {arXiv:0809.0695 [astro-ph]} \BibitemShut
  {NoStop}%
\bibitem [{\citenamefont {Segalis}\ and\ \citenamefont
  {Ori}(2001)}]{Segalis:2001ns}%
  \BibitemOpen
  \bibfield  {author} {\bibinfo {author} {\bibfnamefont {Ehud~B.}\ \bibnamefont
  {Segalis}}\ and\ \bibinfo {author} {\bibfnamefont {Amos}\ \bibnamefont
  {Ori}},\ }\bibfield  {title} {\enquote {\bibinfo {title} {{Emission of
  gravitational radiation from ultrarelativistic sources}},}\ }\href {\doibase
  10.1103/PhysRevD.64.064018} {\bibfield  {journal} {\bibinfo  {journal} {Phys.
  Rev. D}\ }\textbf {\bibinfo {volume} {64}},\ \bibinfo {pages} {064018}
  (\bibinfo {year} {2001})},\ \Eprint {http://arxiv.org/abs/gr-qc/0101117}
  {arXiv:gr-qc/0101117} \BibitemShut {NoStop}%
\bibitem [{\citenamefont {Sago}\ \emph {et~al.}(2004)\citenamefont {Sago},
  \citenamefont {Ioka}, \citenamefont {Nakamura},\ and\ \citenamefont
  {Yamazaki}}]{Sago:2004pn}%
  \BibitemOpen
  \bibfield  {author} {\bibinfo {author} {\bibfnamefont {Norichika}\
  \bibnamefont {Sago}}, \bibinfo {author} {\bibfnamefont {Kunihito}\
  \bibnamefont {Ioka}}, \bibinfo {author} {\bibfnamefont {Takashi}\
  \bibnamefont {Nakamura}}, \ and\ \bibinfo {author} {\bibfnamefont {Ryo}\
  \bibnamefont {Yamazaki}},\ }\bibfield  {title} {\enquote {\bibinfo {title}
  {{Gravitational wave memory of gamma-ray burst jets}},}\ }\href {\doibase
  10.1103/PhysRevD.70.104012} {\bibfield  {journal} {\bibinfo  {journal} {Phys.
  Rev. D}\ }\textbf {\bibinfo {volume} {70}},\ \bibinfo {pages} {104012}
  (\bibinfo {year} {2004})},\ \Eprint {http://arxiv.org/abs/gr-qc/0405067}
  {arXiv:gr-qc/0405067} \BibitemShut {NoStop}%
\bibitem [{\citenamefont {Birnholtz}\ and\ \citenamefont
  {Piran}(2013)}]{Birnholtz:2013bea}%
  \BibitemOpen
  \bibfield  {author} {\bibinfo {author} {\bibfnamefont {Ofek}\ \bibnamefont
  {Birnholtz}}\ and\ \bibinfo {author} {\bibfnamefont {Tsvi}\ \bibnamefont
  {Piran}},\ }\bibfield  {title} {\enquote {\bibinfo {title} {{Gravitational
  wave memory from gamma ray bursts\textquoteright{} jets}},}\ }\href {\doibase
  10.1103/PhysRevD.87.123007} {\bibfield  {journal} {\bibinfo  {journal} {Phys.
  Rev. D}\ }\textbf {\bibinfo {volume} {87}},\ \bibinfo {pages} {123007}
  (\bibinfo {year} {2013})},\ \Eprint {http://arxiv.org/abs/1302.5713}
  {arXiv:1302.5713 [astro-ph.HE]} \BibitemShut {NoStop}%
\bibitem [{\citenamefont {Akiba}\ \emph {et~al.}(2013)\citenamefont {Akiba},
  \citenamefont {Nakada}, \citenamefont {Yamaguchi},\ and\ \citenamefont
  {Iwamoto}}]{Akiba:2013qwa}%
  \BibitemOpen
  \bibfield  {author} {\bibinfo {author} {\bibfnamefont {Shota}\ \bibnamefont
  {Akiba}}, \bibinfo {author} {\bibfnamefont {Megumi}\ \bibnamefont {Nakada}},
  \bibinfo {author} {\bibfnamefont {Chiyo}\ \bibnamefont {Yamaguchi}}, \ and\
  \bibinfo {author} {\bibfnamefont {Koichi}\ \bibnamefont {Iwamoto}},\
  }\bibfield  {title} {\enquote {\bibinfo {title} {{Gravitational Wave Memory
  from the Relativistic Jet of Gamma-Ray Bursts}},}\ }\href {\doibase
  10.1093/pasj/65.3.59} {\bibfield  {journal} {\bibinfo  {journal} {Publ.
  Astron. Soc. Jap.}\ }\textbf {\bibinfo {volume} {65}},\ \bibinfo {pages} {59}
  (\bibinfo {year} {2013})},\ \Eprint {http://arxiv.org/abs/1303.6566}
  {arXiv:1303.6566 [astro-ph.HE]} \BibitemShut {NoStop}%
\bibitem [{\citenamefont {Tolish}\ and\ \citenamefont
  {Wald}(2014)}]{Tolish:2014bka}%
  \BibitemOpen
  \bibfield  {author} {\bibinfo {author} {\bibfnamefont {Alexander}\
  \bibnamefont {Tolish}}\ and\ \bibinfo {author} {\bibfnamefont {Robert~M.}\
  \bibnamefont {Wald}},\ }\bibfield  {title} {\enquote {\bibinfo {title}
  {{Retarded Fields of Null Particles and the Memory Effect}},}\ }\href
  {\doibase 10.1103/PhysRevD.89.064008} {\bibfield  {journal} {\bibinfo
  {journal} {Phys. Rev. D}\ }\textbf {\bibinfo {volume} {89}},\ \bibinfo
  {pages} {064008} (\bibinfo {year} {2014})},\ \Eprint
  {http://arxiv.org/abs/1401.5831} {arXiv:1401.5831 [gr-qc]} \BibitemShut
  {NoStop}%
\bibitem [{\citenamefont {Tolish}\ \emph {et~al.}(2014)\citenamefont {Tolish},
  \citenamefont {Bieri}, \citenamefont {Garfinkle},\ and\ \citenamefont
  {Wald}}]{Tolish:2014oda}%
  \BibitemOpen
  \bibfield  {author} {\bibinfo {author} {\bibfnamefont {Alexander}\
  \bibnamefont {Tolish}}, \bibinfo {author} {\bibfnamefont {Lydia}\
  \bibnamefont {Bieri}}, \bibinfo {author} {\bibfnamefont {David}\ \bibnamefont
  {Garfinkle}}, \ and\ \bibinfo {author} {\bibfnamefont {Robert~M.}\
  \bibnamefont {Wald}},\ }\bibfield  {title} {\enquote {\bibinfo {title}
  {{Examination of a simple example of gravitational wave memory}},}\ }\href
  {\doibase 10.1103/PhysRevD.90.044060} {\bibfield  {journal} {\bibinfo
  {journal} {Phys. Rev. D}\ }\textbf {\bibinfo {volume} {90}},\ \bibinfo
  {pages} {044060} (\bibinfo {year} {2014})},\ \Eprint
  {http://arxiv.org/abs/1405.6396} {arXiv:1405.6396 [gr-qc]} \BibitemShut
  {NoStop}%
\bibitem [{\citenamefont {Allen}(2020)}]{Allen:2019hnd}%
  \BibitemOpen
  \bibfield  {author} {\bibinfo {author} {\bibfnamefont {Bruce}\ \bibnamefont
  {Allen}},\ }\bibfield  {title} {\enquote {\bibinfo {title} {{Gravitational
  wave stochastic background from cosmological particle decay}},}\ }\href
  {\doibase 10.1103/PhysRevResearch.2.012034} {\bibfield  {journal} {\bibinfo
  {journal} {Phys. Rev. Res.}\ }\textbf {\bibinfo {volume} {2}},\ \bibinfo
  {pages} {012034} (\bibinfo {year} {2020})},\ \Eprint
  {http://arxiv.org/abs/1910.08213} {arXiv:1910.08213 [gr-qc]} \BibitemShut
  {NoStop}%
\bibitem [{\citenamefont {Harry}(2010)}]{Harry:2010zz}%
  \BibitemOpen
  \bibfield  {author} {\bibinfo {author} {\bibfnamefont {Gregory~M.}\
  \bibnamefont {Harry}} (\bibinfo {collaboration} {LIGO Scientific
  Collaboration}),\ }\bibfield  {title} {\enquote {\bibinfo {title} {{Advanced
  LIGO: The next generation of gravitational wave detectors}},}\ }\href
  {\doibase 10.1088/0264-9381/27/8/084006} {\bibfield  {journal} {\bibinfo
  {journal} {Class. Quant. Grav.}\ }\textbf {\bibinfo {volume} {27}},\ \bibinfo
  {pages} {084006} (\bibinfo {year} {2010})}\BibitemShut {NoStop}%
\bibitem [{\citenamefont {Aasi}\ \emph {et~al.}(2015)\citenamefont {Aasi} \emph
  {et~al.}}]{TheLIGOScientific:2014jea}%
  \BibitemOpen
  \bibfield  {author} {\bibinfo {author} {\bibfnamefont {J.}~\bibnamefont
  {Aasi}} \emph {et~al.} (\bibinfo {collaboration} {LIGO Scientific
  Collaboration}),\ }\bibfield  {title} {\enquote {\bibinfo {title} {{Advanced
  LIGO}},}\ }\href {\doibase 10.1088/0264-9381/32/7/074001} {\bibfield
  {journal} {\bibinfo  {journal} {Class. Quant. Grav.}\ }\textbf {\bibinfo
  {volume} {32}},\ \bibinfo {pages} {074001} (\bibinfo {year} {2015})},\
  \Eprint {http://arxiv.org/abs/1411.4547} {arXiv:1411.4547 [gr-qc]}
  \BibitemShut {NoStop}%
\bibitem [{\citenamefont {Acernese}\ \emph {et~al.}(2015)\citenamefont
  {Acernese} \emph {et~al.}}]{TheVirgo:2014hva}%
  \BibitemOpen
  \bibfield  {author} {\bibinfo {author} {\bibfnamefont {F.}~\bibnamefont
  {Acernese}} \emph {et~al.} (\bibinfo {collaboration} {Virgo Collaboration}),\
  }\bibfield  {title} {\enquote {\bibinfo {title} {{Advanced Virgo: a
  second-generation interferometric gravitational wave detector}},}\ }\href
  {\doibase 10.1088/0264-9381/32/2/024001} {\bibfield  {journal} {\bibinfo
  {journal} {Class. Quant. Grav.}\ }\textbf {\bibinfo {volume} {32}},\ \bibinfo
  {pages} {024001} (\bibinfo {year} {2015})},\ \Eprint
  {http://arxiv.org/abs/1408.3978} {arXiv:1408.3978 [gr-qc]} \BibitemShut
  {NoStop}%
\bibitem [{\citenamefont {Abbott}\ \emph
  {et~al.}(2021{\natexlab{b}})\citenamefont {Abbott} \emph
  {et~al.}}]{LIGOScientific:2020ibl}%
  \BibitemOpen
  \bibfield  {author} {\bibinfo {author} {\bibfnamefont {R.}~\bibnamefont
  {Abbott}} \emph {et~al.} (\bibinfo {collaboration} {LIGO Scientific
  Collaboration, Virgo Collaboration}),\ }\bibfield  {title} {\enquote
  {\bibinfo {title} {{GWTC-2: Compact Binary Coalescences Observed by LIGO and
  Virgo During the First Half of the Third Observing Run}},}\ }\href {\doibase
  10.1103/PhysRevX.11.021053} {\bibfield  {journal} {\bibinfo  {journal} {Phys.
  Rev. X}\ }\textbf {\bibinfo {volume} {11}},\ \bibinfo {pages} {021053}
  (\bibinfo {year} {2021}{\natexlab{b}})},\ \Eprint
  {http://arxiv.org/abs/2010.14527} {arXiv:2010.14527 [gr-qc]} \BibitemShut
  {NoStop}%
\bibitem [{\citenamefont {Favata}(2009{\natexlab{a}})}]{Favata:2008ti}%
  \BibitemOpen
  \bibfield  {author} {\bibinfo {author} {\bibfnamefont {Marc}\ \bibnamefont
  {Favata}},\ }\bibfield  {title} {\enquote {\bibinfo {title}
  {{Gravitational-wave memory revisited: memory from the merger and recoil of
  binary black holes}},}\ }\href {\doibase 10.1088/1742-6596/154/1/012043}
  {\bibfield  {journal} {\bibinfo  {journal} {J. Phys. Conf. Ser.}\ }\textbf
  {\bibinfo {volume} {154}},\ \bibinfo {pages} {012043} (\bibinfo {year}
  {2009}{\natexlab{a}})},\ \Eprint {http://arxiv.org/abs/0811.3451}
  {arXiv:0811.3451 [astro-ph]} \BibitemShut {NoStop}%
\bibitem [{\citenamefont {Campanelli}\ \emph {et~al.}(2007)\citenamefont
  {Campanelli}, \citenamefont {Lousto}, \citenamefont {Zlochower},\ and\
  \citenamefont {Merritt}}]{Campanelli:2007cga}%
  \BibitemOpen
  \bibfield  {author} {\bibinfo {author} {\bibfnamefont {Manuela}\ \bibnamefont
  {Campanelli}}, \bibinfo {author} {\bibfnamefont {Carlos~O.}\ \bibnamefont
  {Lousto}}, \bibinfo {author} {\bibfnamefont {Yosef}\ \bibnamefont
  {Zlochower}}, \ and\ \bibinfo {author} {\bibfnamefont {David}\ \bibnamefont
  {Merritt}},\ }\bibfield  {title} {\enquote {\bibinfo {title} {{Maximum
  gravitational recoil}},}\ }\href {\doibase 10.1103/PhysRevLett.98.231102}
  {\bibfield  {journal} {\bibinfo  {journal} {Phys. Rev. Lett.}\ }\textbf
  {\bibinfo {volume} {98}},\ \bibinfo {pages} {231102} (\bibinfo {year}
  {2007})},\ \Eprint {http://arxiv.org/abs/gr-qc/0702133} {arXiv:gr-qc/0702133}
  \BibitemShut {NoStop}%
\bibitem [{\citenamefont {Christodoulou}(1991)}]{Christodoulou:1991cr}%
  \BibitemOpen
  \bibfield  {author} {\bibinfo {author} {\bibfnamefont {Demetrios}\
  \bibnamefont {Christodoulou}},\ }\bibfield  {title} {\enquote {\bibinfo
  {title} {{Nonlinear nature of gravitation and gravitational wave
  experiments}},}\ }\href {\doibase 10.1103/PhysRevLett.67.1486} {\bibfield
  {journal} {\bibinfo  {journal} {Phys. Rev. Lett.}\ }\textbf {\bibinfo
  {volume} {67}},\ \bibinfo {pages} {1486--1489} (\bibinfo {year}
  {1991})}\BibitemShut {NoStop}%
\bibitem [{\citenamefont {Blanchet}\ and\ \citenamefont
  {Damour}(1992)}]{Blanchet:1992br}%
  \BibitemOpen
  \bibfield  {author} {\bibinfo {author} {\bibfnamefont {Luc}\ \bibnamefont
  {Blanchet}}\ and\ \bibinfo {author} {\bibfnamefont {Thibault}\ \bibnamefont
  {Damour}},\ }\bibfield  {title} {\enquote {\bibinfo {title} {{Hereditary
  effects in gravitational radiation}},}\ }\href {\doibase
  10.1103/PhysRevD.46.4304} {\bibfield  {journal} {\bibinfo  {journal} {Phys.
  Rev. D}\ }\textbf {\bibinfo {volume} {46}},\ \bibinfo {pages} {4304--4319}
  (\bibinfo {year} {1992})}\BibitemShut {NoStop}%
\bibitem [{\citenamefont {Thorne}(1992)}]{Thorne:1992sdb}%
  \BibitemOpen
  \bibfield  {author} {\bibinfo {author} {\bibfnamefont {Kip~S.}\ \bibnamefont
  {Thorne}},\ }\bibfield  {title} {\enquote {\bibinfo {title}
  {{Gravitational-wave bursts with memory: The Christodoulou effect}},}\ }\href
  {\doibase 10.1103/PhysRevD.45.520} {\bibfield  {journal} {\bibinfo  {journal}
  {Phys. Rev. D}\ }\textbf {\bibinfo {volume} {45}},\ \bibinfo {pages}
  {520--524} (\bibinfo {year} {1992})}\BibitemShut {NoStop}%
\bibitem [{\citenamefont {Favata}(2009{\natexlab{b}})}]{Favata:2009ii}%
  \BibitemOpen
  \bibfield  {author} {\bibinfo {author} {\bibfnamefont {Marc}\ \bibnamefont
  {Favata}},\ }\bibfield  {title} {\enquote {\bibinfo {title} {{Nonlinear
  gravitational-wave memory from binary black hole mergers}},}\ }\href
  {\doibase 10.1088/0004-637X/696/2/L159} {\bibfield  {journal} {\bibinfo
  {journal} {Astrophys. J. Lett.}\ }\textbf {\bibinfo {volume} {696}},\
  \bibinfo {pages} {L159--L162} (\bibinfo {year} {2009}{\natexlab{b}})},\
  \Eprint {http://arxiv.org/abs/0902.3660} {arXiv:0902.3660 [astro-ph.SR]}
  \BibitemShut {NoStop}%
\bibitem [{\citenamefont {Bieri}\ and\ \citenamefont
  {Garfinkle}(2013)}]{Bieri:2013hqa}%
  \BibitemOpen
  \bibfield  {author} {\bibinfo {author} {\bibfnamefont {Lydia}\ \bibnamefont
  {Bieri}}\ and\ \bibinfo {author} {\bibfnamefont {David}\ \bibnamefont
  {Garfinkle}},\ }\bibfield  {title} {\enquote {\bibinfo {title} {{An
  electromagnetic analogue of gravitational wave memory}},}\ }\href {\doibase
  10.1088/0264-9381/30/19/195009} {\bibfield  {journal} {\bibinfo  {journal}
  {Class. Quant. Grav.}\ }\textbf {\bibinfo {volume} {30}},\ \bibinfo {pages}
  {195009} (\bibinfo {year} {2013})},\ \Eprint {http://arxiv.org/abs/1307.5098}
  {arXiv:1307.5098 [gr-qc]} \BibitemShut {NoStop}%
\bibitem [{\citenamefont {Susskind}(2015)}]{Susskind:2015hpa}%
  \BibitemOpen
  \bibfield  {author} {\bibinfo {author} {\bibfnamefont {Leonard}\ \bibnamefont
  {Susskind}},\ }\bibfield  {title} {\enquote {\bibinfo {title}
  {{Electromagnetic Memory}},}\ }\href@noop {} {\  (\bibinfo {year} {2015})},\
  \Eprint {http://arxiv.org/abs/1507.02584} {arXiv:1507.02584 [hep-th]}
  \BibitemShut {NoStop}%
\bibitem [{\citenamefont {Pate}\ \emph {et~al.}(2017)\citenamefont {Pate},
  \citenamefont {Raclariu},\ and\ \citenamefont {Strominger}}]{Pate:2017vwa}%
  \BibitemOpen
  \bibfield  {author} {\bibinfo {author} {\bibfnamefont {Monica}\ \bibnamefont
  {Pate}}, \bibinfo {author} {\bibfnamefont {Ana-Maria}\ \bibnamefont
  {Raclariu}}, \ and\ \bibinfo {author} {\bibfnamefont {Andrew}\ \bibnamefont
  {Strominger}},\ }\bibfield  {title} {\enquote {\bibinfo {title} {{Color
  Memory: A Yang-Mills Analog of Gravitational Wave Memory}},}\ }\href
  {\doibase 10.1103/PhysRevLett.119.261602} {\bibfield  {journal} {\bibinfo
  {journal} {Phys. Rev. Lett.}\ }\textbf {\bibinfo {volume} {119}},\ \bibinfo
  {pages} {261602} (\bibinfo {year} {2017})},\ \Eprint
  {http://arxiv.org/abs/1707.08016} {arXiv:1707.08016 [hep-th]} \BibitemShut
  {NoStop}%
\bibitem [{\citenamefont {He}\ \emph {et~al.}(2015)\citenamefont {He},
  \citenamefont {Lysov}, \citenamefont {Mitra},\ and\ \citenamefont
  {Strominger}}]{He:2014laa}%
  \BibitemOpen
  \bibfield  {author} {\bibinfo {author} {\bibfnamefont {Temple}\ \bibnamefont
  {He}}, \bibinfo {author} {\bibfnamefont {Vyacheslav}\ \bibnamefont {Lysov}},
  \bibinfo {author} {\bibfnamefont {Prahar}\ \bibnamefont {Mitra}}, \ and\
  \bibinfo {author} {\bibfnamefont {Andrew}\ \bibnamefont {Strominger}},\
  }\bibfield  {title} {\enquote {\bibinfo {title} {{BMS supertranslations and
  Weinberg\textquoteright{}s soft graviton theorem}},}\ }\href {\doibase
  10.1007/JHEP05(2015)151} {\bibfield  {journal} {\bibinfo  {journal} {JHEP}\
  }\textbf {\bibinfo {volume} {05}},\ \bibinfo {pages} {151} (\bibinfo {year}
  {2015})},\ \Eprint {http://arxiv.org/abs/1401.7026} {arXiv:1401.7026
  [hep-th]} \BibitemShut {NoStop}%
\bibitem [{\citenamefont {Strominger}\ and\ \citenamefont
  {Zhiboedov}(2016)}]{Strominger:2014pwa}%
  \BibitemOpen
  \bibfield  {author} {\bibinfo {author} {\bibfnamefont {Andrew}\ \bibnamefont
  {Strominger}}\ and\ \bibinfo {author} {\bibfnamefont {Alexander}\
  \bibnamefont {Zhiboedov}},\ }\bibfield  {title} {\enquote {\bibinfo {title}
  {{Gravitational Memory, BMS Supertranslations and Soft Theorems}},}\ }\href
  {\doibase 10.1007/JHEP01(2016)086} {\bibfield  {journal} {\bibinfo  {journal}
  {JHEP}\ }\textbf {\bibinfo {volume} {01}},\ \bibinfo {pages} {086} (\bibinfo
  {year} {2016})},\ \Eprint {http://arxiv.org/abs/1411.5745} {arXiv:1411.5745
  [hep-th]} \BibitemShut {NoStop}%
\bibitem [{\citenamefont {Pasterski}\ \emph {et~al.}(2016)\citenamefont
  {Pasterski}, \citenamefont {Strominger},\ and\ \citenamefont
  {Zhiboedov}}]{Pasterski:2015tva}%
  \BibitemOpen
  \bibfield  {author} {\bibinfo {author} {\bibfnamefont {Sabrina}\ \bibnamefont
  {Pasterski}}, \bibinfo {author} {\bibfnamefont {Andrew}\ \bibnamefont
  {Strominger}}, \ and\ \bibinfo {author} {\bibfnamefont {Alexander}\
  \bibnamefont {Zhiboedov}},\ }\bibfield  {title} {\enquote {\bibinfo {title}
  {{New Gravitational Memories}},}\ }\href {\doibase 10.1007/JHEP12(2016)053}
  {\bibfield  {journal} {\bibinfo  {journal} {JHEP}\ }\textbf {\bibinfo
  {volume} {12}},\ \bibinfo {pages} {053} (\bibinfo {year} {2016})},\ \Eprint
  {http://arxiv.org/abs/1502.06120} {arXiv:1502.06120 [hep-th]} \BibitemShut
  {NoStop}%
\bibitem [{\citenamefont {Pasterski}(2017)}]{Pasterski:2015zua}%
  \BibitemOpen
  \bibfield  {author} {\bibinfo {author} {\bibfnamefont {Sabrina}\ \bibnamefont
  {Pasterski}},\ }\bibfield  {title} {\enquote {\bibinfo {title} {{Asymptotic
  Symmetries and Electromagnetic Memory}},}\ }\href {\doibase
  10.1007/JHEP09(2017)154} {\bibfield  {journal} {\bibinfo  {journal} {JHEP}\
  }\textbf {\bibinfo {volume} {09}},\ \bibinfo {pages} {154} (\bibinfo {year}
  {2017})},\ \Eprint {http://arxiv.org/abs/1505.00716} {arXiv:1505.00716
  [hep-th]} \BibitemShut {NoStop}%
\bibitem [{\citenamefont {Kapec}\ \emph {et~al.}(2017)\citenamefont {Kapec},
  \citenamefont {Lysov}, \citenamefont {Pasterski},\ and\ \citenamefont
  {Strominger}}]{Kapec:2015vwa}%
  \BibitemOpen
  \bibfield  {author} {\bibinfo {author} {\bibfnamefont {Daniel}\ \bibnamefont
  {Kapec}}, \bibinfo {author} {\bibfnamefont {Vyacheslav}\ \bibnamefont
  {Lysov}}, \bibinfo {author} {\bibfnamefont {Sabrina}\ \bibnamefont
  {Pasterski}}, \ and\ \bibinfo {author} {\bibfnamefont {Andrew}\ \bibnamefont
  {Strominger}},\ }\bibfield  {title} {\enquote {\bibinfo {title}
  {{Higher-dimensional supertranslations and Weinberg\textquoteright{}s soft
  graviton theorem}},}\ }\href {\doibase 10.4310/AMSA.2017.v2.n1.a2} {\bibfield
   {journal} {\bibinfo  {journal} {Ann. Math. Sci. Appl.}\ }\textbf {\bibinfo
  {volume} {02}},\ \bibinfo {pages} {69--94} (\bibinfo {year} {2017})},\
  \Eprint {http://arxiv.org/abs/1502.07644} {arXiv:1502.07644 [gr-qc]}
  \BibitemShut {NoStop}%
\bibitem [{\citenamefont {Flanagan}\ and\ \citenamefont
  {Nichols}(2017)}]{Flanagan:2015pxa}%
  \BibitemOpen
  \bibfield  {author} {\bibinfo {author} {\bibfnamefont {\'Eanna~\'E.}\
  \bibnamefont {Flanagan}}\ and\ \bibinfo {author} {\bibfnamefont {David~A.}\
  \bibnamefont {Nichols}},\ }\bibfield  {title} {\enquote {\bibinfo {title}
  {{Conserved charges of the extended Bondi-Metzner-Sachs algebra}},}\ }\href
  {\doibase 10.1103/PhysRevD.95.044002} {\bibfield  {journal} {\bibinfo
  {journal} {Phys. Rev. D}\ }\textbf {\bibinfo {volume} {95}},\ \bibinfo
  {pages} {044002} (\bibinfo {year} {2017})},\ \Eprint
  {http://arxiv.org/abs/1510.03386} {arXiv:1510.03386 [hep-th]} \BibitemShut
  {NoStop}%
\bibitem [{\citenamefont {Strominger}(2018)}]{Strominger:2017zoo}%
  \BibitemOpen
  \bibfield  {author} {\bibinfo {author} {\bibfnamefont {Andrew}\ \bibnamefont
  {Strominger}},\ }\href
  {https://press.princeton.edu/books/hardcover/9780691179506/lectures-on-the-infrared-structure-of-gravity-and-gauge-theory}
  {\emph {\bibinfo {title} {{Lectures on the Infrared Structure of Gravity and
  Gauge Theory}}}}\ (\bibinfo  {publisher} {Princeton University Press},\
  \bibinfo {year} {2018})\ \Eprint {http://arxiv.org/abs/1703.05448}
  {arXiv:1703.05448 [hep-th]} \BibitemShut {NoStop}%
\bibitem [{\citenamefont {Wiseman}\ and\ \citenamefont
  {Will}(1991)}]{Wiseman:1991ss}%
  \BibitemOpen
  \bibfield  {author} {\bibinfo {author} {\bibfnamefont {Alan~G.}\ \bibnamefont
  {Wiseman}}\ and\ \bibinfo {author} {\bibfnamefont {Clifford~M.}\ \bibnamefont
  {Will}},\ }\bibfield  {title} {\enquote {\bibinfo {title} {{Christodoulou's
  nonlinear gravitational wave memory: Evaluation in the quadrupole
  approximation}},}\ }\href {\doibase 10.1103/PhysRevD.44.R2945} {\bibfield
  {journal} {\bibinfo  {journal} {Phys. Rev. D}\ }\textbf {\bibinfo {volume}
  {44}},\ \bibinfo {pages} {2945--2949} (\bibinfo {year} {1991})}\BibitemShut
  {NoStop}%
\bibitem [{\citenamefont {Favata}(2009{\natexlab{c}})}]{Favata:2008yd}%
  \BibitemOpen
  \bibfield  {author} {\bibinfo {author} {\bibfnamefont {Marc}\ \bibnamefont
  {Favata}},\ }\bibfield  {title} {\enquote {\bibinfo {title} {{Post-Newtonian
  corrections to the gravitational-wave memory for quasi-circular, inspiralling
  compact binaries}},}\ }\href {\doibase 10.1103/PhysRevD.80.024002} {\bibfield
   {journal} {\bibinfo  {journal} {Phys. Rev. D}\ }\textbf {\bibinfo {volume}
  {80}},\ \bibinfo {pages} {024002} (\bibinfo {year} {2009}{\natexlab{c}})},\
  \Eprint {http://arxiv.org/abs/0812.0069} {arXiv:0812.0069 [gr-qc]}
  \BibitemShut {NoStop}%
\bibitem [{\citenamefont {Pollney}\ and\ \citenamefont
  {Reisswig}(2011)}]{Pollney:2010hs}%
  \BibitemOpen
  \bibfield  {author} {\bibinfo {author} {\bibfnamefont {Denis}\ \bibnamefont
  {Pollney}}\ and\ \bibinfo {author} {\bibfnamefont {Christian}\ \bibnamefont
  {Reisswig}},\ }\bibfield  {title} {\enquote {\bibinfo {title} {{Gravitational
  memory in binary black hole mergers}},}\ }\href {\doibase
  10.1088/2041-8205/732/1/L13} {\bibfield  {journal} {\bibinfo  {journal}
  {Astrophys. J. Lett.}\ }\textbf {\bibinfo {volume} {732}},\ \bibinfo {pages}
  {L13} (\bibinfo {year} {2011})},\ \Eprint {http://arxiv.org/abs/1004.4209}
  {arXiv:1004.4209 [gr-qc]} \BibitemShut {NoStop}%
\bibitem [{\citenamefont {Favata}(2011)}]{Favata:2011qi}%
  \BibitemOpen
  \bibfield  {author} {\bibinfo {author} {\bibfnamefont {Marc}\ \bibnamefont
  {Favata}},\ }\bibfield  {title} {\enquote {\bibinfo {title} {{The
  Gravitational-wave memory from eccentric binaries}},}\ }\href {\doibase
  10.1103/PhysRevD.84.124013} {\bibfield  {journal} {\bibinfo  {journal} {Phys.
  Rev. D}\ }\textbf {\bibinfo {volume} {84}},\ \bibinfo {pages} {124013}
  (\bibinfo {year} {2011})},\ \Eprint {http://arxiv.org/abs/1108.3121}
  {arXiv:1108.3121 [gr-qc]} \BibitemShut {NoStop}%
\bibitem [{\citenamefont {Nichols}(2017)}]{Nichols:2017rqr}%
  \BibitemOpen
  \bibfield  {author} {\bibinfo {author} {\bibfnamefont {David~A.}\
  \bibnamefont {Nichols}},\ }\bibfield  {title} {\enquote {\bibinfo {title}
  {{Spin memory effect for compact binaries in the post-Newtonian
  approximation}},}\ }\href {\doibase 10.1103/PhysRevD.95.084048} {\bibfield
  {journal} {\bibinfo  {journal} {Phys. Rev. D}\ }\textbf {\bibinfo {volume}
  {95}},\ \bibinfo {pages} {084048} (\bibinfo {year} {2017})},\ \Eprint
  {http://arxiv.org/abs/1702.03300} {arXiv:1702.03300 [gr-qc]} \BibitemShut
  {NoStop}%
\bibitem [{\citenamefont {Talbot}\ \emph {et~al.}(2018)\citenamefont {Talbot},
  \citenamefont {Thrane}, \citenamefont {Lasky},\ and\ \citenamefont
  {Lin}}]{Talbot:2018sgr}%
  \BibitemOpen
  \bibfield  {author} {\bibinfo {author} {\bibfnamefont {Colm}\ \bibnamefont
  {Talbot}}, \bibinfo {author} {\bibfnamefont {Eric}\ \bibnamefont {Thrane}},
  \bibinfo {author} {\bibfnamefont {Paul~D.}\ \bibnamefont {Lasky}}, \ and\
  \bibinfo {author} {\bibfnamefont {Fuhui}\ \bibnamefont {Lin}},\ }\bibfield
  {title} {\enquote {\bibinfo {title} {{Gravitational-wave memory: waveforms
  and phenomenology}},}\ }\href {\doibase 10.1103/PhysRevD.98.064031}
  {\bibfield  {journal} {\bibinfo  {journal} {Phys. Rev. D}\ }\textbf {\bibinfo
  {volume} {98}},\ \bibinfo {pages} {064031} (\bibinfo {year} {2018})},\
  \Eprint {http://arxiv.org/abs/1807.00990} {arXiv:1807.00990 [astro-ph.HE]}
  \BibitemShut {NoStop}%
\bibitem [{\citenamefont {Khera}\ \emph {et~al.}(2021)\citenamefont {Khera},
  \citenamefont {Krishnan}, \citenamefont {Ashtekar},\ and\ \citenamefont
  {De~Lorenzo}}]{Khera:2020mcz}%
  \BibitemOpen
  \bibfield  {author} {\bibinfo {author} {\bibfnamefont {Neev}\ \bibnamefont
  {Khera}}, \bibinfo {author} {\bibfnamefont {Badri}\ \bibnamefont {Krishnan}},
  \bibinfo {author} {\bibfnamefont {Abhay}\ \bibnamefont {Ashtekar}}, \ and\
  \bibinfo {author} {\bibfnamefont {Tommaso}\ \bibnamefont {De~Lorenzo}},\
  }\bibfield  {title} {\enquote {\bibinfo {title} {{Inferring the gravitational
  wave memory for binary coalescence events}},}\ }\href {\doibase
  10.1103/PhysRevD.103.044012} {\bibfield  {journal} {\bibinfo  {journal}
  {Phys. Rev. D}\ }\textbf {\bibinfo {volume} {103}},\ \bibinfo {pages}
  {044012} (\bibinfo {year} {2021})},\ \Eprint
  {http://arxiv.org/abs/2009.06351} {arXiv:2009.06351 [gr-qc]} \BibitemShut
  {NoStop}%
\bibitem [{\citenamefont {Kennefick}(1994)}]{Kennefick:1994nw}%
  \BibitemOpen
  \bibfield  {author} {\bibinfo {author} {\bibfnamefont {D.}~\bibnamefont
  {Kennefick}},\ }\bibfield  {title} {\enquote {\bibinfo {title} {{Prospects
  for detecting the Christodoulou memory of gravitational waves from a
  coalescing compact binary and using it to measure neutron star radii}},}\
  }\href {\doibase 10.1103/PhysRevD.50.3587} {\bibfield  {journal} {\bibinfo
  {journal} {Phys. Rev. D}\ }\textbf {\bibinfo {volume} {50}},\ \bibinfo
  {pages} {3587--3595} (\bibinfo {year} {1994})}\BibitemShut {NoStop}%
\bibitem [{\citenamefont {van Haasteren}\ and\ \citenamefont
  {Levin}(2010)}]{vanHaasteren:2009fy}%
  \BibitemOpen
  \bibfield  {author} {\bibinfo {author} {\bibfnamefont {Rutger}\ \bibnamefont
  {van Haasteren}}\ and\ \bibinfo {author} {\bibfnamefont {Yuri}\ \bibnamefont
  {Levin}},\ }\bibfield  {title} {\enquote {\bibinfo {title}
  {{Gravitational-wave memory and pulsar timing arrays}},}\ }\href {\doibase
  10.1111/j.1365-2966.2009.15885.x} {\bibfield  {journal} {\bibinfo  {journal}
  {Mon. Not. Roy. Astron. Soc.}\ }\textbf {\bibinfo {volume} {401}},\ \bibinfo
  {pages} {2372} (\bibinfo {year} {2010})},\ \Eprint
  {http://arxiv.org/abs/0909.0954} {arXiv:0909.0954 [astro-ph.IM]} \BibitemShut
  {NoStop}%
\bibitem [{\citenamefont {Seto}(2009)}]{Seto:2009nv}%
  \BibitemOpen
  \bibfield  {author} {\bibinfo {author} {\bibfnamefont {Naoki}\ \bibnamefont
  {Seto}},\ }\bibfield  {title} {\enquote {\bibinfo {title} {{Search for Memory
  and Inspiral Gravitational Waves from Super-Massive Binary Black Holes with
  Pulsar Timing Arrays}},}\ }\href {\doibase 10.1111/j.1745-3933.2009.00758.x}
  {\bibfield  {journal} {\bibinfo  {journal} {Mon. Not. Roy. Astron. Soc.}\
  }\textbf {\bibinfo {volume} {400}},\ \bibinfo {pages} {L38} (\bibinfo {year}
  {2009})},\ \Eprint {http://arxiv.org/abs/0909.1379} {arXiv:0909.1379
  [astro-ph.CO]} \BibitemShut {NoStop}%
\bibitem [{\citenamefont {Cordes}\ and\ \citenamefont
  {Jenet}(2012)}]{Cordes:2012zz}%
  \BibitemOpen
  \bibfield  {author} {\bibinfo {author} {\bibfnamefont {J.~M.}\ \bibnamefont
  {Cordes}}\ and\ \bibinfo {author} {\bibfnamefont {F.~A.}\ \bibnamefont
  {Jenet}},\ }\bibfield  {title} {\enquote {\bibinfo {title} {{Detecting
  gravitational wave memory with pulsar timing}},}\ }\href {\doibase
  10.1088/0004-637X/752/1/54} {\bibfield  {journal} {\bibinfo  {journal}
  {Astrophys. J.}\ }\textbf {\bibinfo {volume} {752}},\ \bibinfo {pages} {54}
  (\bibinfo {year} {2012})}\BibitemShut {NoStop}%
\bibitem [{\citenamefont {Madison}\ \emph {et~al.}(2014)\citenamefont
  {Madison}, \citenamefont {Cordes},\ and\ \citenamefont
  {Chatterjee}}]{Madison:2014vca}%
  \BibitemOpen
  \bibfield  {author} {\bibinfo {author} {\bibfnamefont {D.~R.}\ \bibnamefont
  {Madison}}, \bibinfo {author} {\bibfnamefont {J.~M.}\ \bibnamefont {Cordes}},
  \ and\ \bibinfo {author} {\bibfnamefont {S.}~\bibnamefont {Chatterjee}},\
  }\bibfield  {title} {\enquote {\bibinfo {title} {{Assessing Pulsar Timing
  Array Sensitivity to Gravitational Wave Bursts with Memory}},}\ }\href
  {\doibase 10.1088/0004-637X/788/2/141} {\bibfield  {journal} {\bibinfo
  {journal} {Astrophys. J.}\ }\textbf {\bibinfo {volume} {788}},\ \bibinfo
  {pages} {141} (\bibinfo {year} {2014})},\ \Eprint
  {http://arxiv.org/abs/1404.5682} {arXiv:1404.5682 [astro-ph.HE]} \BibitemShut
  {NoStop}%
\bibitem [{\citenamefont {Wang}\ \emph {et~al.}(2015)\citenamefont {Wang},
  \citenamefont {Hobbs}, \citenamefont {Coles}, \citenamefont {Shannon},
  \citenamefont {Zhu}, \citenamefont {Madison}, \citenamefont {Kerr},
  \citenamefont {Ravi}, \citenamefont {Keith}, \citenamefont {Manchester},
  \citenamefont {Levin}, \citenamefont {Bailes}, \citenamefont {Bhat},
  \citenamefont {Burke-Spolaor}, \citenamefont {Dai}, \citenamefont
  {Osłowski}, \citenamefont {van Straten}, \citenamefont {Toomey},
  \citenamefont {Wang},\ and\ \citenamefont {Wen}}]{Wang:2014zls}%
  \BibitemOpen
  \bibfield  {author} {\bibinfo {author} {\bibfnamefont {J.~B.}\ \bibnamefont
  {Wang}}, \bibinfo {author} {\bibfnamefont {G.}~\bibnamefont {Hobbs}},
  \bibinfo {author} {\bibfnamefont {W.}~\bibnamefont {Coles}}, \bibinfo
  {author} {\bibfnamefont {R.~M.}\ \bibnamefont {Shannon}}, \bibinfo {author}
  {\bibfnamefont {X.~J.}\ \bibnamefont {Zhu}}, \bibinfo {author} {\bibfnamefont
  {D.~R.}\ \bibnamefont {Madison}}, \bibinfo {author} {\bibfnamefont
  {M.}~\bibnamefont {Kerr}}, \bibinfo {author} {\bibfnamefont {V.}~\bibnamefont
  {Ravi}}, \bibinfo {author} {\bibfnamefont {M.~J.}\ \bibnamefont {Keith}},
  \bibinfo {author} {\bibfnamefont {R.~N.}\ \bibnamefont {Manchester}},
  \bibinfo {author} {\bibfnamefont {Y.}~\bibnamefont {Levin}}, \bibinfo
  {author} {\bibfnamefont {M.}~\bibnamefont {Bailes}}, \bibinfo {author}
  {\bibfnamefont {N.~D.~R.}\ \bibnamefont {Bhat}}, \bibinfo {author}
  {\bibfnamefont {S.}~\bibnamefont {Burke-Spolaor}}, \bibinfo {author}
  {\bibfnamefont {S.}~\bibnamefont {Dai}}, \bibinfo {author} {\bibfnamefont
  {S.}~\bibnamefont {Osłowski}}, \bibinfo {author} {\bibfnamefont
  {W.}~\bibnamefont {van Straten}}, \bibinfo {author} {\bibfnamefont
  {L.}~\bibnamefont {Toomey}}, \bibinfo {author} {\bibfnamefont
  {N.}~\bibnamefont {Wang}}, \ and\ \bibinfo {author} {\bibfnamefont
  {L.}~\bibnamefont {Wen}},\ }\bibfield  {title} {\enquote {\bibinfo {title}
  {{Searching for gravitational wave memory bursts with the Parkes Pulsar
  Timing Array}},}\ }\href {\doibase 10.1093/mnras/stu2137} {\bibfield
  {journal} {\bibinfo  {journal} {Mon. Not. Roy. Astron. Soc.}\ }\textbf
  {\bibinfo {volume} {446}},\ \bibinfo {pages} {1657--1671} (\bibinfo {year}
  {2015})},\ \Eprint {http://arxiv.org/abs/1410.3323} {arXiv:1410.3323
  [astro-ph.GA]} \BibitemShut {NoStop}%
\bibitem [{\citenamefont {Arzoumanian}\ \emph {et~al.}(2015)\citenamefont
  {Arzoumanian}, \citenamefont {Brazier}, \citenamefont {Burke-Spolaor},
  \citenamefont {Chamberlin}, \citenamefont {Chatterjee}, \citenamefont
  {Christy}, \citenamefont {Cordes}, \citenamefont {Cornish}, \citenamefont
  {Demorest}, \citenamefont {Deng}, \citenamefont {Dolch}, \citenamefont
  {Ellis}, \citenamefont {Ferdman}, \citenamefont {Fonseca}, \citenamefont
  {Garver-Daniels}, \citenamefont {Jenet}, \citenamefont {Jones}, \citenamefont
  {Kaspi}, \citenamefont {Koop}, \citenamefont {Lam}, \citenamefont {Lazio},
  \citenamefont {Levin}, \citenamefont {Lommen}, \citenamefont {Lorimer},
  \citenamefont {Luo}, \citenamefont {Lynch}, \citenamefont {Madison},
  \citenamefont {McLaughlin}, \citenamefont {McWilliams}, \citenamefont {Nice},
  \citenamefont {Palliyaguru}, \citenamefont {Pennucci}, \citenamefont
  {Ransom}, \citenamefont {Siemens}, \citenamefont {Stairs}, \citenamefont
  {Stinebring}, \citenamefont {Stovall}, \citenamefont {Swiggum}, \citenamefont
  {Vallisneri}, \citenamefont {van Haasteren}, \citenamefont {Wang},\ and\
  \citenamefont {Zhu}}]{Arzoumanian:2015cxr}%
  \BibitemOpen
  \bibfield  {author} {\bibinfo {author} {\bibfnamefont {Z.}~\bibnamefont
  {Arzoumanian}}, \bibinfo {author} {\bibfnamefont {A.}~\bibnamefont
  {Brazier}}, \bibinfo {author} {\bibfnamefont {S.}~\bibnamefont
  {Burke-Spolaor}}, \bibinfo {author} {\bibfnamefont {S.~J.}\ \bibnamefont
  {Chamberlin}}, \bibinfo {author} {\bibfnamefont {S.}~\bibnamefont
  {Chatterjee}}, \bibinfo {author} {\bibfnamefont {B.}~\bibnamefont {Christy}},
  \bibinfo {author} {\bibfnamefont {J.~M.}\ \bibnamefont {Cordes}}, \bibinfo
  {author} {\bibfnamefont {N.~J.}\ \bibnamefont {Cornish}}, \bibinfo {author}
  {\bibfnamefont {P.~B.}\ \bibnamefont {Demorest}}, \bibinfo {author}
  {\bibfnamefont {X.}~\bibnamefont {Deng}}, \bibinfo {author} {\bibfnamefont
  {T.}~\bibnamefont {Dolch}}, \bibinfo {author} {\bibfnamefont {J.~A.}\
  \bibnamefont {Ellis}}, \bibinfo {author} {\bibfnamefont {R.~D.}\ \bibnamefont
  {Ferdman}}, \bibinfo {author} {\bibfnamefont {E.}~\bibnamefont {Fonseca}},
  \bibinfo {author} {\bibfnamefont {N.}~\bibnamefont {Garver-Daniels}},
  \bibinfo {author} {\bibfnamefont {F.}~\bibnamefont {Jenet}}, \bibinfo
  {author} {\bibfnamefont {G.}~\bibnamefont {Jones}}, \bibinfo {author}
  {\bibfnamefont {V.~M.}\ \bibnamefont {Kaspi}}, \bibinfo {author}
  {\bibfnamefont {M.}~\bibnamefont {Koop}}, \bibinfo {author} {\bibfnamefont
  {M.~T.}\ \bibnamefont {Lam}}, \bibinfo {author} {\bibfnamefont {T.~J.~W.}\
  \bibnamefont {Lazio}}, \bibinfo {author} {\bibfnamefont {L.}~\bibnamefont
  {Levin}}, \bibinfo {author} {\bibfnamefont {A.~N.}\ \bibnamefont {Lommen}},
  \bibinfo {author} {\bibfnamefont {D.~R.}\ \bibnamefont {Lorimer}}, \bibinfo
  {author} {\bibfnamefont {J.}~\bibnamefont {Luo}}, \bibinfo {author}
  {\bibfnamefont {R.~S.}\ \bibnamefont {Lynch}}, \bibinfo {author}
  {\bibfnamefont {D.~R.}\ \bibnamefont {Madison}}, \bibinfo {author}
  {\bibfnamefont {M.~A.}\ \bibnamefont {McLaughlin}}, \bibinfo {author}
  {\bibfnamefont {S.~T.}\ \bibnamefont {McWilliams}}, \bibinfo {author}
  {\bibfnamefont {D.~J.}\ \bibnamefont {Nice}}, \bibinfo {author}
  {\bibfnamefont {N.}~\bibnamefont {Palliyaguru}}, \bibinfo {author}
  {\bibfnamefont {T.~T.}\ \bibnamefont {Pennucci}}, \bibinfo {author}
  {\bibfnamefont {S.~M.}\ \bibnamefont {Ransom}}, \bibinfo {author}
  {\bibfnamefont {X.}~\bibnamefont {Siemens}}, \bibinfo {author} {\bibfnamefont
  {I.~H.}\ \bibnamefont {Stairs}}, \bibinfo {author} {\bibfnamefont {D.~R.}\
  \bibnamefont {Stinebring}}, \bibinfo {author} {\bibfnamefont
  {K.}~\bibnamefont {Stovall}}, \bibinfo {author} {\bibfnamefont
  {J.}~\bibnamefont {Swiggum}}, \bibinfo {author} {\bibfnamefont
  {M.}~\bibnamefont {Vallisneri}}, \bibinfo {author} {\bibfnamefont
  {R.}~\bibnamefont {van Haasteren}}, \bibinfo {author} {\bibfnamefont
  {Y.}~\bibnamefont {Wang}}, \ and\ \bibinfo {author} {\bibfnamefont {W.~W.}\
  \bibnamefont {Zhu}} (\bibinfo {collaboration} {NANOGrav Collaboration}),\
  }\bibfield  {title} {\enquote {\bibinfo {title} {{NANOGrav Constraints on
  Gravitational Wave Bursts with Memory}},}\ }\href {\doibase
  10.1088/0004-637X/810/2/150} {\bibfield  {journal} {\bibinfo  {journal}
  {Astrophys. J.}\ }\textbf {\bibinfo {volume} {810}},\ \bibinfo {pages} {150}
  (\bibinfo {year} {2015})},\ \Eprint {http://arxiv.org/abs/1501.05343}
  {arXiv:1501.05343 [astro-ph.GA]} \BibitemShut {NoStop}%
\bibitem [{\citenamefont {Lasky}\ \emph
  {et~al.}(2016{\natexlab{a}})\citenamefont {Lasky}, \citenamefont {Thrane},
  \citenamefont {Levin}, \citenamefont {Blackman},\ and\ \citenamefont
  {Chen}}]{Lasky:2016knh}%
  \BibitemOpen
  \bibfield  {author} {\bibinfo {author} {\bibfnamefont {Paul~D.}\ \bibnamefont
  {Lasky}}, \bibinfo {author} {\bibfnamefont {Eric}\ \bibnamefont {Thrane}},
  \bibinfo {author} {\bibfnamefont {Yuri}\ \bibnamefont {Levin}}, \bibinfo
  {author} {\bibfnamefont {Jonathan}\ \bibnamefont {Blackman}}, \ and\ \bibinfo
  {author} {\bibfnamefont {Yanbei}\ \bibnamefont {Chen}},\ }\bibfield  {title}
  {\enquote {\bibinfo {title} {{Detecting gravitational-wave memory with LIGO:
  implications of GW150914}},}\ }\href {\doibase
  10.1103/PhysRevLett.117.061102} {\bibfield  {journal} {\bibinfo  {journal}
  {Phys. Rev. Lett.}\ }\textbf {\bibinfo {volume} {117}},\ \bibinfo {pages}
  {061102} (\bibinfo {year} {2016}{\natexlab{a}})},\ \Eprint
  {http://arxiv.org/abs/1605.01415} {arXiv:1605.01415 [astro-ph.HE]}
  \BibitemShut {NoStop}%
\bibitem [{\citenamefont {Yang}\ and\ \citenamefont
  {Martynov}(2018)}]{Yang:2018ceq}%
  \BibitemOpen
  \bibfield  {author} {\bibinfo {author} {\bibfnamefont {Huan}\ \bibnamefont
  {Yang}}\ and\ \bibinfo {author} {\bibfnamefont {Denis}\ \bibnamefont
  {Martynov}},\ }\bibfield  {title} {\enquote {\bibinfo {title} {{Testing
  Gravitational Memory Generation with Compact Binary Mergers}},}\ }\href
  {\doibase 10.1103/PhysRevLett.121.071102} {\bibfield  {journal} {\bibinfo
  {journal} {Phys. Rev. Lett.}\ }\textbf {\bibinfo {volume} {121}},\ \bibinfo
  {pages} {071102} (\bibinfo {year} {2018})},\ \Eprint
  {http://arxiv.org/abs/1803.02429} {arXiv:1803.02429 [gr-qc]} \BibitemShut
  {NoStop}%
\bibitem [{\citenamefont {Johnson}\ \emph {et~al.}(2019)\citenamefont
  {Johnson}, \citenamefont {Kapadia}, \citenamefont {Osborne}, \citenamefont
  {Hixon},\ and\ \citenamefont {Kennefick}}]{Johnson:2018xly}%
  \BibitemOpen
  \bibfield  {author} {\bibinfo {author} {\bibfnamefont {Aaron~D.}\
  \bibnamefont {Johnson}}, \bibinfo {author} {\bibfnamefont {Shasvath~J.}\
  \bibnamefont {Kapadia}}, \bibinfo {author} {\bibfnamefont {Andrew}\
  \bibnamefont {Osborne}}, \bibinfo {author} {\bibfnamefont {Alex}\
  \bibnamefont {Hixon}}, \ and\ \bibinfo {author} {\bibfnamefont {Daniel}\
  \bibnamefont {Kennefick}},\ }\bibfield  {title} {\enquote {\bibinfo {title}
  {{Prospects of detecting the nonlinear gravitational wave memory}},}\ }\href
  {\doibase 10.1103/PhysRevD.99.044045} {\bibfield  {journal} {\bibinfo
  {journal} {Phys. Rev. D}\ }\textbf {\bibinfo {volume} {99}},\ \bibinfo
  {pages} {044045} (\bibinfo {year} {2019})},\ \Eprint
  {http://arxiv.org/abs/1810.09563} {arXiv:1810.09563 [gr-qc]} \BibitemShut
  {NoStop}%
\bibitem [{\citenamefont {Islo}\ \emph {et~al.}(2019)\citenamefont {Islo},
  \citenamefont {Simon}, \citenamefont {Burke-Spolaor},\ and\ \citenamefont
  {Siemens}}]{Islo:2019qht}%
  \BibitemOpen
  \bibfield  {author} {\bibinfo {author} {\bibfnamefont {Kristina}\
  \bibnamefont {Islo}}, \bibinfo {author} {\bibfnamefont {Joseph}\ \bibnamefont
  {Simon}}, \bibinfo {author} {\bibfnamefont {Sarah}\ \bibnamefont
  {Burke-Spolaor}}, \ and\ \bibinfo {author} {\bibfnamefont {Xavier}\
  \bibnamefont {Siemens}},\ }\bibfield  {title} {\enquote {\bibinfo {title}
  {{Prospects for Memory Detection with Low-Frequency Gravitational Wave
  Detectors}},}\ }\href@noop {} {\  (\bibinfo {year} {2019})},\ \Eprint
  {http://arxiv.org/abs/1906.11936} {arXiv:1906.11936 [astro-ph.HE]}
  \BibitemShut {NoStop}%
\bibitem [{\citenamefont {Divakarla}\ \emph {et~al.}(2020)\citenamefont
  {Divakarla}, \citenamefont {Thrane}, \citenamefont {Lasky},\ and\
  \citenamefont {Whiting}}]{Divakarla:2019zjj}%
  \BibitemOpen
  \bibfield  {author} {\bibinfo {author} {\bibfnamefont {Atul~K.}\ \bibnamefont
  {Divakarla}}, \bibinfo {author} {\bibfnamefont {Eric}\ \bibnamefont
  {Thrane}}, \bibinfo {author} {\bibfnamefont {Paul~D.}\ \bibnamefont {Lasky}},
  \ and\ \bibinfo {author} {\bibfnamefont {Bernard~F.}\ \bibnamefont
  {Whiting}},\ }\bibfield  {title} {\enquote {\bibinfo {title} {{Memory Effect
  or Cosmic String? Classifying Gravitational-Wave Bursts with Bayesian
  Inference}},}\ }\href {\doibase 10.1103/PhysRevD.102.023010} {\bibfield
  {journal} {\bibinfo  {journal} {Phys. Rev. D}\ }\textbf {\bibinfo {volume}
  {102}},\ \bibinfo {pages} {023010} (\bibinfo {year} {2020})},\ \Eprint
  {http://arxiv.org/abs/1911.07998} {arXiv:1911.07998 [gr-qc]} \BibitemShut
  {NoStop}%
\bibitem [{\citenamefont {Aggarwal}\ \emph {et~al.}(2020)\citenamefont
  {Aggarwal}, \citenamefont {Arzoumanian}, \citenamefont {Baker}, \citenamefont
  {Brazier}, \citenamefont {Brook}, \citenamefont {Burke-Spolaor},
  \citenamefont {Chatterjee}, \citenamefont {Cordes}, \citenamefont {Cornish},
  \citenamefont {Crawford}, \citenamefont {Cromartie}, \citenamefont {Crowter},
  \citenamefont {DeCesar}, \citenamefont {Demorest}, \citenamefont {Dolch},
  \citenamefont {Ellis}, \citenamefont {Ferdman}, \citenamefont {Ferrara},
  \citenamefont {Fonseca}, \citenamefont {Garver-Daniels}, \citenamefont
  {Gentile}, \citenamefont {Good}, \citenamefont {Hazboun}, \citenamefont
  {Holgado}, \citenamefont {Huerta}, \citenamefont {Islo}, \citenamefont
  {Jennings}, \citenamefont {Jones}, \citenamefont {Jones}, \citenamefont
  {Kaplan}, \citenamefont {Kelley}, \citenamefont {Key}, \citenamefont {Lam},
  \citenamefont {Lazio}, \citenamefont {Levin}, \citenamefont {Lorimer},
  \citenamefont {Luo}, \citenamefont {Lynch}, \citenamefont {Madison},
  \citenamefont {McLaughlin}, \citenamefont {McWilliams}, \citenamefont
  {Mingarelli}, \citenamefont {Ng}, \citenamefont {Nice}, \citenamefont
  {Pennucci}, \citenamefont {Pol}, \citenamefont {Ransom}, \citenamefont {Ray},
  \citenamefont {Siemens}, \citenamefont {Simon}, \citenamefont {Spiewak},
  \citenamefont {Stairs}, \citenamefont {Stinebring}, \citenamefont {Stovall},
  \citenamefont {Swiggum}, \citenamefont {Taylor}, \citenamefont {Vallisneri},
  \citenamefont {van Haasteren}, \citenamefont {Vigeland}, \citenamefont
  {Witt},\ and\ \citenamefont {Zhu}}]{Aggarwal:2019ypr}%
  \BibitemOpen
  \bibfield  {author} {\bibinfo {author} {\bibfnamefont {K.}~\bibnamefont
  {Aggarwal}}, \bibinfo {author} {\bibfnamefont {Z.}~\bibnamefont
  {Arzoumanian}}, \bibinfo {author} {\bibfnamefont {P.~T.}\ \bibnamefont
  {Baker}}, \bibinfo {author} {\bibfnamefont {A.}~\bibnamefont {Brazier}},
  \bibinfo {author} {\bibfnamefont {P.~R.}\ \bibnamefont {Brook}}, \bibinfo
  {author} {\bibfnamefont {S.}~\bibnamefont {Burke-Spolaor}}, \bibinfo {author}
  {\bibfnamefont {S.}~\bibnamefont {Chatterjee}}, \bibinfo {author}
  {\bibfnamefont {J.~M.}\ \bibnamefont {Cordes}}, \bibinfo {author}
  {\bibfnamefont {N.~J.}\ \bibnamefont {Cornish}}, \bibinfo {author}
  {\bibfnamefont {F.}~\bibnamefont {Crawford}}, \bibinfo {author}
  {\bibfnamefont {H.~T.}\ \bibnamefont {Cromartie}}, \bibinfo {author}
  {\bibfnamefont {K.}~\bibnamefont {Crowter}}, \bibinfo {author} {\bibfnamefont
  {M.}~\bibnamefont {DeCesar}}, \bibinfo {author} {\bibfnamefont {P.~B.}\
  \bibnamefont {Demorest}}, \bibinfo {author} {\bibfnamefont {T.}~\bibnamefont
  {Dolch}}, \bibinfo {author} {\bibfnamefont {J.~A.}\ \bibnamefont {Ellis}},
  \bibinfo {author} {\bibfnamefont {R.~D.}\ \bibnamefont {Ferdman}}, \bibinfo
  {author} {\bibfnamefont {E.~C.}\ \bibnamefont {Ferrara}}, \bibinfo {author}
  {\bibfnamefont {E.}~\bibnamefont {Fonseca}}, \bibinfo {author} {\bibfnamefont
  {N.}~\bibnamefont {Garver-Daniels}}, \bibinfo {author} {\bibfnamefont
  {P.}~\bibnamefont {Gentile}}, \bibinfo {author} {\bibfnamefont
  {D.}~\bibnamefont {Good}}, \bibinfo {author} {\bibfnamefont {J.~S.}\
  \bibnamefont {Hazboun}}, \bibinfo {author} {\bibfnamefont {A.~M.}\
  \bibnamefont {Holgado}}, \bibinfo {author} {\bibfnamefont {E.~A.}\
  \bibnamefont {Huerta}}, \bibinfo {author} {\bibfnamefont {K.}~\bibnamefont
  {Islo}}, \bibinfo {author} {\bibfnamefont {R.}~\bibnamefont {Jennings}},
  \bibinfo {author} {\bibfnamefont {G.}~\bibnamefont {Jones}}, \bibinfo
  {author} {\bibfnamefont {M.~L.}\ \bibnamefont {Jones}}, \bibinfo {author}
  {\bibfnamefont {D.~L.}\ \bibnamefont {Kaplan}}, \bibinfo {author}
  {\bibfnamefont {L.~Z.}\ \bibnamefont {Kelley}}, \bibinfo {author}
  {\bibfnamefont {J.~S.}\ \bibnamefont {Key}}, \bibinfo {author} {\bibfnamefont
  {M.~T.}\ \bibnamefont {Lam}}, \bibinfo {author} {\bibfnamefont {T.~J.~W.}\
  \bibnamefont {Lazio}}, \bibinfo {author} {\bibfnamefont {L.}~\bibnamefont
  {Levin}}, \bibinfo {author} {\bibfnamefont {D.~R.}\ \bibnamefont {Lorimer}},
  \bibinfo {author} {\bibfnamefont {J.}~\bibnamefont {Luo}}, \bibinfo {author}
  {\bibfnamefont {R.~S.}\ \bibnamefont {Lynch}}, \bibinfo {author}
  {\bibfnamefont {D.~R.}\ \bibnamefont {Madison}}, \bibinfo {author}
  {\bibfnamefont {M.~A.}\ \bibnamefont {McLaughlin}}, \bibinfo {author}
  {\bibfnamefont {S.~T.}\ \bibnamefont {McWilliams}}, \bibinfo {author}
  {\bibfnamefont {C.~M.~F.}\ \bibnamefont {Mingarelli}}, \bibinfo {author}
  {\bibfnamefont {C.}~\bibnamefont {Ng}}, \bibinfo {author} {\bibfnamefont
  {D.~J.}\ \bibnamefont {Nice}}, \bibinfo {author} {\bibfnamefont {T.~T.}\
  \bibnamefont {Pennucci}}, \bibinfo {author} {\bibfnamefont {N.~S.}\
  \bibnamefont {Pol}}, \bibinfo {author} {\bibfnamefont {S.~M.}\ \bibnamefont
  {Ransom}}, \bibinfo {author} {\bibfnamefont {P.~S.}\ \bibnamefont {Ray}},
  \bibinfo {author} {\bibfnamefont {X.}~\bibnamefont {Siemens}}, \bibinfo
  {author} {\bibfnamefont {J.}~\bibnamefont {Simon}}, \bibinfo {author}
  {\bibfnamefont {R.}~\bibnamefont {Spiewak}}, \bibinfo {author} {\bibfnamefont
  {I.~H.}\ \bibnamefont {Stairs}}, \bibinfo {author} {\bibfnamefont {D.~R.}\
  \bibnamefont {Stinebring}}, \bibinfo {author} {\bibfnamefont
  {K.}~\bibnamefont {Stovall}}, \bibinfo {author} {\bibfnamefont {J.~K.}\
  \bibnamefont {Swiggum}}, \bibinfo {author} {\bibfnamefont {S.~R.}\
  \bibnamefont {Taylor}}, \bibinfo {author} {\bibfnamefont {M.}~\bibnamefont
  {Vallisneri}}, \bibinfo {author} {\bibfnamefont {R.}~\bibnamefont {van
  Haasteren}}, \bibinfo {author} {\bibfnamefont {S.~J.}\ \bibnamefont
  {Vigeland}}, \bibinfo {author} {\bibfnamefont {C.~A.}\ \bibnamefont {Witt}},
  \ and\ \bibinfo {author} {\bibfnamefont {W.~W.}\ \bibnamefont {Zhu}}
  (\bibinfo {collaboration} {NANOGrav Collaboration}),\ }\bibfield  {title}
  {\enquote {\bibinfo {title} {{The NANOGrav 11 yr Data Set: Limits on
  Gravitational Wave Memory}},}\ }\href {\doibase 10.3847/1538-4357/ab6083}
  {\bibfield  {journal} {\bibinfo  {journal} {Astrophys. J.}\ }\textbf
  {\bibinfo {volume} {889}},\ \bibinfo {pages} {38} (\bibinfo {year} {2020})},\
  \Eprint {http://arxiv.org/abs/1911.08488} {arXiv:1911.08488 [astro-ph.HE]}
  \BibitemShut {NoStop}%
\bibitem [{\citenamefont {H\"ubner}\ \emph {et~al.}(2020)\citenamefont
  {H\"ubner}, \citenamefont {Talbot}, \citenamefont {Lasky},\ and\
  \citenamefont {Thrane}}]{Hubner:2019sly}%
  \BibitemOpen
  \bibfield  {author} {\bibinfo {author} {\bibfnamefont {Moritz}\ \bibnamefont
  {H\"ubner}}, \bibinfo {author} {\bibfnamefont {Colm}\ \bibnamefont {Talbot}},
  \bibinfo {author} {\bibfnamefont {Paul~D.}\ \bibnamefont {Lasky}}, \ and\
  \bibinfo {author} {\bibfnamefont {Eric}\ \bibnamefont {Thrane}},\ }\bibfield
  {title} {\enquote {\bibinfo {title} {{Measuring gravitational-wave memory in
  the first LIGO/Virgo gravitational-wave transient catalog}},}\ }\href
  {\doibase 10.1103/PhysRevD.101.023011} {\bibfield  {journal} {\bibinfo
  {journal} {Phys. Rev. D}\ }\textbf {\bibinfo {volume} {101}},\ \bibinfo
  {pages} {023011} (\bibinfo {year} {2020})},\ \Eprint
  {http://arxiv.org/abs/1911.12496} {arXiv:1911.12496 [astro-ph.HE]}
  \BibitemShut {NoStop}%
\bibitem [{\citenamefont {Boersma}\ \emph {et~al.}(2020)\citenamefont
  {Boersma}, \citenamefont {Nichols},\ and\ \citenamefont
  {Schmidt}}]{Boersma:2020gxx}%
  \BibitemOpen
  \bibfield  {author} {\bibinfo {author} {\bibfnamefont {Oliver~M.}\
  \bibnamefont {Boersma}}, \bibinfo {author} {\bibfnamefont {David~A.}\
  \bibnamefont {Nichols}}, \ and\ \bibinfo {author} {\bibfnamefont {Patricia}\
  \bibnamefont {Schmidt}},\ }\bibfield  {title} {\enquote {\bibinfo {title}
  {{Forecasts for detecting the gravitational-wave memory effect with Advanced
  LIGO and Virgo}},}\ }\href {\doibase 10.1103/PhysRevD.101.083026} {\bibfield
  {journal} {\bibinfo  {journal} {Phys. Rev. D}\ }\textbf {\bibinfo {volume}
  {101}},\ \bibinfo {pages} {083026} (\bibinfo {year} {2020})},\ \Eprint
  {http://arxiv.org/abs/2002.01821} {arXiv:2002.01821 [astro-ph.HE]}
  \BibitemShut {NoStop}%
\bibitem [{\citenamefont {Ebersold}\ and\ \citenamefont
  {Tiwari}(2020)}]{Ebersold:2020zah}%
  \BibitemOpen
  \bibfield  {author} {\bibinfo {author} {\bibfnamefont {Michael}\ \bibnamefont
  {Ebersold}}\ and\ \bibinfo {author} {\bibfnamefont {Shubhanshu}\ \bibnamefont
  {Tiwari}},\ }\bibfield  {title} {\enquote {\bibinfo {title} {{Search for
  nonlinear memory from subsolar mass compact binary mergers}},}\ }\href
  {\doibase 10.1103/PhysRevD.101.104041} {\bibfield  {journal} {\bibinfo
  {journal} {Phys. Rev. D}\ }\textbf {\bibinfo {volume} {101}},\ \bibinfo
  {pages} {104041} (\bibinfo {year} {2020})},\ \Eprint
  {http://arxiv.org/abs/2005.03306} {arXiv:2005.03306 [gr-qc]} \BibitemShut
  {NoStop}%
\bibitem [{\citenamefont {Burko}\ and\ \citenamefont
  {Khanna}(2020)}]{Burko:2020gse}%
  \BibitemOpen
  \bibfield  {author} {\bibinfo {author} {\bibfnamefont {Lior~M.}\ \bibnamefont
  {Burko}}\ and\ \bibinfo {author} {\bibfnamefont {Gaurav}\ \bibnamefont
  {Khanna}},\ }\bibfield  {title} {\enquote {\bibinfo {title} {{Climbing up the
  memory staircase: Equatorial zoom-whirl orbits}},}\ }\href {\doibase
  10.1103/PhysRevD.102.084035} {\bibfield  {journal} {\bibinfo  {journal}
  {Phys. Rev. D}\ }\textbf {\bibinfo {volume} {102}},\ \bibinfo {pages}
  {084035} (\bibinfo {year} {2020})},\ \Eprint
  {http://arxiv.org/abs/2007.12545} {arXiv:2007.12545 [gr-qc]} \BibitemShut
  {NoStop}%
\bibitem [{\citenamefont {Kibble}(1976)}]{Kibble:1976sj}%
  \BibitemOpen
  \bibfield  {author} {\bibinfo {author} {\bibfnamefont {T.~W.~B.}\
  \bibnamefont {Kibble}},\ }\bibfield  {title} {\enquote {\bibinfo {title}
  {{Topology of Cosmic Domains and Strings}},}\ }\href {\doibase
  10.1088/0305-4470/9/8/029} {\bibfield  {journal} {\bibinfo  {journal} {J.
  Phys. A}\ }\textbf {\bibinfo {volume} {9}},\ \bibinfo {pages} {1387--1398}
  (\bibinfo {year} {1976})}\BibitemShut {NoStop}%
\bibitem [{\citenamefont {Vilenkin}(1985)}]{Vilenkin:1984ib}%
  \BibitemOpen
  \bibfield  {author} {\bibinfo {author} {\bibfnamefont {Alexander}\
  \bibnamefont {Vilenkin}},\ }\bibfield  {title} {\enquote {\bibinfo {title}
  {{Cosmic Strings and Domain Walls}},}\ }\href {\doibase
  10.1016/0370-1573(85)90033-X} {\bibfield  {journal} {\bibinfo  {journal}
  {Phys. Rept.}\ }\textbf {\bibinfo {volume} {121}},\ \bibinfo {pages}
  {263--315} (\bibinfo {year} {1985})}\BibitemShut {NoStop}%
\bibitem [{\citenamefont {Hindmarsh}\ and\ \citenamefont
  {Kibble}(1995)}]{Hindmarsh:1994re}%
  \BibitemOpen
  \bibfield  {author} {\bibinfo {author} {\bibfnamefont {M.~B.}\ \bibnamefont
  {Hindmarsh}}\ and\ \bibinfo {author} {\bibfnamefont {T.~W.~B.}\ \bibnamefont
  {Kibble}},\ }\bibfield  {title} {\enquote {\bibinfo {title} {{Cosmic
  strings}},}\ }\href {\doibase 10.1088/0034-4885/58/5/001} {\bibfield
  {journal} {\bibinfo  {journal} {Rept. Prog. Phys.}\ }\textbf {\bibinfo
  {volume} {58}},\ \bibinfo {pages} {477--562} (\bibinfo {year} {1995})},\
  \Eprint {http://arxiv.org/abs/hep-ph/9411342} {arXiv:hep-ph/9411342}
  \BibitemShut {NoStop}%
\bibitem [{\citenamefont {Vilenkin}\ and\ \citenamefont
  {Shellard}(2000)}]{Vilenkin:2000jqa}%
  \BibitemOpen
  \bibfield  {author} {\bibinfo {author} {\bibfnamefont {A.}~\bibnamefont
  {Vilenkin}}\ and\ \bibinfo {author} {\bibfnamefont {E.~P.~S.}\ \bibnamefont
  {Shellard}},\ }\href
  {https://www.cambridge.org/vi/academic/subjects/physics/theoretical-physics-and-mathematical-physics/cosmic-strings-and-other-topological-defects?format=PB}
  {\emph {\bibinfo {title} {{Cosmic Strings and Other Topological Defects}}}},\
  Cambridge Monographs on Mathematical Physics\ (\bibinfo  {publisher}
  {Cambridge University Press},\ \bibinfo {year} {2000})\BibitemShut {NoStop}%
\bibitem [{\citenamefont {Jeannerot}\ \emph {et~al.}(2003)\citenamefont
  {Jeannerot}, \citenamefont {Rocher},\ and\ \citenamefont
  {Sakellariadou}}]{Jeannerot:2003qv}%
  \BibitemOpen
  \bibfield  {author} {\bibinfo {author} {\bibfnamefont {Rachel}\ \bibnamefont
  {Jeannerot}}, \bibinfo {author} {\bibfnamefont {Jonathan}\ \bibnamefont
  {Rocher}}, \ and\ \bibinfo {author} {\bibfnamefont {Mairi}\ \bibnamefont
  {Sakellariadou}},\ }\bibfield  {title} {\enquote {\bibinfo {title} {{How
  generic is cosmic string formation in SUSY GUTs}},}\ }\href {\doibase
  10.1103/PhysRevD.68.103514} {\bibfield  {journal} {\bibinfo  {journal} {Phys.
  Rev. D}\ }\textbf {\bibinfo {volume} {68}},\ \bibinfo {pages} {103514}
  (\bibinfo {year} {2003})},\ \Eprint {http://arxiv.org/abs/hep-ph/0308134}
  {arXiv:hep-ph/0308134} \BibitemShut {NoStop}%
\bibitem [{\citenamefont {Kibble}\ and\ \citenamefont
  {Turok}(1982)}]{Kibble:1982cb}%
  \BibitemOpen
  \bibfield  {author} {\bibinfo {author} {\bibfnamefont {T.~W.~B.}\
  \bibnamefont {Kibble}}\ and\ \bibinfo {author} {\bibfnamefont {Neil}\
  \bibnamefont {Turok}},\ }\bibfield  {title} {\enquote {\bibinfo {title}
  {{Selfintersection of Cosmic Strings}},}\ }\href {\doibase
  10.1016/0370-2693(82)90993-5} {\bibfield  {journal} {\bibinfo  {journal}
  {Phys. Lett. B}\ }\textbf {\bibinfo {volume} {116}},\ \bibinfo {pages}
  {141--143} (\bibinfo {year} {1982})}\BibitemShut {NoStop}%
\bibitem [{\citenamefont {Turok}(1984)}]{Turok:1984cn}%
  \BibitemOpen
  \bibfield  {author} {\bibinfo {author} {\bibfnamefont {Neil}\ \bibnamefont
  {Turok}},\ }\bibfield  {title} {\enquote {\bibinfo {title} {{Grand Unified
  Strings and Galaxy Formation}},}\ }\href {\doibase
  10.1016/0550-3213(84)90407-3} {\bibfield  {journal} {\bibinfo  {journal}
  {Nucl. Phys. B}\ }\textbf {\bibinfo {volume} {242}},\ \bibinfo {pages}
  {520--541} (\bibinfo {year} {1984})}\BibitemShut {NoStop}%
\bibitem [{\citenamefont {Damour}\ and\ \citenamefont
  {Vilenkin}(2001)}]{Damour:2001bk}%
  \BibitemOpen
  \bibfield  {author} {\bibinfo {author} {\bibfnamefont {Thibault}\
  \bibnamefont {Damour}}\ and\ \bibinfo {author} {\bibfnamefont {Alexander}\
  \bibnamefont {Vilenkin}},\ }\bibfield  {title} {\enquote {\bibinfo {title}
  {{Gravitational wave bursts from cusps and kinks on cosmic strings}},}\
  }\href {\doibase 10.1103/PhysRevD.64.064008} {\bibfield  {journal} {\bibinfo
  {journal} {Phys. Rev. D}\ }\textbf {\bibinfo {volume} {64}},\ \bibinfo
  {pages} {064008} (\bibinfo {year} {2001})},\ \Eprint
  {http://arxiv.org/abs/gr-qc/0104026} {arXiv:gr-qc/0104026} \BibitemShut
  {NoStop}%
\bibitem [{\citenamefont {Abbott}\ \emph {et~al.}(2009)\citenamefont {Abbott}
  \emph {et~al.}}]{Abbott:2009rr}%
  \BibitemOpen
  \bibfield  {author} {\bibinfo {author} {\bibfnamefont {B.~P.}\ \bibnamefont
  {Abbott}} \emph {et~al.} (\bibinfo {collaboration} {LIGO Scientific
  Collaboration}),\ }\bibfield  {title} {\enquote {\bibinfo {title} {{First
  LIGO search for gravitational wave bursts from cosmic (super)strings}},}\
  }\href {\doibase 10.1103/PhysRevD.80.062002} {\bibfield  {journal} {\bibinfo
  {journal} {Phys. Rev. D}\ }\textbf {\bibinfo {volume} {80}},\ \bibinfo
  {pages} {062002} (\bibinfo {year} {2009})},\ \Eprint
  {http://arxiv.org/abs/0904.4718} {arXiv:0904.4718 [astro-ph.CO]} \BibitemShut
  {NoStop}%
\bibitem [{\citenamefont {Abbott}\ \emph {et~al.}(2018)\citenamefont {Abbott}
  \emph {et~al.}}]{Abbott:2017mem}%
  \BibitemOpen
  \bibfield  {author} {\bibinfo {author} {\bibfnamefont {B.~P.}\ \bibnamefont
  {Abbott}} \emph {et~al.} (\bibinfo {collaboration} {LIGO Scientific
  Collaboration, Virgo Collaboration}),\ }\bibfield  {title} {\enquote
  {\bibinfo {title} {{Constraints on cosmic strings using data from the first
  Advanced LIGO observing run}},}\ }\href {\doibase 10.1103/PhysRevD.97.102002}
  {\bibfield  {journal} {\bibinfo  {journal} {Phys. Rev. D}\ }\textbf {\bibinfo
  {volume} {97}},\ \bibinfo {pages} {102002} (\bibinfo {year} {2018})},\
  \Eprint {http://arxiv.org/abs/1712.01168} {arXiv:1712.01168 [gr-qc]}
  \BibitemShut {NoStop}%
\bibitem [{\citenamefont {Abbott}\ \emph {et~al.}(2019)\citenamefont {Abbott}
  \emph {et~al.}}]{LIGOScientific:2019vic}%
  \BibitemOpen
  \bibfield  {author} {\bibinfo {author} {\bibfnamefont {B.~P.}\ \bibnamefont
  {Abbott}} \emph {et~al.} (\bibinfo {collaboration} {LIGO Scientific
  Collaboration, Virgo Collaboration}),\ }\bibfield  {title} {\enquote
  {\bibinfo {title} {{Search for the isotropic stochastic background using data
  from Advanced LIGO's second observing run}},}\ }\href {\doibase
  10.1103/PhysRevD.100.061101} {\bibfield  {journal} {\bibinfo  {journal}
  {Phys. Rev. D}\ }\textbf {\bibinfo {volume} {100}},\ \bibinfo {pages}
  {061101} (\bibinfo {year} {2019})},\ \Eprint
  {http://arxiv.org/abs/1903.02886} {arXiv:1903.02886 [gr-qc]} \BibitemShut
  {NoStop}%
\bibitem [{\citenamefont {Abbott}\ \emph
  {et~al.}(2021{\natexlab{c}})\citenamefont {Abbott} \emph
  {et~al.}}]{Abbott:2021xxi}%
  \BibitemOpen
  \bibfield  {author} {\bibinfo {author} {\bibfnamefont {R.}~\bibnamefont
  {Abbott}} \emph {et~al.} (\bibinfo {collaboration} {LIGO Scientific
  Collaboration, Virgo Collaboration, KAGRA Collaboration}),\ }\bibfield
  {title} {\enquote {\bibinfo {title} {{Upper Limits on the Isotropic
  Gravitational-Wave Background from Advanced LIGO's and Advanced Virgo's Third
  Observing Run}},}\ }\href@noop {} {\  (\bibinfo {year}
  {2021}{\natexlab{c}})},\ \Eprint {http://arxiv.org/abs/2101.12130}
  {arXiv:2101.12130 [gr-qc]} \BibitemShut {NoStop}%
\bibitem [{\citenamefont {Abbott}\ \emph
  {et~al.}(2021{\natexlab{d}})\citenamefont {Abbott} \emph
  {et~al.}}]{LIGOScientific:2021nrg}%
  \BibitemOpen
  \bibfield  {author} {\bibinfo {author} {\bibfnamefont {R.}~\bibnamefont
  {Abbott}} \emph {et~al.} (\bibinfo {collaboration} {LIGO Scientific
  Collaboration, Virgo Collaboration, KAGRA Collaboration}),\ }\bibfield
  {title} {\enquote {\bibinfo {title} {{Constraints on Cosmic Strings Using
  Data from the Third Advanced LIGO\textendash{}Virgo Observing Run}},}\ }\href
  {\doibase 10.1103/PhysRevLett.126.241102} {\bibfield  {journal} {\bibinfo
  {journal} {Phys. Rev. Lett.}\ }\textbf {\bibinfo {volume} {126}},\ \bibinfo
  {pages} {241102} (\bibinfo {year} {2021}{\natexlab{d}})},\ \Eprint
  {http://arxiv.org/abs/2101.12248} {arXiv:2101.12248 [gr-qc]} \BibitemShut
  {NoStop}%
\bibitem [{\citenamefont {Lasky}\ \emph
  {et~al.}(2016{\natexlab{b}})\citenamefont {Lasky}, \citenamefont
  {Mingarelli}, \citenamefont {Smith}, \citenamefont {Giblin}, \citenamefont
  {Thrane}, \citenamefont {Reardon}, \citenamefont {Caldwell}, \citenamefont
  {Bailes}, \citenamefont {Bhat}, \citenamefont {Burke-Spolaor}, \citenamefont
  {Dai}, \citenamefont {Dempsey}, \citenamefont {Hobbs}, \citenamefont {Kerr},
  \citenamefont {Levin}, \citenamefont {Manchester}, \citenamefont
  {Os\l{}owski}, \citenamefont {Ravi}, \citenamefont {Rosado}, \citenamefont
  {Shannon}, \citenamefont {Spiewak}, \citenamefont {van Straten},
  \citenamefont {Toomey}, \citenamefont {Wang}, \citenamefont {Wen},
  \citenamefont {You},\ and\ \citenamefont {Zhu}}]{Lasky:2015lej}%
  \BibitemOpen
  \bibfield  {author} {\bibinfo {author} {\bibfnamefont {Paul~D.}\ \bibnamefont
  {Lasky}}, \bibinfo {author} {\bibfnamefont {Chiara M.~F.}\ \bibnamefont
  {Mingarelli}}, \bibinfo {author} {\bibfnamefont {Tristan~L.}\ \bibnamefont
  {Smith}}, \bibinfo {author} {\bibfnamefont {John~T.}\ \bibnamefont {Giblin}},
  \bibinfo {author} {\bibfnamefont {Eric}\ \bibnamefont {Thrane}}, \bibinfo
  {author} {\bibfnamefont {Daniel~J.}\ \bibnamefont {Reardon}}, \bibinfo
  {author} {\bibfnamefont {Robert}\ \bibnamefont {Caldwell}}, \bibinfo {author}
  {\bibfnamefont {Matthew}\ \bibnamefont {Bailes}}, \bibinfo {author}
  {\bibfnamefont {N.~D.~Ramesh}\ \bibnamefont {Bhat}}, \bibinfo {author}
  {\bibfnamefont {Sarah}\ \bibnamefont {Burke-Spolaor}}, \bibinfo {author}
  {\bibfnamefont {Shi}\ \bibnamefont {Dai}}, \bibinfo {author} {\bibfnamefont
  {James}\ \bibnamefont {Dempsey}}, \bibinfo {author} {\bibfnamefont {George}\
  \bibnamefont {Hobbs}}, \bibinfo {author} {\bibfnamefont {Matthew}\
  \bibnamefont {Kerr}}, \bibinfo {author} {\bibfnamefont {Yuri}\ \bibnamefont
  {Levin}}, \bibinfo {author} {\bibfnamefont {Richard~N.}\ \bibnamefont
  {Manchester}}, \bibinfo {author} {\bibfnamefont {Stefan}\ \bibnamefont
  {Os\l{}owski}}, \bibinfo {author} {\bibfnamefont {Vikram}\ \bibnamefont
  {Ravi}}, \bibinfo {author} {\bibfnamefont {Pablo~A.}\ \bibnamefont {Rosado}},
  \bibinfo {author} {\bibfnamefont {Ryan~M.}\ \bibnamefont {Shannon}}, \bibinfo
  {author} {\bibfnamefont {Ren\'ee}\ \bibnamefont {Spiewak}}, \bibinfo {author}
  {\bibfnamefont {Willem}\ \bibnamefont {van Straten}}, \bibinfo {author}
  {\bibfnamefont {Lawrence}\ \bibnamefont {Toomey}}, \bibinfo {author}
  {\bibfnamefont {Jingbo}\ \bibnamefont {Wang}}, \bibinfo {author}
  {\bibfnamefont {Linqing}\ \bibnamefont {Wen}}, \bibinfo {author}
  {\bibfnamefont {Xiaopeng}\ \bibnamefont {You}}, \ and\ \bibinfo {author}
  {\bibfnamefont {Xingjiang}\ \bibnamefont {Zhu}},\ }\bibfield  {title}
  {\enquote {\bibinfo {title} {{Gravitational-wave cosmology across 29 decades
  in frequency}},}\ }\href {\doibase 10.1103/PhysRevX.6.011035} {\bibfield
  {journal} {\bibinfo  {journal} {Phys. Rev. X}\ }\textbf {\bibinfo {volume}
  {6}},\ \bibinfo {pages} {011035} (\bibinfo {year} {2016}{\natexlab{b}})},\
  \Eprint {http://arxiv.org/abs/1511.05994} {arXiv:1511.05994 [astro-ph.CO]}
  \BibitemShut {NoStop}%
\bibitem [{\citenamefont {Blanco-Pillado}\ \emph
  {et~al.}(2018{\natexlab{a}})\citenamefont {Blanco-Pillado}, \citenamefont
  {Olum},\ and\ \citenamefont {Siemens}}]{Blanco-Pillado:2017rnf}%
  \BibitemOpen
  \bibfield  {author} {\bibinfo {author} {\bibfnamefont {Jose~J.}\ \bibnamefont
  {Blanco-Pillado}}, \bibinfo {author} {\bibfnamefont {Ken~D.}\ \bibnamefont
  {Olum}}, \ and\ \bibinfo {author} {\bibfnamefont {Xavier}\ \bibnamefont
  {Siemens}},\ }\bibfield  {title} {\enquote {\bibinfo {title} {{New limits on
  cosmic strings from gravitational wave observation}},}\ }\href {\doibase
  10.1016/j.physletb.2018.01.050} {\bibfield  {journal} {\bibinfo  {journal}
  {Phys. Lett. B}\ }\textbf {\bibinfo {volume} {778}},\ \bibinfo {pages}
  {392--396} (\bibinfo {year} {2018}{\natexlab{a}})},\ \Eprint
  {http://arxiv.org/abs/1709.02434} {arXiv:1709.02434 [astro-ph.CO]}
  \BibitemShut {NoStop}%
\bibitem [{\citenamefont {Yonemaru}\ \emph {et~al.}(2021)\citenamefont
  {Yonemaru}, \citenamefont {Kuroyanagi}, \citenamefont {Hobbs}, \citenamefont
  {Takahashi}, \citenamefont {Zhu}, \citenamefont {Coles}, \citenamefont {Dai},
  \citenamefont {Howard}, \citenamefont {Manchester}, \citenamefont {Reardon},
  \citenamefont {Russell}, \citenamefont {Shannon}, \citenamefont
  {Thyagarajan}, \citenamefont {Spiewak},\ and\ \citenamefont
  {Wang}}]{Yonemaru:2020bmr}%
  \BibitemOpen
  \bibfield  {author} {\bibinfo {author} {\bibfnamefont {N.}~\bibnamefont
  {Yonemaru}}, \bibinfo {author} {\bibfnamefont {S.}~\bibnamefont
  {Kuroyanagi}}, \bibinfo {author} {\bibfnamefont {G.}~\bibnamefont {Hobbs}},
  \bibinfo {author} {\bibfnamefont {K.}~\bibnamefont {Takahashi}}, \bibinfo
  {author} {\bibfnamefont {X.-J.}\ \bibnamefont {Zhu}}, \bibinfo {author}
  {\bibfnamefont {W.~A.}\ \bibnamefont {Coles}}, \bibinfo {author}
  {\bibfnamefont {S.}~\bibnamefont {Dai}}, \bibinfo {author} {\bibfnamefont
  {E.}~\bibnamefont {Howard}}, \bibinfo {author} {\bibfnamefont
  {R.}~\bibnamefont {Manchester}}, \bibinfo {author} {\bibfnamefont
  {D.}~\bibnamefont {Reardon}}, \bibinfo {author} {\bibfnamefont
  {C.}~\bibnamefont {Russell}}, \bibinfo {author} {\bibfnamefont {R.~M.}\
  \bibnamefont {Shannon}}, \bibinfo {author} {\bibfnamefont {N.}~\bibnamefont
  {Thyagarajan}}, \bibinfo {author} {\bibfnamefont {R.}~\bibnamefont
  {Spiewak}}, \ and\ \bibinfo {author} {\bibfnamefont {J.-B.}\ \bibnamefont
  {Wang}},\ }\bibfield  {title} {\enquote {\bibinfo {title} {{Searching for
  gravitational wave bursts from cosmic string cusps with the Parkes Pulsar
  Timing Array}},}\ }\href {\doibase 10.1093/mnras/staa3721} {\bibfield
  {journal} {\bibinfo  {journal} {Mon. Not. Roy. Astron. Soc.}\ }\textbf
  {\bibinfo {volume} {501}},\ \bibinfo {pages} {701--712} (\bibinfo {year}
  {2021})},\ \Eprint {http://arxiv.org/abs/2011.13490} {arXiv:2011.13490
  [gr-qc]} \BibitemShut {NoStop}%
\bibitem [{\citenamefont {Arzoumanian}\ \emph {et~al.}(2020)\citenamefont
  {Arzoumanian}, \citenamefont {Baker}, \citenamefont {Blumer}, \citenamefont
  {Bécsy}, \citenamefont {Brazier}, \citenamefont {Brook}, \citenamefont
  {Burke-Spolaor}, \citenamefont {Chatterjee}, \citenamefont {Chen},
  \citenamefont {Cordes}, \citenamefont {Cornish}, \citenamefont {Crawford},
  \citenamefont {Cromartie}, \citenamefont {DeCesar}, \citenamefont {Demorest},
  \citenamefont {Dolch}, \citenamefont {Ellis}, \citenamefont {Ferrara},
  \citenamefont {Fiore}, \citenamefont {Fonseca}, \citenamefont
  {Garver-Daniels}, \citenamefont {Gentile}, \citenamefont {Good},
  \citenamefont {Hazboun}, \citenamefont {Holgado}, \citenamefont {Islo},
  \citenamefont {Jennings}, \citenamefont {Jones}, \citenamefont {Kaiser},
  \citenamefont {Kaplan}, \citenamefont {Kelley}, \citenamefont {Key},
  \citenamefont {Laal}, \citenamefont {Lam}, \citenamefont {Lazio},
  \citenamefont {Lorimer}, \citenamefont {Luo}, \citenamefont {Lynch},
  \citenamefont {Madison}, \citenamefont {McLaughlin}, \citenamefont
  {Mingarelli}, \citenamefont {Ng}, \citenamefont {Nice}, \citenamefont
  {Pennucci}, \citenamefont {Pol}, \citenamefont {Ransom}, \citenamefont {Ray},
  \citenamefont {Shapiro-Albert}, \citenamefont {Siemens}, \citenamefont
  {Simon}, \citenamefont {Spiewak}, \citenamefont {Stairs}, \citenamefont
  {Stinebring}, \citenamefont {Stovall}, \citenamefont {Sun}, \citenamefont
  {Swiggum}, \citenamefont {Taylor}, \citenamefont {Turner}, \citenamefont
  {Vallisneri}, \citenamefont {Vigeland},\ and\ \citenamefont
  {Witt}}]{Arzoumanian:2020vkk}%
  \BibitemOpen
  \bibfield  {author} {\bibinfo {author} {\bibfnamefont {Zaven}\ \bibnamefont
  {Arzoumanian}}, \bibinfo {author} {\bibfnamefont {Paul~T.}\ \bibnamefont
  {Baker}}, \bibinfo {author} {\bibfnamefont {Harsha}\ \bibnamefont {Blumer}},
  \bibinfo {author} {\bibfnamefont {Bence}\ \bibnamefont {Bécsy}}, \bibinfo
  {author} {\bibfnamefont {Adam}\ \bibnamefont {Brazier}}, \bibinfo {author}
  {\bibfnamefont {Paul~R.}\ \bibnamefont {Brook}}, \bibinfo {author}
  {\bibfnamefont {Sarah}\ \bibnamefont {Burke-Spolaor}}, \bibinfo {author}
  {\bibfnamefont {Shami}\ \bibnamefont {Chatterjee}}, \bibinfo {author}
  {\bibfnamefont {Siyuan}\ \bibnamefont {Chen}}, \bibinfo {author}
  {\bibfnamefont {James~M.}\ \bibnamefont {Cordes}}, \bibinfo {author}
  {\bibfnamefont {Neil~J.}\ \bibnamefont {Cornish}}, \bibinfo {author}
  {\bibfnamefont {Fronefield}\ \bibnamefont {Crawford}}, \bibinfo {author}
  {\bibfnamefont {H.~Thankful}\ \bibnamefont {Cromartie}}, \bibinfo {author}
  {\bibfnamefont {Megan~E.}\ \bibnamefont {DeCesar}}, \bibinfo {author}
  {\bibfnamefont {Paul~B.}\ \bibnamefont {Demorest}}, \bibinfo {author}
  {\bibfnamefont {Timothy}\ \bibnamefont {Dolch}}, \bibinfo {author}
  {\bibfnamefont {Justin~A.}\ \bibnamefont {Ellis}}, \bibinfo {author}
  {\bibfnamefont {Elizabeth~C.}\ \bibnamefont {Ferrara}}, \bibinfo {author}
  {\bibfnamefont {William}\ \bibnamefont {Fiore}}, \bibinfo {author}
  {\bibfnamefont {Emmanuel}\ \bibnamefont {Fonseca}}, \bibinfo {author}
  {\bibfnamefont {Nathan}\ \bibnamefont {Garver-Daniels}}, \bibinfo {author}
  {\bibfnamefont {Peter~A.}\ \bibnamefont {Gentile}}, \bibinfo {author}
  {\bibfnamefont {Deborah~C.}\ \bibnamefont {Good}}, \bibinfo {author}
  {\bibfnamefont {Jeffrey~S.}\ \bibnamefont {Hazboun}}, \bibinfo {author}
  {\bibfnamefont {A.~Miguel}\ \bibnamefont {Holgado}}, \bibinfo {author}
  {\bibfnamefont {Kristina}\ \bibnamefont {Islo}}, \bibinfo {author}
  {\bibfnamefont {Ross~J.}\ \bibnamefont {Jennings}}, \bibinfo {author}
  {\bibfnamefont {Megan~L.}\ \bibnamefont {Jones}}, \bibinfo {author}
  {\bibfnamefont {Andrew~R.}\ \bibnamefont {Kaiser}}, \bibinfo {author}
  {\bibfnamefont {David~L.}\ \bibnamefont {Kaplan}}, \bibinfo {author}
  {\bibfnamefont {Luke~Zoltan}\ \bibnamefont {Kelley}}, \bibinfo {author}
  {\bibfnamefont {Joey~Shapiro}\ \bibnamefont {Key}}, \bibinfo {author}
  {\bibfnamefont {Nima}\ \bibnamefont {Laal}}, \bibinfo {author} {\bibfnamefont
  {Michael~T.}\ \bibnamefont {Lam}}, \bibinfo {author} {\bibfnamefont
  {T.~Joseph~W.}\ \bibnamefont {Lazio}}, \bibinfo {author} {\bibfnamefont
  {Duncan~R.}\ \bibnamefont {Lorimer}}, \bibinfo {author} {\bibfnamefont
  {Jing}\ \bibnamefont {Luo}}, \bibinfo {author} {\bibfnamefont {Ryan~S.}\
  \bibnamefont {Lynch}}, \bibinfo {author} {\bibfnamefont {Dustin~R.}\
  \bibnamefont {Madison}}, \bibinfo {author} {\bibfnamefont {Maura~A.}\
  \bibnamefont {McLaughlin}}, \bibinfo {author} {\bibfnamefont {Chiara M.~F.}\
  \bibnamefont {Mingarelli}}, \bibinfo {author} {\bibfnamefont {Cherry}\
  \bibnamefont {Ng}}, \bibinfo {author} {\bibfnamefont {David~J.}\ \bibnamefont
  {Nice}}, \bibinfo {author} {\bibfnamefont {Timothy~T.}\ \bibnamefont
  {Pennucci}}, \bibinfo {author} {\bibfnamefont {Nihan~S.}\ \bibnamefont
  {Pol}}, \bibinfo {author} {\bibfnamefont {Scott~M.}\ \bibnamefont {Ransom}},
  \bibinfo {author} {\bibfnamefont {Paul~S.}\ \bibnamefont {Ray}}, \bibinfo
  {author} {\bibfnamefont {Brent~J.}\ \bibnamefont {Shapiro-Albert}}, \bibinfo
  {author} {\bibfnamefont {Xavier}\ \bibnamefont {Siemens}}, \bibinfo {author}
  {\bibfnamefont {Joseph}\ \bibnamefont {Simon}}, \bibinfo {author}
  {\bibfnamefont {Renée}\ \bibnamefont {Spiewak}}, \bibinfo {author}
  {\bibfnamefont {Ingrid~H.}\ \bibnamefont {Stairs}}, \bibinfo {author}
  {\bibfnamefont {Daniel~R.}\ \bibnamefont {Stinebring}}, \bibinfo {author}
  {\bibfnamefont {Kevin}\ \bibnamefont {Stovall}}, \bibinfo {author}
  {\bibfnamefont {Jerry~P.}\ \bibnamefont {Sun}}, \bibinfo {author}
  {\bibfnamefont {Joseph~K.}\ \bibnamefont {Swiggum}}, \bibinfo {author}
  {\bibfnamefont {Stephen~R.}\ \bibnamefont {Taylor}}, \bibinfo {author}
  {\bibfnamefont {Jacob~E.}\ \bibnamefont {Turner}}, \bibinfo {author}
  {\bibfnamefont {Michele}\ \bibnamefont {Vallisneri}}, \bibinfo {author}
  {\bibfnamefont {Sarah~J.}\ \bibnamefont {Vigeland}}, \ and\ \bibinfo {author}
  {\bibfnamefont {Caitlin~A.}\ \bibnamefont {Witt}} (\bibinfo {collaboration}
  {NANOGrav Collaboration}),\ }\bibfield  {title} {\enquote {\bibinfo {title}
  {{The NANOGrav 12.5~yr Data Set: Search for an Isotropic Stochastic
  Gravitational-wave Background}},}\ }\href {\doibase 10.3847/2041-8213/abd401}
  {\bibfield  {journal} {\bibinfo  {journal} {Astrophys. J. Lett.}\ }\textbf
  {\bibinfo {volume} {905}},\ \bibinfo {pages} {L34} (\bibinfo {year}
  {2020})},\ \Eprint {http://arxiv.org/abs/2009.04496} {arXiv:2009.04496
  [astro-ph.HE]} \BibitemShut {NoStop}%
\bibitem [{\citenamefont {Ellis}\ and\ \citenamefont
  {Lewicki}(2021)}]{Ellis:2020ena}%
  \BibitemOpen
  \bibfield  {author} {\bibinfo {author} {\bibfnamefont {John}\ \bibnamefont
  {Ellis}}\ and\ \bibinfo {author} {\bibfnamefont {Marek}\ \bibnamefont
  {Lewicki}},\ }\bibfield  {title} {\enquote {\bibinfo {title} {{Cosmic String
  Interpretation of NANOGrav Pulsar Timing Data}},}\ }\href {\doibase
  10.1103/PhysRevLett.126.041304} {\bibfield  {journal} {\bibinfo  {journal}
  {Phys. Rev. Lett.}\ }\textbf {\bibinfo {volume} {126}},\ \bibinfo {pages}
  {041304} (\bibinfo {year} {2021})},\ \Eprint
  {http://arxiv.org/abs/2009.06555} {arXiv:2009.06555 [astro-ph.CO]}
  \BibitemShut {NoStop}%
\bibitem [{\citenamefont {Blasi}\ \emph {et~al.}(2021)\citenamefont {Blasi},
  \citenamefont {Brdar},\ and\ \citenamefont {Schmitz}}]{Blasi:2020mfx}%
  \BibitemOpen
  \bibfield  {author} {\bibinfo {author} {\bibfnamefont {Simone}\ \bibnamefont
  {Blasi}}, \bibinfo {author} {\bibfnamefont {Vedran}\ \bibnamefont {Brdar}}, \
  and\ \bibinfo {author} {\bibfnamefont {Kai}\ \bibnamefont {Schmitz}},\
  }\bibfield  {title} {\enquote {\bibinfo {title} {{Has NANOGrav found first
  evidence for cosmic strings?}}}\ }\href {\doibase
  10.1103/PhysRevLett.126.041305} {\bibfield  {journal} {\bibinfo  {journal}
  {Phys. Rev. Lett.}\ }\textbf {\bibinfo {volume} {126}},\ \bibinfo {pages}
  {041305} (\bibinfo {year} {2021})},\ \Eprint
  {http://arxiv.org/abs/2009.06607} {arXiv:2009.06607 [astro-ph.CO]}
  \BibitemShut {NoStop}%
\bibitem [{\citenamefont {Buchmuller}\ \emph {et~al.}(2020)\citenamefont
  {Buchmuller}, \citenamefont {Domcke},\ and\ \citenamefont
  {Schmitz}}]{Buchmuller:2020lbh}%
  \BibitemOpen
  \bibfield  {author} {\bibinfo {author} {\bibfnamefont {Wilfried}\
  \bibnamefont {Buchmuller}}, \bibinfo {author} {\bibfnamefont {Valerie}\
  \bibnamefont {Domcke}}, \ and\ \bibinfo {author} {\bibfnamefont {Kai}\
  \bibnamefont {Schmitz}},\ }\bibfield  {title} {\enquote {\bibinfo {title}
  {{From NANOGrav to LIGO with metastable cosmic strings}},}\ }\href {\doibase
  10.1016/j.physletb.2020.135914} {\bibfield  {journal} {\bibinfo  {journal}
  {Phys. Lett. B}\ }\textbf {\bibinfo {volume} {811}},\ \bibinfo {pages}
  {135914} (\bibinfo {year} {2020})},\ \Eprint
  {http://arxiv.org/abs/2009.10649} {arXiv:2009.10649 [astro-ph.CO]}
  \BibitemShut {NoStop}%
\bibitem [{\citenamefont {Bian}\ \emph {et~al.}(2021)\citenamefont {Bian},
  \citenamefont {Cai}, \citenamefont {Liu}, \citenamefont {Yang},\ and\
  \citenamefont {Zhou}}]{Bian:2020urb}%
  \BibitemOpen
  \bibfield  {author} {\bibinfo {author} {\bibfnamefont {Ligong}\ \bibnamefont
  {Bian}}, \bibinfo {author} {\bibfnamefont {Rong-Gen}\ \bibnamefont {Cai}},
  \bibinfo {author} {\bibfnamefont {Jing}\ \bibnamefont {Liu}}, \bibinfo
  {author} {\bibfnamefont {Xing-Yu}\ \bibnamefont {Yang}}, \ and\ \bibinfo
  {author} {\bibfnamefont {Ruiyu}\ \bibnamefont {Zhou}},\ }\bibfield  {title}
  {\enquote {\bibinfo {title} {{Evidence for different gravitational-wave
  sources in the NANOGrav dataset}},}\ }\href {\doibase
  10.1103/PhysRevD.103.L081301} {\bibfield  {journal} {\bibinfo  {journal}
  {Phys. Rev. D}\ }\textbf {\bibinfo {volume} {103}},\ \bibinfo {pages}
  {L081301} (\bibinfo {year} {2021})},\ \Eprint
  {http://arxiv.org/abs/2009.13893} {arXiv:2009.13893 [astro-ph.CO]}
  \BibitemShut {NoStop}%
\bibitem [{\citenamefont {Blanco-Pillado}\ \emph {et~al.}(2021)\citenamefont
  {Blanco-Pillado}, \citenamefont {Olum},\ and\ \citenamefont
  {Wachter}}]{Blanco-Pillado:2021ygr}%
  \BibitemOpen
  \bibfield  {author} {\bibinfo {author} {\bibfnamefont {Jose~J.}\ \bibnamefont
  {Blanco-Pillado}}, \bibinfo {author} {\bibfnamefont {Ken~D.}\ \bibnamefont
  {Olum}}, \ and\ \bibinfo {author} {\bibfnamefont {Jeremy~M.}\ \bibnamefont
  {Wachter}},\ }\bibfield  {title} {\enquote {\bibinfo {title} {{Comparison of
  cosmic string and superstring models to NANOGrav 12.5-year results}},}\
  }\href {\doibase 10.1103/PhysRevD.103.103512} {\bibfield  {journal} {\bibinfo
   {journal} {Phys. Rev. D}\ }\textbf {\bibinfo {volume} {103}},\ \bibinfo
  {pages} {103512} (\bibinfo {year} {2021})},\ \Eprint
  {http://arxiv.org/abs/2102.08194} {arXiv:2102.08194 [astro-ph.CO]}
  \BibitemShut {NoStop}%
\bibitem [{\citenamefont {Aurrekoetxea}\ \emph {et~al.}(2020)\citenamefont
  {Aurrekoetxea}, \citenamefont {Helfer},\ and\ \citenamefont
  {Lim}}]{Aurrekoetxea:2020tuw}%
  \BibitemOpen
  \bibfield  {author} {\bibinfo {author} {\bibfnamefont {Josu~C.}\ \bibnamefont
  {Aurrekoetxea}}, \bibinfo {author} {\bibfnamefont {Thomas}\ \bibnamefont
  {Helfer}}, \ and\ \bibinfo {author} {\bibfnamefont {Eugene~A.}\ \bibnamefont
  {Lim}},\ }\bibfield  {title} {\enquote {\bibinfo {title} {{Coherent
  Gravitational Waveforms and Memory from Cosmic String Loops}},}\ }\href
  {\doibase 10.1088/1361-6382/aba28b} {\bibfield  {journal} {\bibinfo
  {journal} {Class. Quant. Grav.}\ }\textbf {\bibinfo {volume} {37}},\ \bibinfo
  {pages} {204001} (\bibinfo {year} {2020})},\ \Eprint
  {http://arxiv.org/abs/2002.05177} {arXiv:2002.05177 [gr-qc]} \BibitemShut
  {NoStop}%
\bibitem [{\citenamefont {McNeill}\ \emph {et~al.}(2017)\citenamefont
  {McNeill}, \citenamefont {Thrane},\ and\ \citenamefont
  {Lasky}}]{McNeill:2017uvq}%
  \BibitemOpen
  \bibfield  {author} {\bibinfo {author} {\bibfnamefont {Lucy~O.}\ \bibnamefont
  {McNeill}}, \bibinfo {author} {\bibfnamefont {Eric}\ \bibnamefont {Thrane}},
  \ and\ \bibinfo {author} {\bibfnamefont {Paul~D.}\ \bibnamefont {Lasky}},\
  }\bibfield  {title} {\enquote {\bibinfo {title} {{Detecting Gravitational
  Wave Memory without Parent Signals}},}\ }\href {\doibase
  10.1103/PhysRevLett.118.181103} {\bibfield  {journal} {\bibinfo  {journal}
  {Phys. Rev. Lett.}\ }\textbf {\bibinfo {volume} {118}},\ \bibinfo {pages}
  {181103} (\bibinfo {year} {2017})},\ \Eprint
  {http://arxiv.org/abs/1702.01759} {arXiv:1702.01759 [astro-ph.IM]}
  \BibitemShut {NoStop}%
\bibitem [{\citenamefont {Jenkins}\ and\ \citenamefont
  {Sakellariadou}(2020)}]{Jenkins:2020ctp}%
  \BibitemOpen
  \bibfield  {author} {\bibinfo {author} {\bibfnamefont {Alexander~C.}\
  \bibnamefont {Jenkins}}\ and\ \bibinfo {author} {\bibfnamefont {Mairi}\
  \bibnamefont {Sakellariadou}},\ }\bibfield  {title} {\enquote {\bibinfo
  {title} {{Primordial black holes from cusp collapse on cosmic strings}},}\
  }\href@noop {} {\  (\bibinfo {year} {2020})},\ \Eprint
  {http://arxiv.org/abs/2006.16249} {arXiv:2006.16249 [astro-ph.CO]}
  \BibitemShut {NoStop}%
\bibitem [{\citenamefont {Maggiore}(2007)}]{Maggiore:1900zz}%
  \BibitemOpen
  \bibfield  {author} {\bibinfo {author} {\bibfnamefont {Michele}\ \bibnamefont
  {Maggiore}},\ }\href {\doibase 10.1093/acprof:oso/9780198570745.001.0001}
  {\emph {\bibinfo {title} {{Gravitational Waves. Vol. 1: Theory and
  Experiments}}}},\ Oxford Master Series in Physics\ (\bibinfo  {publisher}
  {Oxford University Press},\ \bibinfo {year} {2007})\BibitemShut {NoStop}%
\bibitem [{\citenamefont {Isaacson}(1968{\natexlab{a}})}]{Isaacson:1967zz}%
  \BibitemOpen
  \bibfield  {author} {\bibinfo {author} {\bibfnamefont {Richard~A.}\
  \bibnamefont {Isaacson}},\ }\bibfield  {title} {\enquote {\bibinfo {title}
  {{Gravitational Radiation in the Limit of High Frequency. I. The Linear
  Approximation and Geometrical Optics}},}\ }\href {\doibase
  10.1103/PhysRev.166.1263} {\bibfield  {journal} {\bibinfo  {journal} {Phys.
  Rev.}\ }\textbf {\bibinfo {volume} {166}},\ \bibinfo {pages} {1263--1271}
  (\bibinfo {year} {1968}{\natexlab{a}})}\BibitemShut {NoStop}%
\bibitem [{\citenamefont {Isaacson}(1968{\natexlab{b}})}]{Isaacson:1968zza}%
  \BibitemOpen
  \bibfield  {author} {\bibinfo {author} {\bibfnamefont {Richard~A.}\
  \bibnamefont {Isaacson}},\ }\bibfield  {title} {\enquote {\bibinfo {title}
  {{Gravitational Radiation in the Limit of High Frequency. II. Nonlinear Terms
  and the Ef fective Stress Tensor}},}\ }\href {\doibase
  10.1103/PhysRev.166.1272} {\bibfield  {journal} {\bibinfo  {journal} {Phys.
  Rev.}\ }\textbf {\bibinfo {volume} {166}},\ \bibinfo {pages} {1272--1279}
  (\bibinfo {year} {1968}{\natexlab{b}})}\BibitemShut {NoStop}%
\bibitem [{\citenamefont {Saulson}(1995)}]{Saulson:1995zi}%
  \BibitemOpen
  \bibfield  {author} {\bibinfo {author} {\bibfnamefont {Peter~R.}\
  \bibnamefont {Saulson}},\ }\href
  {https://www.worldscientific.com/doi/abs/10.1142/10116} {\emph {\bibinfo
  {title} {{Fundamentals of interferometric gravitational wave detectors}}}}\
  (\bibinfo  {publisher} {World Scientific},\ \bibinfo {year}
  {1995})\BibitemShut {NoStop}%
\bibitem [{\citenamefont {Ringeval}\ and\ \citenamefont
  {Suyama}(2017)}]{Ringeval:2017eww}%
  \BibitemOpen
  \bibfield  {author} {\bibinfo {author} {\bibfnamefont {Christophe}\
  \bibnamefont {Ringeval}}\ and\ \bibinfo {author} {\bibfnamefont {Teruaki}\
  \bibnamefont {Suyama}},\ }\bibfield  {title} {\enquote {\bibinfo {title}
  {{Stochastic gravitational waves from cosmic string loops in scaling}},}\
  }\href {\doibase 10.1088/1475-7516/2017/12/027} {\bibfield  {journal}
  {\bibinfo  {journal} {JCAP}\ }\textbf {\bibinfo {volume} {12}},\ \bibinfo
  {pages} {027} (\bibinfo {year} {2017})},\ \Eprint
  {http://arxiv.org/abs/1709.03845} {arXiv:1709.03845 [astro-ph.CO]}
  \BibitemShut {NoStop}%
\bibitem [{\citenamefont {Srednicki}\ and\ \citenamefont
  {Theisen}(1987)}]{Srednicki:1986xg}%
  \BibitemOpen
  \bibfield  {author} {\bibinfo {author} {\bibfnamefont {Mark}\ \bibnamefont
  {Srednicki}}\ and\ \bibinfo {author} {\bibfnamefont {Stefan}\ \bibnamefont
  {Theisen}},\ }\bibfield  {title} {\enquote {\bibinfo {title}
  {{Nongravitational Decay of Cosmic Strings}},}\ }\href {\doibase
  10.1016/0370-2693(87)90648-4} {\bibfield  {journal} {\bibinfo  {journal}
  {Phys. Lett. B}\ }\textbf {\bibinfo {volume} {189}},\ \bibinfo {pages} {397}
  (\bibinfo {year} {1987})}\BibitemShut {NoStop}%
\bibitem [{\citenamefont {Helfer}\ \emph {et~al.}(2019)\citenamefont {Helfer},
  \citenamefont {Aurrekoetxea},\ and\ \citenamefont {Lim}}]{Helfer:2018qgv}%
  \BibitemOpen
  \bibfield  {author} {\bibinfo {author} {\bibfnamefont {Thomas}\ \bibnamefont
  {Helfer}}, \bibinfo {author} {\bibfnamefont {Josu~C.}\ \bibnamefont
  {Aurrekoetxea}}, \ and\ \bibinfo {author} {\bibfnamefont {Eugene~A.}\
  \bibnamefont {Lim}},\ }\bibfield  {title} {\enquote {\bibinfo {title}
  {{Cosmic String Loop Collapse in Full General Relativity}},}\ }\href
  {\doibase 10.1103/PhysRevD.99.104028} {\bibfield  {journal} {\bibinfo
  {journal} {Phys. Rev. D}\ }\textbf {\bibinfo {volume} {99}},\ \bibinfo
  {pages} {104028} (\bibinfo {year} {2019})},\ \Eprint
  {http://arxiv.org/abs/1808.06678} {arXiv:1808.06678 [gr-qc]} \BibitemShut
  {NoStop}%
\bibitem [{\citenamefont {Wagoner}(1979)}]{Wagoner:1979dd}%
  \BibitemOpen
  \bibfield  {author} {\bibinfo {author} {\bibfnamefont {Robert~V.}\
  \bibnamefont {Wagoner}},\ }\bibfield  {title} {\enquote {\bibinfo {title}
  {{Low Frequency Gravitational Radtiation From Collapsing Systems}},}\ }\href
  {\doibase 10.1103/PhysRevD.19.2897} {\bibfield  {journal} {\bibinfo
  {journal} {Phys. Rev. D}\ }\textbf {\bibinfo {volume} {19}},\ \bibinfo
  {pages} {2897--2901} (\bibinfo {year} {1979})}\BibitemShut {NoStop}%
\bibitem [{\citenamefont {Thompson}(1988)}]{Thompson:1988yj}%
  \BibitemOpen
  \bibfield  {author} {\bibinfo {author} {\bibfnamefont {Arthur~Christopher}\
  \bibnamefont {Thompson}},\ }\bibfield  {title} {\enquote {\bibinfo {title}
  {{Dynamics of Cosmic String}},}\ }\href {\doibase 10.1103/PhysRevD.37.283}
  {\bibfield  {journal} {\bibinfo  {journal} {Phys. Rev. D}\ }\textbf {\bibinfo
  {volume} {37}},\ \bibinfo {pages} {283--297} (\bibinfo {year}
  {1988})}\BibitemShut {NoStop}%
\bibitem [{\citenamefont {Quashnock}\ and\ \citenamefont
  {Spergel}(1990)}]{Quashnock:1990wv}%
  \BibitemOpen
  \bibfield  {author} {\bibinfo {author} {\bibfnamefont {Jean~M.}\ \bibnamefont
  {Quashnock}}\ and\ \bibinfo {author} {\bibfnamefont {David~N.}\ \bibnamefont
  {Spergel}},\ }\bibfield  {title} {\enquote {\bibinfo {title} {{Gravitational
  Selfinteractions of Cosmic Strings}},}\ }\href {\doibase
  10.1103/PhysRevD.42.2505} {\bibfield  {journal} {\bibinfo  {journal} {Phys.
  Rev. D}\ }\textbf {\bibinfo {volume} {42}},\ \bibinfo {pages} {2505--2520}
  (\bibinfo {year} {1990})}\BibitemShut {NoStop}%
\bibitem [{\citenamefont {Copeland}\ \emph {et~al.}(1990)\citenamefont
  {Copeland}, \citenamefont {Haws},\ and\ \citenamefont
  {Hindmarsh}}]{Copeland:1990qu}%
  \BibitemOpen
  \bibfield  {author} {\bibinfo {author} {\bibfnamefont {Edmund~J.}\
  \bibnamefont {Copeland}}, \bibinfo {author} {\bibfnamefont {D.}~\bibnamefont
  {Haws}}, \ and\ \bibinfo {author} {\bibfnamefont {M.}~\bibnamefont
  {Hindmarsh}},\ }\bibfield  {title} {\enquote {\bibinfo {title} {{Classical
  theory of radiating strings}},}\ }\href {\doibase 10.1103/PhysRevD.42.726}
  {\bibfield  {journal} {\bibinfo  {journal} {Phys. Rev. D}\ }\textbf {\bibinfo
  {volume} {42}},\ \bibinfo {pages} {726--730} (\bibinfo {year}
  {1990})}\BibitemShut {NoStop}%
\bibitem [{\citenamefont {Battye}\ and\ \citenamefont
  {Shellard}(1995)}]{Battye:1994qa}%
  \BibitemOpen
  \bibfield  {author} {\bibinfo {author} {\bibfnamefont {R.~A.}\ \bibnamefont
  {Battye}}\ and\ \bibinfo {author} {\bibfnamefont {E.~P.~S.}\ \bibnamefont
  {Shellard}},\ }\bibfield  {title} {\enquote {\bibinfo {title} {{String
  radiative back reaction}},}\ }\href {\doibase 10.1103/PhysRevLett.75.4354}
  {\bibfield  {journal} {\bibinfo  {journal} {Phys. Rev. Lett.}\ }\textbf
  {\bibinfo {volume} {75}},\ \bibinfo {pages} {4354--4357} (\bibinfo {year}
  {1995})},\ \Eprint {http://arxiv.org/abs/astro-ph/9408078}
  {arXiv:astro-ph/9408078} \BibitemShut {NoStop}%
\bibitem [{\citenamefont {Buonanno}\ and\ \citenamefont
  {Damour}(1999)}]{Buonanno:1998is}%
  \BibitemOpen
  \bibfield  {author} {\bibinfo {author} {\bibfnamefont {Alessandra}\
  \bibnamefont {Buonanno}}\ and\ \bibinfo {author} {\bibfnamefont {Thibault}\
  \bibnamefont {Damour}},\ }\bibfield  {title} {\enquote {\bibinfo {title}
  {{Gravitational, dilatonic and axionic radiative damping of cosmic
  strings}},}\ }\href {\doibase 10.1103/PhysRevD.60.023517} {\bibfield
  {journal} {\bibinfo  {journal} {Phys. Rev. D}\ }\textbf {\bibinfo {volume}
  {60}},\ \bibinfo {pages} {023517} (\bibinfo {year} {1999})},\ \Eprint
  {http://arxiv.org/abs/gr-qc/9801105} {arXiv:gr-qc/9801105} \BibitemShut
  {NoStop}%
\bibitem [{\citenamefont {Carter}\ and\ \citenamefont
  {Battye}(1998)}]{Carter:1998ix}%
  \BibitemOpen
  \bibfield  {author} {\bibinfo {author} {\bibfnamefont {Brandon}\ \bibnamefont
  {Carter}}\ and\ \bibinfo {author} {\bibfnamefont {Richard~A.}\ \bibnamefont
  {Battye}},\ }\bibfield  {title} {\enquote {\bibinfo {title} {{Nondivergence
  of gravitational selfinteractions for Goto-Nambu strings}},}\ }\href
  {\doibase 10.1016/S0370-2693(98)00496-1} {\bibfield  {journal} {\bibinfo
  {journal} {Phys. Lett. B}\ }\textbf {\bibinfo {volume} {430}},\ \bibinfo
  {pages} {49--53} (\bibinfo {year} {1998})},\ \Eprint
  {http://arxiv.org/abs/hep-th/9803012} {arXiv:hep-th/9803012} \BibitemShut
  {NoStop}%
\bibitem [{\citenamefont {Wachter}\ and\ \citenamefont
  {Olum}(2017{\natexlab{a}})}]{Wachter:2016hgi}%
  \BibitemOpen
  \bibfield  {author} {\bibinfo {author} {\bibfnamefont {Jeremy~M.}\
  \bibnamefont {Wachter}}\ and\ \bibinfo {author} {\bibfnamefont {Ken~D.}\
  \bibnamefont {Olum}},\ }\bibfield  {title} {\enquote {\bibinfo {title}
  {{Gravitational smoothing of kinks on cosmic string loops}},}\ }\href
  {\doibase 10.1103/PhysRevLett.118.051301} {\bibfield  {journal} {\bibinfo
  {journal} {Phys. Rev. Lett.}\ }\textbf {\bibinfo {volume} {118}},\ \bibinfo
  {pages} {051301} (\bibinfo {year} {2017}{\natexlab{a}})},\ \bibinfo {note}
  {[Erratum: Phys.Rev.Lett. 121, 149901 (2018)]},\ \Eprint
  {http://arxiv.org/abs/1609.01153} {arXiv:1609.01153 [gr-qc]} \BibitemShut
  {NoStop}%
\bibitem [{\citenamefont {Wachter}\ and\ \citenamefont
  {Olum}(2017{\natexlab{b}})}]{Wachter:2016rwc}%
  \BibitemOpen
  \bibfield  {author} {\bibinfo {author} {\bibfnamefont {Jeremy~M.}\
  \bibnamefont {Wachter}}\ and\ \bibinfo {author} {\bibfnamefont {Ken~D.}\
  \bibnamefont {Olum}},\ }\bibfield  {title} {\enquote {\bibinfo {title}
  {{Gravitational backreaction on piecewise linear cosmic string loops}},}\
  }\href {\doibase 10.1103/PhysRevD.95.023519} {\bibfield  {journal} {\bibinfo
  {journal} {Phys. Rev. D}\ }\textbf {\bibinfo {volume} {95}},\ \bibinfo
  {pages} {023519} (\bibinfo {year} {2017}{\natexlab{b}})},\ \Eprint
  {http://arxiv.org/abs/1609.01685} {arXiv:1609.01685 [gr-qc]} \BibitemShut
  {NoStop}%
\bibitem [{\citenamefont {Blanco-Pillado}\ \emph
  {et~al.}(2018{\natexlab{b}})\citenamefont {Blanco-Pillado}, \citenamefont
  {Olum},\ and\ \citenamefont {Wachter}}]{Blanco-Pillado:2018ael}%
  \BibitemOpen
  \bibfield  {author} {\bibinfo {author} {\bibfnamefont {Jose~J.}\ \bibnamefont
  {Blanco-Pillado}}, \bibinfo {author} {\bibfnamefont {Ken~D.}\ \bibnamefont
  {Olum}}, \ and\ \bibinfo {author} {\bibfnamefont {Jeremy~M.}\ \bibnamefont
  {Wachter}},\ }\bibfield  {title} {\enquote {\bibinfo {title} {{Gravitational
  backreaction near cosmic string kinks and cusps}},}\ }\href {\doibase
  10.1103/PhysRevD.98.123507} {\bibfield  {journal} {\bibinfo  {journal} {Phys.
  Rev. D}\ }\textbf {\bibinfo {volume} {98}},\ \bibinfo {pages} {123507}
  (\bibinfo {year} {2018}{\natexlab{b}})},\ \Eprint
  {http://arxiv.org/abs/1808.08254} {arXiv:1808.08254 [gr-qc]} \BibitemShut
  {NoStop}%
\bibitem [{\citenamefont {Chernoff}\ \emph {et~al.}(2019)\citenamefont
  {Chernoff}, \citenamefont {Flanagan},\ and\ \citenamefont
  {Wardell}}]{Chernoff:2018evo}%
  \BibitemOpen
  \bibfield  {author} {\bibinfo {author} {\bibfnamefont {David~F.}\
  \bibnamefont {Chernoff}}, \bibinfo {author} {\bibfnamefont
  {{\'E}anna~{\'E}.}\ \bibnamefont {Flanagan}}, \ and\ \bibinfo {author}
  {\bibfnamefont {Barry}\ \bibnamefont {Wardell}},\ }\bibfield  {title}
  {\enquote {\bibinfo {title} {{Gravitational backreaction on a cosmic string:
  Formalism}},}\ }\href {\doibase 10.1103/PhysRevD.99.084036} {\bibfield
  {journal} {\bibinfo  {journal} {Phys. Rev. D}\ }\textbf {\bibinfo {volume}
  {99}},\ \bibinfo {pages} {084036} (\bibinfo {year} {2019})},\ \Eprint
  {http://arxiv.org/abs/1808.08631} {arXiv:1808.08631 [gr-qc]} \BibitemShut
  {NoStop}%
\bibitem [{\citenamefont {Blanco-Pillado}\ \emph {et~al.}(2019)\citenamefont
  {Blanco-Pillado}, \citenamefont {Olum},\ and\ \citenamefont
  {Wachter}}]{Blanco-Pillado:2019nto}%
  \BibitemOpen
  \bibfield  {author} {\bibinfo {author} {\bibfnamefont {Jose~J.}\ \bibnamefont
  {Blanco-Pillado}}, \bibinfo {author} {\bibfnamefont {Ken~D.}\ \bibnamefont
  {Olum}}, \ and\ \bibinfo {author} {\bibfnamefont {Jeremy~M.}\ \bibnamefont
  {Wachter}},\ }\bibfield  {title} {\enquote {\bibinfo {title} {{Gravitational
  backreaction simulations of simple cosmic string loops}},}\ }\href {\doibase
  10.1103/PhysRevD.100.023535} {\bibfield  {journal} {\bibinfo  {journal}
  {Phys. Rev. D}\ }\textbf {\bibinfo {volume} {100}},\ \bibinfo {pages}
  {023535} (\bibinfo {year} {2019})},\ \Eprint
  {http://arxiv.org/abs/1903.06079} {arXiv:1903.06079 [gr-qc]} \BibitemShut
  {NoStop}%
\bibitem [{\citenamefont {Blanco-Pillado}\ \emph {et~al.}(2014)\citenamefont
  {Blanco-Pillado}, \citenamefont {Olum},\ and\ \citenamefont
  {Shlaer}}]{Blanco-Pillado:2013qja}%
  \BibitemOpen
  \bibfield  {author} {\bibinfo {author} {\bibfnamefont {Jose~J.}\ \bibnamefont
  {Blanco-Pillado}}, \bibinfo {author} {\bibfnamefont {Ken~D.}\ \bibnamefont
  {Olum}}, \ and\ \bibinfo {author} {\bibfnamefont {Benjamin}\ \bibnamefont
  {Shlaer}},\ }\bibfield  {title} {\enquote {\bibinfo {title} {{The number of
  cosmic string loops}},}\ }\href {\doibase 10.1103/PhysRevD.89.023512}
  {\bibfield  {journal} {\bibinfo  {journal} {Phys. Rev. D}\ }\textbf {\bibinfo
  {volume} {89}},\ \bibinfo {pages} {023512} (\bibinfo {year} {2014})},\
  \Eprint {http://arxiv.org/abs/1309.6637} {arXiv:1309.6637 [astro-ph.CO]}
  \BibitemShut {NoStop}%
\bibitem [{\citenamefont {Thrane}\ and\ \citenamefont
  {Romano}(2013{\natexlab{a}})}]{Thrane:2013oya}%
  \BibitemOpen
  \bibfield  {author} {\bibinfo {author} {\bibfnamefont {Eric}\ \bibnamefont
  {Thrane}}\ and\ \bibinfo {author} {\bibfnamefont {Joseph~D.}\ \bibnamefont
  {Romano}},\ }\bibfield  {title} {\enquote {\bibinfo {title} {{Sensitivity
  curves for searches for gravitational-wave backgrounds}},}\ }\href {\doibase
  10.1103/PhysRevD.88.124032} {\bibfield  {journal} {\bibinfo  {journal} {Phys.
  Rev. D}\ }\textbf {\bibinfo {volume} {88}},\ \bibinfo {pages} {124032}
  (\bibinfo {year} {2013}{\natexlab{a}})},\ \Eprint
  {http://arxiv.org/abs/1310.5300} {arXiv:1310.5300 [astro-ph.IM]} \BibitemShut
  {NoStop}%
\bibitem [{\citenamefont {Abbott}\ \emph
  {et~al.}(2021{\natexlab{e}})\citenamefont {Abbott} \emph
  {et~al.}}]{LIGOScientific:2021PI}%
  \BibitemOpen
  \bibfield  {author} {\bibinfo {author} {\bibfnamefont {R.}~\bibnamefont
  {Abbott}} \emph {et~al.} (\bibinfo {collaboration} {LIGO Scientific
  Collaboration, Virgo Collaboration, KAGRA Collaboration}),\ }\href@noop {}
  {\enquote {\bibinfo {title} {{Data for `Upper Limits on the Isotropic
  Gravitational-Wave Background from Advanced LIGO's and Advanced Virgo's Third
  Observing Run'}},}\ }\bibinfo {howpublished}
  {\url{https://dcc.ligo.org/G2001287/public}} (\bibinfo {year}
  {2021}{\natexlab{e}})\BibitemShut {NoStop}%
\bibitem [{\citenamefont {Shannon}\ \emph {et~al.}(2015)\citenamefont
  {Shannon}, \citenamefont {Ravi}, \citenamefont {Lentati}, \citenamefont
  {Lasky}, \citenamefont {Hobbs}, \citenamefont {Kerr}, \citenamefont
  {Manchester}, \citenamefont {Coles}, \citenamefont {Levin}, \citenamefont
  {Bailes}, \citenamefont {Bhat}, \citenamefont {Burke-Spolaor}, \citenamefont
  {Dai}, \citenamefont {Keith}, \citenamefont {Os{\l}owski}, \citenamefont
  {Reardon}, \citenamefont {van Straten}, \citenamefont {Toomey}, \citenamefont
  {Wang}, \citenamefont {Wen}, \citenamefont {Wyithe},\ and\ \citenamefont
  {Zhu}}]{Shannon:2015ect}%
  \BibitemOpen
  \bibfield  {author} {\bibinfo {author} {\bibfnamefont {R.~M.}\ \bibnamefont
  {Shannon}}, \bibinfo {author} {\bibfnamefont {V.}~\bibnamefont {Ravi}},
  \bibinfo {author} {\bibfnamefont {L.~T.}\ \bibnamefont {Lentati}}, \bibinfo
  {author} {\bibfnamefont {P.~D.}\ \bibnamefont {Lasky}}, \bibinfo {author}
  {\bibfnamefont {G.}~\bibnamefont {Hobbs}}, \bibinfo {author} {\bibfnamefont
  {M.}~\bibnamefont {Kerr}}, \bibinfo {author} {\bibfnamefont {R.~N.}\
  \bibnamefont {Manchester}}, \bibinfo {author} {\bibfnamefont {W.~A.}\
  \bibnamefont {Coles}}, \bibinfo {author} {\bibfnamefont {Y.}~\bibnamefont
  {Levin}}, \bibinfo {author} {\bibfnamefont {M.}~\bibnamefont {Bailes}},
  \bibinfo {author} {\bibfnamefont {N.~D.~R.}\ \bibnamefont {Bhat}}, \bibinfo
  {author} {\bibfnamefont {S.}~\bibnamefont {Burke-Spolaor}}, \bibinfo {author}
  {\bibfnamefont {S.}~\bibnamefont {Dai}}, \bibinfo {author} {\bibfnamefont
  {M.~J.}\ \bibnamefont {Keith}}, \bibinfo {author} {\bibfnamefont
  {S.}~\bibnamefont {Os{\l}owski}}, \bibinfo {author} {\bibfnamefont {D.~J.}\
  \bibnamefont {Reardon}}, \bibinfo {author} {\bibfnamefont {W.}~\bibnamefont
  {van Straten}}, \bibinfo {author} {\bibfnamefont {L.}~\bibnamefont {Toomey}},
  \bibinfo {author} {\bibfnamefont {J.-B.}\ \bibnamefont {Wang}}, \bibinfo
  {author} {\bibfnamefont {L.}~\bibnamefont {Wen}}, \bibinfo {author}
  {\bibfnamefont {J.~S.~B.}\ \bibnamefont {Wyithe}}, \ and\ \bibinfo {author}
  {\bibfnamefont {X.-J.}\ \bibnamefont {Zhu}},\ }\bibfield  {title} {\enquote
  {\bibinfo {title} {{Gravitational waves from binary supermassive black holes
  missing in pulsar observations}},}\ }\href {\doibase 10.1126/science.aab1910}
  {\bibfield  {journal} {\bibinfo  {journal} {Science}\ }\textbf {\bibinfo
  {volume} {349}},\ \bibinfo {pages} {1522--1525} (\bibinfo {year} {2015})},\
  \Eprint {http://arxiv.org/abs/1509.07320} {arXiv:1509.07320 [astro-ph.CO]}
  \BibitemShut {NoStop}%
\bibitem [{\citenamefont {Verbiest}\ \emph {et~al.}(2016)\citenamefont
  {Verbiest}, \citenamefont {Lentati}, \citenamefont {Hobbs}, \citenamefont
  {van Haasteren}, \citenamefont {Demorest}, \citenamefont {Janssen},
  \citenamefont {Wang}, \citenamefont {Desvignes}, \citenamefont {Caballero},
  \citenamefont {Keith}, \citenamefont {Champion}, \citenamefont {Arzoumanian},
  \citenamefont {Babak}, \citenamefont {Bassa}, \citenamefont {Bhat},
  \citenamefont {Brazier}, \citenamefont {Brem}, \citenamefont {Burgay},
  \citenamefont {Burke-Spolaor}, \citenamefont {Chamberlin}, \citenamefont
  {Chatterjee}, \citenamefont {Christy}, \citenamefont {Cognard}, \citenamefont
  {Cordes}, \citenamefont {Dai}, \citenamefont {Dolch}, \citenamefont {Ellis},
  \citenamefont {Ferdman}, \citenamefont {Fonseca}, \citenamefont {Gair},
  \citenamefont {Garver-Daniels}, \citenamefont {Gentile}, \citenamefont
  {Gonzalez}, \citenamefont {Graikou}, \citenamefont {Guillemot}, \citenamefont
  {Hessels}, \citenamefont {Jones}, \citenamefont {Karuppusamy}, \citenamefont
  {Kerr}, \citenamefont {Kramer}, \citenamefont {Lam}, \citenamefont {Lasky},
  \citenamefont {Lassus}, \citenamefont {Lazarus}, \citenamefont {Lazio},
  \citenamefont {Lee}, \citenamefont {Levin}, \citenamefont {Liu},
  \citenamefont {Lynch}, \citenamefont {Lyne}, \citenamefont {Mckee},
  \citenamefont {McLaughlin}, \citenamefont {McWilliams}, \citenamefont
  {Madison}, \citenamefont {Manchester}, \citenamefont {Mingarelli},
  \citenamefont {Nice}, \citenamefont {Osłowski}, \citenamefont {Palliyaguru},
  \citenamefont {Pennucci}, \citenamefont {Perera}, \citenamefont {Perrodin},
  \citenamefont {Possenti}, \citenamefont {Petiteau}, \citenamefont {Ransom},
  \citenamefont {Reardon}, \citenamefont {Rosado}, \citenamefont {Sanidas},
  \citenamefont {Sesana}, \citenamefont {Shaifullah}, \citenamefont {Shannon},
  \citenamefont {Siemens}, \citenamefont {Simon}, \citenamefont {Smits},
  \citenamefont {Spiewak}, \citenamefont {Stairs}, \citenamefont {Stappers},
  \citenamefont {Stinebring}, \citenamefont {Stovall}, \citenamefont {Swiggum},
  \citenamefont {Taylor}, \citenamefont {Theureau}, \citenamefont {Tiburzi},
  \citenamefont {Toomey}, \citenamefont {Vallisneri}, \citenamefont {van
  Straten}, \citenamefont {Vecchio}, \citenamefont {Wang}, \citenamefont {Wen},
  \citenamefont {You}, \citenamefont {Zhu},\ and\ \citenamefont
  {Zhu}}]{Verbiest:2016vem}%
  \BibitemOpen
  \bibfield  {author} {\bibinfo {author} {\bibfnamefont {J.~P.~W.}\
  \bibnamefont {Verbiest}}, \bibinfo {author} {\bibfnamefont {L.}~\bibnamefont
  {Lentati}}, \bibinfo {author} {\bibfnamefont {G.}~\bibnamefont {Hobbs}},
  \bibinfo {author} {\bibfnamefont {R.}~\bibnamefont {van Haasteren}}, \bibinfo
  {author} {\bibfnamefont {P.~B.}\ \bibnamefont {Demorest}}, \bibinfo {author}
  {\bibfnamefont {G.~H.}\ \bibnamefont {Janssen}}, \bibinfo {author}
  {\bibfnamefont {J.-B.}\ \bibnamefont {Wang}}, \bibinfo {author}
  {\bibfnamefont {G.}~\bibnamefont {Desvignes}}, \bibinfo {author}
  {\bibfnamefont {R.~N.}\ \bibnamefont {Caballero}}, \bibinfo {author}
  {\bibfnamefont {M.~J.}\ \bibnamefont {Keith}}, \bibinfo {author}
  {\bibfnamefont {D.~J.}\ \bibnamefont {Champion}}, \bibinfo {author}
  {\bibfnamefont {Z.}~\bibnamefont {Arzoumanian}}, \bibinfo {author}
  {\bibfnamefont {S.}~\bibnamefont {Babak}}, \bibinfo {author} {\bibfnamefont
  {C.~G.}\ \bibnamefont {Bassa}}, \bibinfo {author} {\bibfnamefont {N.~D.~R.}\
  \bibnamefont {Bhat}}, \bibinfo {author} {\bibfnamefont {A.}~\bibnamefont
  {Brazier}}, \bibinfo {author} {\bibfnamefont {P.}~\bibnamefont {Brem}},
  \bibinfo {author} {\bibfnamefont {M.}~\bibnamefont {Burgay}}, \bibinfo
  {author} {\bibfnamefont {S.}~\bibnamefont {Burke-Spolaor}}, \bibinfo {author}
  {\bibfnamefont {S.~J.}\ \bibnamefont {Chamberlin}}, \bibinfo {author}
  {\bibfnamefont {S.}~\bibnamefont {Chatterjee}}, \bibinfo {author}
  {\bibfnamefont {B.}~\bibnamefont {Christy}}, \bibinfo {author} {\bibfnamefont
  {I.}~\bibnamefont {Cognard}}, \bibinfo {author} {\bibfnamefont {J.~M.}\
  \bibnamefont {Cordes}}, \bibinfo {author} {\bibfnamefont {S.}~\bibnamefont
  {Dai}}, \bibinfo {author} {\bibfnamefont {T.}~\bibnamefont {Dolch}}, \bibinfo
  {author} {\bibfnamefont {J.~A.}\ \bibnamefont {Ellis}}, \bibinfo {author}
  {\bibfnamefont {R.~D.}\ \bibnamefont {Ferdman}}, \bibinfo {author}
  {\bibfnamefont {E.}~\bibnamefont {Fonseca}}, \bibinfo {author} {\bibfnamefont
  {J.~R.}\ \bibnamefont {Gair}}, \bibinfo {author} {\bibfnamefont {N.~E.}\
  \bibnamefont {Garver-Daniels}}, \bibinfo {author} {\bibfnamefont
  {P.}~\bibnamefont {Gentile}}, \bibinfo {author} {\bibfnamefont {M.~E.}\
  \bibnamefont {Gonzalez}}, \bibinfo {author} {\bibfnamefont {E.}~\bibnamefont
  {Graikou}}, \bibinfo {author} {\bibfnamefont {L.}~\bibnamefont {Guillemot}},
  \bibinfo {author} {\bibfnamefont {J.~W.~T.}\ \bibnamefont {Hessels}},
  \bibinfo {author} {\bibfnamefont {G.}~\bibnamefont {Jones}}, \bibinfo
  {author} {\bibfnamefont {R.}~\bibnamefont {Karuppusamy}}, \bibinfo {author}
  {\bibfnamefont {M.}~\bibnamefont {Kerr}}, \bibinfo {author} {\bibfnamefont
  {M.}~\bibnamefont {Kramer}}, \bibinfo {author} {\bibfnamefont {M.~T.}\
  \bibnamefont {Lam}}, \bibinfo {author} {\bibfnamefont {P.~D.}\ \bibnamefont
  {Lasky}}, \bibinfo {author} {\bibfnamefont {A.}~\bibnamefont {Lassus}},
  \bibinfo {author} {\bibfnamefont {P.}~\bibnamefont {Lazarus}}, \bibinfo
  {author} {\bibfnamefont {T.~J.~W.}\ \bibnamefont {Lazio}}, \bibinfo {author}
  {\bibfnamefont {K.~J.}\ \bibnamefont {Lee}}, \bibinfo {author} {\bibfnamefont
  {L.}~\bibnamefont {Levin}}, \bibinfo {author} {\bibfnamefont
  {K.}~\bibnamefont {Liu}}, \bibinfo {author} {\bibfnamefont {R.~S.}\
  \bibnamefont {Lynch}}, \bibinfo {author} {\bibfnamefont {A.~G.}\ \bibnamefont
  {Lyne}}, \bibinfo {author} {\bibfnamefont {J.}~\bibnamefont {Mckee}},
  \bibinfo {author} {\bibfnamefont {M.~A.}\ \bibnamefont {McLaughlin}},
  \bibinfo {author} {\bibfnamefont {S.~T.}\ \bibnamefont {McWilliams}},
  \bibinfo {author} {\bibfnamefont {D.~R.}\ \bibnamefont {Madison}}, \bibinfo
  {author} {\bibfnamefont {R.~N.}\ \bibnamefont {Manchester}}, \bibinfo
  {author} {\bibfnamefont {C.~M.~F.}\ \bibnamefont {Mingarelli}}, \bibinfo
  {author} {\bibfnamefont {D.~J.}\ \bibnamefont {Nice}}, \bibinfo {author}
  {\bibfnamefont {S.}~\bibnamefont {Osłowski}}, \bibinfo {author}
  {\bibfnamefont {N.~T.}\ \bibnamefont {Palliyaguru}}, \bibinfo {author}
  {\bibfnamefont {T.~T.}\ \bibnamefont {Pennucci}}, \bibinfo {author}
  {\bibfnamefont {B.~B.~P.}\ \bibnamefont {Perera}}, \bibinfo {author}
  {\bibfnamefont {D.}~\bibnamefont {Perrodin}}, \bibinfo {author}
  {\bibfnamefont {A.}~\bibnamefont {Possenti}}, \bibinfo {author}
  {\bibfnamefont {A.}~\bibnamefont {Petiteau}}, \bibinfo {author}
  {\bibfnamefont {S.~M.}\ \bibnamefont {Ransom}}, \bibinfo {author}
  {\bibfnamefont {D.}~\bibnamefont {Reardon}}, \bibinfo {author} {\bibfnamefont
  {P.~A.}\ \bibnamefont {Rosado}}, \bibinfo {author} {\bibfnamefont {S.~A.}\
  \bibnamefont {Sanidas}}, \bibinfo {author} {\bibfnamefont {A.}~\bibnamefont
  {Sesana}}, \bibinfo {author} {\bibfnamefont {G.}~\bibnamefont {Shaifullah}},
  \bibinfo {author} {\bibfnamefont {R.~M.}\ \bibnamefont {Shannon}}, \bibinfo
  {author} {\bibfnamefont {X.}~\bibnamefont {Siemens}}, \bibinfo {author}
  {\bibfnamefont {J.}~\bibnamefont {Simon}}, \bibinfo {author} {\bibfnamefont
  {R.}~\bibnamefont {Smits}}, \bibinfo {author} {\bibfnamefont
  {R.}~\bibnamefont {Spiewak}}, \bibinfo {author} {\bibfnamefont {I.~H.}\
  \bibnamefont {Stairs}}, \bibinfo {author} {\bibfnamefont {B.~W.}\
  \bibnamefont {Stappers}}, \bibinfo {author} {\bibfnamefont {D.~R.}\
  \bibnamefont {Stinebring}}, \bibinfo {author} {\bibfnamefont
  {K.}~\bibnamefont {Stovall}}, \bibinfo {author} {\bibfnamefont {J.~K.}\
  \bibnamefont {Swiggum}}, \bibinfo {author} {\bibfnamefont {S.~R.}\
  \bibnamefont {Taylor}}, \bibinfo {author} {\bibfnamefont {G.}~\bibnamefont
  {Theureau}}, \bibinfo {author} {\bibfnamefont {C.}~\bibnamefont {Tiburzi}},
  \bibinfo {author} {\bibfnamefont {L.}~\bibnamefont {Toomey}}, \bibinfo
  {author} {\bibfnamefont {M.}~\bibnamefont {Vallisneri}}, \bibinfo {author}
  {\bibfnamefont {W.}~\bibnamefont {van Straten}}, \bibinfo {author}
  {\bibfnamefont {A.}~\bibnamefont {Vecchio}}, \bibinfo {author} {\bibfnamefont
  {Y.}~\bibnamefont {Wang}}, \bibinfo {author} {\bibfnamefont {L.}~\bibnamefont
  {Wen}}, \bibinfo {author} {\bibfnamefont {X.~P.}\ \bibnamefont {You}},
  \bibinfo {author} {\bibfnamefont {W.~W.}\ \bibnamefont {Zhu}}, \ and\
  \bibinfo {author} {\bibfnamefont {X.-J.}\ \bibnamefont {Zhu}},\ }\bibfield
  {title} {\enquote {\bibinfo {title} {{The International Pulsar Timing Array:
  First Data Release}},}\ }\href {\doibase 10.1093/mnras/stw347} {\bibfield
  {journal} {\bibinfo  {journal} {Mon. Not. Roy. Astron. Soc.}\ }\textbf
  {\bibinfo {volume} {458}},\ \bibinfo {pages} {1267--1288} (\bibinfo {year}
  {2016})},\ \Eprint {http://arxiv.org/abs/1602.03640} {arXiv:1602.03640
  [astro-ph.IM]} \BibitemShut {NoStop}%
\bibitem [{\citenamefont {Thrane}\ and\ \citenamefont
  {Romano}(2013{\natexlab{b}})}]{Thrane:2013PI}%
  \BibitemOpen
  \bibfield  {author} {\bibinfo {author} {\bibfnamefont {Eric}\ \bibnamefont
  {Thrane}}\ and\ \bibinfo {author} {\bibfnamefont {Joseph~D.}\ \bibnamefont
  {Romano}},\ }\href@noop {} {\enquote {\bibinfo {title} {{Sensitivity curves
  for searches for gravitational-wave backgrounds}},}\ }\bibinfo {howpublished}
  {\url{https://dcc.ligo.org/LIGO-P1300115/public}} (\bibinfo {year}
  {2013}{\natexlab{b}})\BibitemShut {NoStop}%
\bibitem [{\citenamefont {Amaro-Seoane}\ \emph {et~al.}(2017)\citenamefont
  {Amaro-Seoane} \emph {et~al.}}]{Audley:2017drz}%
  \BibitemOpen
  \bibfield  {author} {\bibinfo {author} {\bibfnamefont {Pau}\ \bibnamefont
  {Amaro-Seoane}} \emph {et~al.} (\bibinfo {collaboration} {LISA}),\ }\bibfield
   {title} {\enquote {\bibinfo {title} {{Laser Interferometer Space
  Antenna}},}\ }\href@noop {} {\  (\bibinfo {year} {2017})},\ \Eprint
  {http://arxiv.org/abs/1702.00786} {arXiv:1702.00786 [astro-ph.IM]}
  \BibitemShut {NoStop}%
\bibitem [{\citenamefont {Caprini}\ \emph {et~al.}(2019)\citenamefont
  {Caprini}, \citenamefont {Figueroa}, \citenamefont {Flauger}, \citenamefont
  {Nardini}, \citenamefont {Peloso}, \citenamefont {Pieroni}, \citenamefont
  {Ricciardone},\ and\ \citenamefont {Tasinato}}]{Caprini:2019pxz}%
  \BibitemOpen
  \bibfield  {author} {\bibinfo {author} {\bibfnamefont {Chiara}\ \bibnamefont
  {Caprini}}, \bibinfo {author} {\bibfnamefont {Daniel~G.}\ \bibnamefont
  {Figueroa}}, \bibinfo {author} {\bibfnamefont {Raphael}\ \bibnamefont
  {Flauger}}, \bibinfo {author} {\bibfnamefont {Germano}\ \bibnamefont
  {Nardini}}, \bibinfo {author} {\bibfnamefont {Marco}\ \bibnamefont {Peloso}},
  \bibinfo {author} {\bibfnamefont {Mauro}\ \bibnamefont {Pieroni}}, \bibinfo
  {author} {\bibfnamefont {Angelo}\ \bibnamefont {Ricciardone}}, \ and\
  \bibinfo {author} {\bibfnamefont {Gianmassimo}\ \bibnamefont {Tasinato}},\
  }\bibfield  {title} {\enquote {\bibinfo {title} {{Reconstructing the spectral
  shape of a stochastic gravitational wave background with LISA}},}\ }\href
  {\doibase 10.1088/1475-7516/2019/11/017} {\bibfield  {journal} {\bibinfo
  {journal} {JCAP}\ }\textbf {\bibinfo {volume} {11}},\ \bibinfo {pages} {017}
  (\bibinfo {year} {2019})},\ \Eprint {http://arxiv.org/abs/1906.09244}
  {arXiv:1906.09244 [astro-ph.CO]} \BibitemShut {NoStop}%
\bibitem [{\citenamefont {Smith}\ and\ \citenamefont
  {Caldwell}(2019)}]{Smith:2019wny}%
  \BibitemOpen
  \bibfield  {author} {\bibinfo {author} {\bibfnamefont {Tristan~L.}\
  \bibnamefont {Smith}}\ and\ \bibinfo {author} {\bibfnamefont {Robert}\
  \bibnamefont {Caldwell}},\ }\bibfield  {title} {\enquote {\bibinfo {title}
  {{LISA for Cosmologists: Calculating the Signal-to-Noise Ratio for Stochastic
  and Deterministic Sources}},}\ }\href {\doibase 10.1103/PhysRevD.100.104055}
  {\bibfield  {journal} {\bibinfo  {journal} {Phys. Rev. D}\ }\textbf {\bibinfo
  {volume} {100}},\ \bibinfo {pages} {104055} (\bibinfo {year} {2019})},\
  \Eprint {http://arxiv.org/abs/1908.00546} {arXiv:1908.00546 [astro-ph.CO]}
  \BibitemShut {NoStop}%
\bibitem [{\citenamefont {Punturo}\ \emph {et~al.}(2010)\citenamefont {Punturo}
  \emph {et~al.}}]{Punturo:2010zz}%
  \BibitemOpen
  \bibfield  {author} {\bibinfo {author} {\bibfnamefont {M.}~\bibnamefont
  {Punturo}} \emph {et~al.},\ }\bibfield  {title} {\enquote {\bibinfo {title}
  {{The Einstein Telescope: A third-generation gravitational wave
  observatory}},}\ }\href {\doibase 10.1088/0264-9381/27/19/194002} {\bibfield
  {journal} {\bibinfo  {journal} {Class. Quant. Grav.}\ }\textbf {\bibinfo
  {volume} {27}},\ \bibinfo {pages} {194002} (\bibinfo {year}
  {2010})}\BibitemShut {NoStop}%
\bibitem [{\citenamefont {{Reitze}}\ \emph {et~al.}(2019)\citenamefont
  {{Reitze}}, \citenamefont {{Adhikari}}, \citenamefont {{Ballmer}},
  \citenamefont {{Barish}}, \citenamefont {{Barsotti}}, \citenamefont
  {{Billingsley}}, \citenamefont {{Brown}}, \citenamefont {{Chen}},
  \citenamefont {{Coyne}}, \citenamefont {{Eisenstein}}, \citenamefont
  {{Evans}}, \citenamefont {{Fritschel}}, \citenamefont {{Hall}}, \citenamefont
  {{Lazzarini}}, \citenamefont {{Lovelace}}, \citenamefont {{Read}},
  \citenamefont {{Sathyaprakash}}, \citenamefont {{Shoemaker}}, \citenamefont
  {{Smith}}, \citenamefont {{Torrie}}, \citenamefont {{Vitale}}, \citenamefont
  {{Weiss}}, \citenamefont {{Wipf}},\ and\ \citenamefont
  {{Zucker}}}]{Reitze:2019iox}%
  \BibitemOpen
  \bibfield  {author} {\bibinfo {author} {\bibfnamefont {David}\ \bibnamefont
  {{Reitze}}}, \bibinfo {author} {\bibfnamefont {Rana~X.}\ \bibnamefont
  {{Adhikari}}}, \bibinfo {author} {\bibfnamefont {Stefan}\ \bibnamefont
  {{Ballmer}}}, \bibinfo {author} {\bibfnamefont {Barry}\ \bibnamefont
  {{Barish}}}, \bibinfo {author} {\bibfnamefont {Lisa}\ \bibnamefont
  {{Barsotti}}}, \bibinfo {author} {\bibfnamefont {GariLynn}\ \bibnamefont
  {{Billingsley}}}, \bibinfo {author} {\bibfnamefont {Duncan~A.}\ \bibnamefont
  {{Brown}}}, \bibinfo {author} {\bibfnamefont {Yanbei}\ \bibnamefont
  {{Chen}}}, \bibinfo {author} {\bibfnamefont {Dennis}\ \bibnamefont
  {{Coyne}}}, \bibinfo {author} {\bibfnamefont {Robert}\ \bibnamefont
  {{Eisenstein}}}, \bibinfo {author} {\bibfnamefont {Matthew}\ \bibnamefont
  {{Evans}}}, \bibinfo {author} {\bibfnamefont {Peter}\ \bibnamefont
  {{Fritschel}}}, \bibinfo {author} {\bibfnamefont {Evan~D.}\ \bibnamefont
  {{Hall}}}, \bibinfo {author} {\bibfnamefont {Albert}\ \bibnamefont
  {{Lazzarini}}}, \bibinfo {author} {\bibfnamefont {Geoffrey}\ \bibnamefont
  {{Lovelace}}}, \bibinfo {author} {\bibfnamefont {Jocelyn}\ \bibnamefont
  {{Read}}}, \bibinfo {author} {\bibfnamefont {B.~S.}\ \bibnamefont
  {{Sathyaprakash}}}, \bibinfo {author} {\bibfnamefont {David}\ \bibnamefont
  {{Shoemaker}}}, \bibinfo {author} {\bibfnamefont {Joshua}\ \bibnamefont
  {{Smith}}}, \bibinfo {author} {\bibfnamefont {Calum}\ \bibnamefont
  {{Torrie}}}, \bibinfo {author} {\bibfnamefont {Salvatore}\ \bibnamefont
  {{Vitale}}}, \bibinfo {author} {\bibfnamefont {Rainer}\ \bibnamefont
  {{Weiss}}}, \bibinfo {author} {\bibfnamefont {Christopher}\ \bibnamefont
  {{Wipf}}}, \ and\ \bibinfo {author} {\bibfnamefont {Michael}\ \bibnamefont
  {{Zucker}}},\ }\bibfield  {title} {\enquote {\bibinfo {title} {{Cosmic
  Explorer: The U.S. Contribution to Gravitational-Wave Astronomy beyond
  LIGO}},}\ }\href@noop {} {\bibfield  {journal} {\bibinfo  {journal} {Bull.
  Am. Astron. Soc.}\ }\textbf {\bibinfo {volume} {51}},\ \bibinfo {pages} {035}
  (\bibinfo {year} {2019})},\ \Eprint {http://arxiv.org/abs/1907.04833}
  {arXiv:1907.04833 [astro-ph.IM]} \BibitemShut {NoStop}%
\bibitem [{\citenamefont {Hild}\ \emph {et~al.}(2011)\citenamefont {Hild} \emph
  {et~al.}}]{Hild:2010id}%
  \BibitemOpen
  \bibfield  {author} {\bibinfo {author} {\bibfnamefont {S.}~\bibnamefont
  {Hild}} \emph {et~al.},\ }\bibfield  {title} {\enquote {\bibinfo {title}
  {{Sensitivity Studies for Third-Generation Gravitational Wave
  Observatories}},}\ }\href {\doibase 10.1088/0264-9381/28/9/094013} {\bibfield
   {journal} {\bibinfo  {journal} {Class. Quant. Grav.}\ }\textbf {\bibinfo
  {volume} {28}},\ \bibinfo {pages} {094013} (\bibinfo {year} {2011})},\
  \Eprint {http://arxiv.org/abs/1012.0908} {arXiv:1012.0908 [gr-qc]}
  \BibitemShut {NoStop}%
\bibitem [{\citenamefont {{Hall}}\ \emph {et~al.}(2021)\citenamefont {{Hall}},
  \citenamefont {{Kuns}}, \citenamefont {{Smith}}, \citenamefont {{Bai}},
  \citenamefont {{Wipf}}, \citenamefont {{Biscans}}, \citenamefont
  {{Adhikari}}, \citenamefont {{Arai}}, \citenamefont {{Ballmer}},
  \citenamefont {{Barsotti}}, \citenamefont {{Chen}}, \citenamefont {{Evans}},
  \citenamefont {{Fritschel}}, \citenamefont {{Harms}}, \citenamefont
  {{Kamai}}, \citenamefont {{Graef Rollins}}, \citenamefont {{Shoemaker}},
  \citenamefont {{Slagmolen}}, \citenamefont {{Weiss}},\ and\ \citenamefont
  {{Yamamoto}}}]{Hall:2020dps}%
  \BibitemOpen
  \bibfield  {author} {\bibinfo {author} {\bibfnamefont {Evan~D.}\ \bibnamefont
  {{Hall}}}, \bibinfo {author} {\bibfnamefont {Kevin}\ \bibnamefont {{Kuns}}},
  \bibinfo {author} {\bibfnamefont {Joshua~R.}\ \bibnamefont {{Smith}}},
  \bibinfo {author} {\bibfnamefont {Yuntao}\ \bibnamefont {{Bai}}}, \bibinfo
  {author} {\bibfnamefont {Christopher}\ \bibnamefont {{Wipf}}}, \bibinfo
  {author} {\bibfnamefont {Sebastien}\ \bibnamefont {{Biscans}}}, \bibinfo
  {author} {\bibfnamefont {Rana~X}\ \bibnamefont {{Adhikari}}}, \bibinfo
  {author} {\bibfnamefont {Koji}\ \bibnamefont {{Arai}}}, \bibinfo {author}
  {\bibfnamefont {Stefan}\ \bibnamefont {{Ballmer}}}, \bibinfo {author}
  {\bibfnamefont {Lisa}\ \bibnamefont {{Barsotti}}}, \bibinfo {author}
  {\bibfnamefont {Yanbei}\ \bibnamefont {{Chen}}}, \bibinfo {author}
  {\bibfnamefont {Matthew}\ \bibnamefont {{Evans}}}, \bibinfo {author}
  {\bibfnamefont {Peter}\ \bibnamefont {{Fritschel}}}, \bibinfo {author}
  {\bibfnamefont {Jan}\ \bibnamefont {{Harms}}}, \bibinfo {author}
  {\bibfnamefont {Brittany}\ \bibnamefont {{Kamai}}}, \bibinfo {author}
  {\bibfnamefont {Jameson}\ \bibnamefont {{Graef Rollins}}}, \bibinfo {author}
  {\bibfnamefont {David}\ \bibnamefont {{Shoemaker}}}, \bibinfo {author}
  {\bibfnamefont {Bram}\ \bibnamefont {{Slagmolen}}}, \bibinfo {author}
  {\bibfnamefont {Rainer}\ \bibnamefont {{Weiss}}}, \ and\ \bibinfo {author}
  {\bibfnamefont {Hiro}\ \bibnamefont {{Yamamoto}}},\ }\bibfield  {title}
  {\enquote {\bibinfo {title} {{Gravitational-wave physics with Cosmic
  Explorer: Limits to low-frequency sensitivity}},}\ }\href {\doibase
  10.1103/PhysRevD.103.122004} {\bibfield  {journal} {\bibinfo  {journal}
  {Phys. Rev. D}\ }\textbf {\bibinfo {volume} {103}},\ \bibinfo {pages}
  {122004} (\bibinfo {year} {2021})},\ \Eprint
  {http://arxiv.org/abs/2012.03608} {arXiv:2012.03608 [gr-qc]} \BibitemShut
  {NoStop}%
\bibitem [{\citenamefont {Ringeval}\ \emph {et~al.}(2007)\citenamefont
  {Ringeval}, \citenamefont {Sakellariadou},\ and\ \citenamefont
  {Bouchet}}]{Ringeval:2005kr}%
  \BibitemOpen
  \bibfield  {author} {\bibinfo {author} {\bibfnamefont {Christophe}\
  \bibnamefont {Ringeval}}, \bibinfo {author} {\bibfnamefont {Mairi}\
  \bibnamefont {Sakellariadou}}, \ and\ \bibinfo {author} {\bibfnamefont
  {Francois}\ \bibnamefont {Bouchet}},\ }\bibfield  {title} {\enquote {\bibinfo
  {title} {{Cosmological evolution of cosmic string loops}},}\ }\href {\doibase
  10.1088/1475-7516/2007/02/023} {\bibfield  {journal} {\bibinfo  {journal}
  {JCAP}\ }\textbf {\bibinfo {volume} {02}},\ \bibinfo {pages} {023} (\bibinfo
  {year} {2007})},\ \Eprint {http://arxiv.org/abs/astro-ph/0511646}
  {arXiv:astro-ph/0511646} \BibitemShut {NoStop}%
\bibitem [{\citenamefont {Lorenz}\ \emph {et~al.}(2010)\citenamefont {Lorenz},
  \citenamefont {Ringeval},\ and\ \citenamefont
  {Sakellariadou}}]{Lorenz:2010sm}%
  \BibitemOpen
  \bibfield  {author} {\bibinfo {author} {\bibfnamefont {Larissa}\ \bibnamefont
  {Lorenz}}, \bibinfo {author} {\bibfnamefont {Christophe}\ \bibnamefont
  {Ringeval}}, \ and\ \bibinfo {author} {\bibfnamefont {Mairi}\ \bibnamefont
  {Sakellariadou}},\ }\bibfield  {title} {\enquote {\bibinfo {title} {{Cosmic
  string loop distribution on all length scales and at any redshift}},}\ }\href
  {\doibase 10.1088/1475-7516/2010/10/003} {\bibfield  {journal} {\bibinfo
  {journal} {JCAP}\ }\textbf {\bibinfo {volume} {10}},\ \bibinfo {pages} {003}
  (\bibinfo {year} {2010})},\ \Eprint {http://arxiv.org/abs/1006.0931}
  {arXiv:1006.0931 [astro-ph.CO]} \BibitemShut {NoStop}%
\bibitem [{\citenamefont {Auclair}\ \emph
  {et~al.}(2020{\natexlab{a}})\citenamefont {Auclair}, \citenamefont
  {Blanco-Pillado}, \citenamefont {Figueroa}, \citenamefont {Jenkins},
  \citenamefont {Lewicki}, \citenamefont {Sakellariadou}, \citenamefont
  {Sanidas}, \citenamefont {Sousa}, \citenamefont {Steer}, \citenamefont
  {Wachter},\ and\ \citenamefont {Kuroyanagi}}]{Auclair:2019wcv}%
  \BibitemOpen
  \bibfield  {author} {\bibinfo {author} {\bibfnamefont {Pierre}\ \bibnamefont
  {Auclair}}, \bibinfo {author} {\bibfnamefont {Jose~J.}\ \bibnamefont
  {Blanco-Pillado}}, \bibinfo {author} {\bibfnamefont {Daniel~G.}\ \bibnamefont
  {Figueroa}}, \bibinfo {author} {\bibfnamefont {Alexander~C.}\ \bibnamefont
  {Jenkins}}, \bibinfo {author} {\bibfnamefont {Marek}\ \bibnamefont
  {Lewicki}}, \bibinfo {author} {\bibfnamefont {Mairi}\ \bibnamefont
  {Sakellariadou}}, \bibinfo {author} {\bibfnamefont {Sotiris}\ \bibnamefont
  {Sanidas}}, \bibinfo {author} {\bibfnamefont {Lara}\ \bibnamefont {Sousa}},
  \bibinfo {author} {\bibfnamefont {Danièle~A.}\ \bibnamefont {Steer}},
  \bibinfo {author} {\bibfnamefont {Jeremy~M.}\ \bibnamefont {Wachter}}, \ and\
  \bibinfo {author} {\bibfnamefont {Sachiko}\ \bibnamefont {Kuroyanagi}},\
  }\bibfield  {title} {\enquote {\bibinfo {title} {{Probing the gravitational
  wave background from cosmic strings with LISA}},}\ }\href {\doibase
  10.1088/1475-7516/2020/04/034} {\bibfield  {journal} {\bibinfo  {journal}
  {JCAP}\ }\textbf {\bibinfo {volume} {04}},\ \bibinfo {pages} {034} (\bibinfo
  {year} {2020}{\natexlab{a}})},\ \Eprint {http://arxiv.org/abs/1909.00819}
  {arXiv:1909.00819 [astro-ph.CO]} \BibitemShut {NoStop}%
\bibitem [{\citenamefont {Vilenkin}(1981)}]{Vilenkin:1981bx}%
  \BibitemOpen
  \bibfield  {author} {\bibinfo {author} {\bibfnamefont {A.}~\bibnamefont
  {Vilenkin}},\ }\bibfield  {title} {\enquote {\bibinfo {title} {{Gravitational
  radiation from cosmic strings}},}\ }\href {\doibase
  10.1016/0370-2693(81)91144-8} {\bibfield  {journal} {\bibinfo  {journal}
  {Phys. Lett. B}\ }\textbf {\bibinfo {volume} {107}},\ \bibinfo {pages}
  {47--50} (\bibinfo {year} {1981})}\BibitemShut {NoStop}%
\bibitem [{\citenamefont {Allen}(1996)}]{Allen:1996vm}%
  \BibitemOpen
  \bibfield  {author} {\bibinfo {author} {\bibfnamefont {Bruce}\ \bibnamefont
  {Allen}},\ }\bibfield  {title} {\enquote {\bibinfo {title} {{The Stochastic
  gravity wave background: Sources and detection}},}\ }in\ \href@noop {} {\emph
  {\bibinfo {booktitle} {{Les Houches School of Physics: Astrophysical Sources
  of Gravitational Radiation}}}}\ (\bibinfo {year} {1996})\ pp.\ \bibinfo
  {pages} {373--417},\ \Eprint {http://arxiv.org/abs/gr-qc/9604033}
  {arXiv:gr-qc/9604033} \BibitemShut {NoStop}%
\bibitem [{\citenamefont {Maggiore}(2000)}]{Maggiore:1999vm}%
  \BibitemOpen
  \bibfield  {author} {\bibinfo {author} {\bibfnamefont {Michele}\ \bibnamefont
  {Maggiore}},\ }\bibfield  {title} {\enquote {\bibinfo {title} {{Gravitational
  wave experiments and early universe cosmology}},}\ }\href {\doibase
  10.1016/S0370-1573(99)00102-7} {\bibfield  {journal} {\bibinfo  {journal}
  {Phys. Rept.}\ }\textbf {\bibinfo {volume} {331}},\ \bibinfo {pages}
  {283--367} (\bibinfo {year} {2000})},\ \Eprint
  {http://arxiv.org/abs/gr-qc/9909001} {arXiv:gr-qc/9909001} \BibitemShut
  {NoStop}%
\bibitem [{\citenamefont {Siemens}\ \emph {et~al.}(2007)\citenamefont
  {Siemens}, \citenamefont {Mandic},\ and\ \citenamefont
  {Creighton}}]{Siemens:2006yp}%
  \BibitemOpen
  \bibfield  {author} {\bibinfo {author} {\bibfnamefont {Xavier}\ \bibnamefont
  {Siemens}}, \bibinfo {author} {\bibfnamefont {Vuk}\ \bibnamefont {Mandic}}, \
  and\ \bibinfo {author} {\bibfnamefont {Jolien}\ \bibnamefont {Creighton}},\
  }\bibfield  {title} {\enquote {\bibinfo {title} {{Gravitational wave
  stochastic background from cosmic (super)strings}},}\ }\href {\doibase
  10.1103/PhysRevLett.98.111101} {\bibfield  {journal} {\bibinfo  {journal}
  {Phys. Rev. Lett.}\ }\textbf {\bibinfo {volume} {98}},\ \bibinfo {pages}
  {111101} (\bibinfo {year} {2007})},\ \Eprint
  {http://arxiv.org/abs/astro-ph/0610920} {arXiv:astro-ph/0610920} \BibitemShut
  {NoStop}%
\bibitem [{\citenamefont {Caprini}\ and\ \citenamefont
  {Figueroa}(2018)}]{Caprini:2018mtu}%
  \BibitemOpen
  \bibfield  {author} {\bibinfo {author} {\bibfnamefont {Chiara}\ \bibnamefont
  {Caprini}}\ and\ \bibinfo {author} {\bibfnamefont {Daniel~G.}\ \bibnamefont
  {Figueroa}},\ }\bibfield  {title} {\enquote {\bibinfo {title} {{Cosmological
  Backgrounds of Gravitational Waves}},}\ }\href {\doibase
  10.1088/1361-6382/aac608} {\bibfield  {journal} {\bibinfo  {journal} {Class.
  Quant. Grav.}\ }\textbf {\bibinfo {volume} {35}},\ \bibinfo {pages} {163001}
  (\bibinfo {year} {2018})},\ \Eprint {http://arxiv.org/abs/1801.04268}
  {arXiv:1801.04268 [astro-ph.CO]} \BibitemShut {NoStop}%
\bibitem [{\citenamefont {Jenkins}\ and\ \citenamefont
  {Sakellariadou}(2018)}]{Jenkins:2018nty}%
  \BibitemOpen
  \bibfield  {author} {\bibinfo {author} {\bibfnamefont {Alexander~C.}\
  \bibnamefont {Jenkins}}\ and\ \bibinfo {author} {\bibfnamefont {Mairi}\
  \bibnamefont {Sakellariadou}},\ }\bibfield  {title} {\enquote {\bibinfo
  {title} {{Anisotropies in the stochastic gravitational-wave background:
  Formalism and the cosmic string case}},}\ }\href {\doibase
  10.1103/PhysRevD.98.063509} {\bibfield  {journal} {\bibinfo  {journal} {Phys.
  Rev. D}\ }\textbf {\bibinfo {volume} {98}},\ \bibinfo {pages} {063509}
  (\bibinfo {year} {2018})},\ \Eprint {http://arxiv.org/abs/1802.06046}
  {arXiv:1802.06046 [astro-ph.CO]} \BibitemShut {NoStop}%
\bibitem [{\citenamefont {Auclair}\ \emph
  {et~al.}(2020{\natexlab{b}})\citenamefont {Auclair}, \citenamefont {Steer},\
  and\ \citenamefont {Vachaspati}}]{Auclair:2019jip}%
  \BibitemOpen
  \bibfield  {author} {\bibinfo {author} {\bibfnamefont {Pierre}\ \bibnamefont
  {Auclair}}, \bibinfo {author} {\bibfnamefont {Dani\`ele~A.}\ \bibnamefont
  {Steer}}, \ and\ \bibinfo {author} {\bibfnamefont {Tanmay}\ \bibnamefont
  {Vachaspati}},\ }\bibfield  {title} {\enquote {\bibinfo {title} {{Particle
  emission and gravitational radiation from cosmic strings: observational
  constraints}},}\ }\href {\doibase 10.1103/PhysRevD.101.083511} {\bibfield
  {journal} {\bibinfo  {journal} {Phys. Rev. D}\ }\textbf {\bibinfo {volume}
  {101}},\ \bibinfo {pages} {083511} (\bibinfo {year} {2020}{\natexlab{b}})},\
  \Eprint {http://arxiv.org/abs/1911.12066} {arXiv:1911.12066 [hep-ph]}
  \BibitemShut {NoStop}%
\end{thebibliography}%
\end{document}